\documentclass[english,10pt]{article}
\usepackage[utf8]{inputenc}
\usepackage{graphicx}
\usepackage{color}
\usepackage{float}
\usepackage{slashed}
\usepackage{subfigure}
\usepackage{amsthm,amsmath,amssymb,mathrsfs}
\usepackage{setspace}
\allowdisplaybreaks
\usepackage{jheppub}
\usepackage{xcolor} 
\usepackage{babel}
\usepackage{booktabs}
\usepackage{mathtools}
\usepackage{braket}

\usepackage[normalem]{ulem}

\newcommand{\veck}{\mathbf{k}}

\newcommand{\vecp}{\mathbf{p}}
\newcommand{\vecph}{\mathbf{\hat{p}}}
\newcommand{\vecq}{\mathbf{q}}
\newcommand{\vecqh}{\mathbf{\hat{q}}}
\newcommand{\vecr}{\mathbf{r}}
\newcommand{\vecrh}{\mathbf{\hat{r}}}
\newcommand{\alphaeff}{\alpha_{\rm eff}}
\usepackage[compat=1.1.0]{tikz-feynman}
\begin{document}

\title{Sommerfeld Enhancement from Quantum Forces for Dark Matter}

\author{Steven Ferrante,}
\emailAdd{sef87@cornell.edu}
\author{Maxim Perelstein,}
\emailAdd{m.perelstein@cornell.edu}
\author{Bingrong Yu}
\emailAdd{bingrong.yu@cornell.edu}
\affiliation{Department of Physics, LEPP, Cornell University, Ithaca, NY 14853, USA}

\abstract{Quantum forces are long-range interactions that arise only at the loop level. In this work, we study the Sommerfeld enhancement of dark matter (DM) annihilation cross sections caused by quantum forces.
One notable feature of quantum forces is that they are subject to coherent enhancement in the presence of a background of mediator particles, which occurs in many situations in cosmology. We show that this effect
has important implications for the Sommerfeld enhancement and DM physics. 
For the first time, we calculate the Sommerfeld factor induced by quantum forces for both bosonic and fermionic mediators, including the background corrections. We observe several novel features of the Sommerfeld factor that do not exist in the case of the Yukawa potential, such as temperature-induced resonance peaks for massless mediators, and having both enhancement and suppression effects in the same model with different DM masses. As direct applications, we discuss the DM phenomenology affected by the Sommerfeld enhancement from quantum forces, including thermal freeze-out, CMB spectral distortion from DM annihilation, and DM indirect detection. We highlight one particularly interesting effect relevant to indirect detection caused by the Sommerfeld enhancement in a non-thermal background of bosonic mediators in the galaxy, in which case the DM mass is shifted due to the background correction and the effective cross section for DM annihilation can be either enhanced or suppressed. This may be important for DM searches in the Milky Way or its satellite galaxies.
}

\maketitle

\section{Introduction}
\label{sec:intro}

Long-range interactions exist in the Standard Model (SM) as well as in many new physics theories beyond the SM. A long-range interaction between two particles that approach each other with non-relativistic velocity will deform the particle wave functions and thus significantly change the cross section of the particle scattering. This non-perturbative effect is known as the Sommerfeld enhancement~\cite{Sommerfeld:1931qaf}.

In the past two decades, the Sommerfeld enhancement has been shown to have important implications in dark matter (DM) physics~\cite{Hisano:2002fk,Hisano:2003ec,Hisano:2004ds,Hisano:2005ec,Hisano:2006nn,Cirelli:2008pk,Arkani-Hamed:2008hhe,Pospelov:2008jd,Kamionkowski:2008gj,Fox:2008kb,Lattanzi:2008qa,Iengo:2009ni,Iengo:2009xf,Cassel:2009wt,Bedaque:2009ri,Dent:2009bv,Zavala:2009mi,Slatyer:2009vg,Feng:2009hw,Feng:2010zp,Hannestad:2010zt,Iminniyaz:2010hy,McDonald:2012nc,Liu:2013vha,Bellazzini:2013foa,Blum:2016nrz,Ovanesyan:2016vkk,Kahlhoefer:2017umn,Baumgart:2017nsr,Agrawal:2020lea,Chaffey:2021tmj,Beneke:2019qaa,Beneke:2020vff,Urban:2021cdu,Coy:2022cpt,Biondini:2023zcz,Bottaro:2023wjv,Biondini:2024aan,Parikh:2024mwa,Biondini:2025jvp}. Generally speaking, when two DM particles move slowly, the interaction between them can be factorized into two parts: the short-range part that can be computed using perturbative quantum field theory and the long-range part described by the wave functions.\footnote{This factorization is a general result of non-relativistic effective field theory~\cite{Bodwin:1994jh}, see \cite{Hisano:2002fk,Hisano:2003ec,Hisano:2004ds} for the first application in the context of neutralino DM.} When there exist attractive (repulsive) long-range interactions, the incoming wave functions will be deformed from a plane wave, causing an enhancement (suppression) of the DM cross section. The contribution from the long-range part is non-perturbative because, at low velocities, the interaction energy is not necessarily small compared to the kinetic energy. Technically, this non-perturbative effect can be captured by resumming the ladder diagrams with infinite particle exchanges; alternatively, it can also be calculated by solving the Schr\"{o}dinger equation with the long-range non-relativistic potential taken into account.

\begin{figure}[t]
\centering
\includegraphics[scale=0.45]
{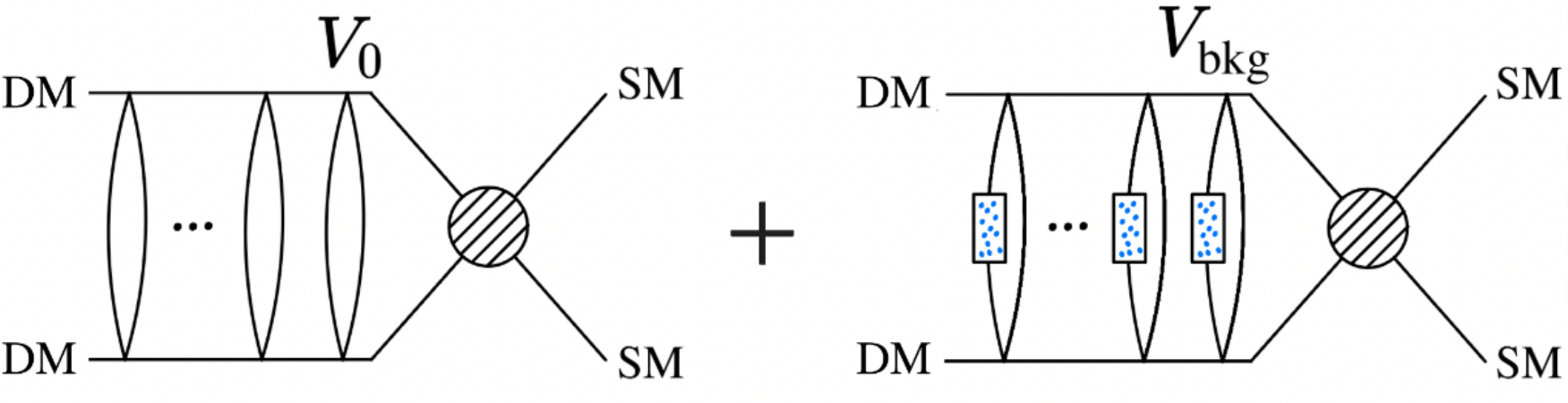}
\caption{\label{fig:Schematic} A schematic diagram of the Sommerfeld enhancement induced by quantum forces. \emph{Left}: Quantum forces in vacuum. \emph{Right}: Quantum forces in the presence of a background of mediator particles. The small blue dots on the internal lines represent the background particles, which is equivalent to putting one of the propagators on-shell. When there exists a background, the total effective potential that contributes to the Sommerfeld enhancement is given by the sum of the vacuum potential $V_0$ and the background potential $V_{\rm bkg}$.}
\end{figure}

The Sommerfeld enhancement with a tree-level Yukawa potential (classical force) has been studied extensively; see, e.g., \cite{Iengo:2009ni,Iengo:2009xf,Cassel:2009wt} for the calculation of arbitrary partial waves. In this work, we go beyond the tree-level exchange potential and study the Sommerfeld enhancement from quantum forces (see the left panel of Fig.~\ref{fig:Schematic}), that is, the long-range potentials generated at the one-loop level via exchanging two mediators~\cite{Fichet:2017bng,Brax:2017xho}. Naively, this effect seems to be suppressed compared to the tree-level exchange. For example, the one-loop radiative correction to the Yukawa potential for the Sommerfeld enhancement in the context of certain ultraviolet (UV) models was studied in \cite{Beneke:2019qaa,Beneke:2020vff,Urban:2021cdu,Coy:2022cpt,Bottaro:2023wjv,Parikh:2024mwa}, where the typical effect of the one-loop correction was found to be at the percent level. However, in the following, we highlight two crucial differences between the quantum forces studied in this work and the one-loop correction in  \cite{Beneke:2019qaa,Beneke:2020vff,Urban:2021cdu,Coy:2022cpt,Bottaro:2023wjv,Parikh:2024mwa}, which also comprise the motivations of this work.

First, there are many simple scenarios where the leading couplings between DM and mediator particles are quadratic, such as the scalar mediator with a $\mathbb{Z}_2$ symmetry $\bar{\chi}\chi \phi^2$, the fermionic mediator $\bar{\chi}\chi \bar{\psi}\psi$, and the  spin-independent component of the interaction with a pseudo-scalar mediator $\bar{\chi}\chi (\partial_\mu a)^2$ (note that the one-pseudo-scalar exchange potential is always spin-dependent). Here, $\chi$ is the DM particle, $\phi$, $\psi$, and $a$ denote mediator particles with different quantum numbers, and we have suppressed the possible Lorentz structure for brevity. In these scenarios, there is no relevant tree-level potential as a result of some conserved quantum numbers, and the quantum forces at the one-loop level are the leading long-range interactions between two DM particles. Therefore, the quantum forces considered in this work are not necessarily radiative corrections to tree-level potentials. In particular, our focus is to investigate how the \emph{long-range} part of quantum forces can contribute to the Sommerfeld effect. To this end, we perform calculations starting from a set of effective operators, which makes our results insensitive to the unknown UV physics in certain regions of parameter space.

Second, quantum forces have a generic feature that does not exist for tree-level potentials: they can be coherently enhanced in the presence of a background of mediator particles~\cite{Horowitz:1993kw,Ferrer:1998ju,Ferrer:1998rw,Ferrer:1999ad,Ferrer:2000hm,Hees:2018fpg,Fukuda:2021drn,Banerjee:2022sqg,Ghosh:2022nzo,VanTilburg:2024xib,Barbosa:2024pkl,Ghosh:2024qai,Grossman:2025cov,Cheng:2025fak,Gan:2025nlu}. This can be understood in the following way. In vacuum, the interaction between two $\chi$ particles is mediated by exchanging two mediators, both of which are off-shell. However, when there exists a background of mediators, one can put one of the mediators on-shell, and the interaction can also be mediated through the coherent scattering between $\chi$ and background particles (see the right panel of Fig.~\ref{fig:Schematic}). As a result, the effective potential in the background contains two terms:
\begin{align}
    V_{\rm eff}(r) = V_0(r) +V_{\rm bkg}(r)\;,
    \label{eq:Veff}
\end{align}
where $V_0$ is the quantum force in vacuum and $V_{\rm bkg}$ denotes the contribution from the background. Note that $V_{\rm bkg}$ is proportional to the number density of background particles, indicating that this is a coherent effect. Moreover, due to the effective regularization from the background, $V_{\rm bkg}$ always scales as $1/r$ in the coherent region where $r$ is small compared to the wavelength of background particles~\cite{Ghosh:2022nzo,VanTilburg:2024xib}. 
This effect can be very significant if background particles are in a condensate state that has a large occupation number~\cite{VanTilburg:2024xib,Grossman:2025cov}. There has been growing interest in studying the background-induced force $V_{\rm bkg}$ mediated by light mediators in recent years~\cite{Ghosh:2022nzo,VanTilburg:2024xib,Barbosa:2024pkl,Ghosh:2024qai,Grossman:2025cov,Cheng:2025fak,Gan:2025nlu}. While these papers mostly focused on the implications for fifth-force detection experiments, the study of this effect in the early universe cosmology is still lacking. However, it is possible that during the cosmic evolution of $\chi$, it couples to some light mediator that has a large background at that time. In this case, the quantum force is significantly enhanced by the background, and so is the Sommerfeld factor. In this work, we study the Sommerfeld enhancement from $V_{\rm bkg}$ for the first time. We present a  comprehensive analysis of this effect, including both scenarios where the background is thermal or non-thermal.

The remaining part of this paper is organized as follows. In Sec.~\ref{sec:operators}, we classify the relevant effective operators with both scalar and fermionic mediators and calculate the quantum forces induced by them, including both the vacuum potential $V_0$ and the background potential $V_{\rm bkg}$. 
In Sec.~\ref{sec:vacuum}, we calculate the Sommerfeld factor from $V_0$, assuming there is no background. We study the background effect on Sommerfeld enhancement in the next three sections: the case of a thermal background with scalar and fermionic mediators is discussed in Sec.~\ref{sec:thermal-scalar} and Sec.~\ref{sec:thermal-fermion}, respectively, while the case of a non-thermal background is discussed in Sec.~\ref{sec:non-thermal}. In Sec.~\ref{sec:pheno}, we study some phenomenologically relevant effects in DM physics caused by the Sommerfeld effect from quantum forces. We summarize our main conclusions in Sec.~\ref{sec:conclusion}. We present some technical details in four appendices: Appendix~\ref{app:quantum-force} provides calculation details of the vacuum and background quantum forces; Appendix~\ref{app:S-calculation} reviews the general strategy to calculate the Sommerfeld factor; Appendix~\ref{app:unitarity} discusses the unitarity bound on the Sommerfeld factor from quantum forces; and in Appendix~\ref{app:regular-dependence} we show how the Sommerfeld enhancement from quantum forces depends on the regulator.

\section{Effective operators, quantum forces, and Sommerfeld review}
\label{sec:operators}
In this section, we specify the effective operators involved in this work and calculate the quantum forces induced from them. We also briefly review the method for calculating the Sommerfeld factor with a general potential.

\subsection{Effective operators}
A comprehensive classification of effective operators that can lead to quantum forces at the one-loop level can be found in \cite{Fichet:2017bng}. In this work, we focus on the following three typical operators:\footnote{The effective operators in Eqs.~(\ref{eq:OS})-(\ref{eq:OFtilde}) are responsible for the long-range part of the DM annihilation to SM particles (corresponding to the ladders in Fig.~\ref{fig:Schematic}); the short-range part (denoted by the shaded bubble in Fig.~\ref{fig:Schematic}), on the other hand, is determined by the interaction between DM and SM particles. These two parts of interaction do not necessarily share the same fundamental physics.}
\begin{align}
    \mathcal{O}_{S} &= 
    \frac{1}{\Lambda}
    \overline{\chi}\chi\frac{\phi^{2}}{2}\;,\label{eq:OS}
    \\
      \mathcal{O}_{F} &= 
    \frac{1}{\Lambda^{2}}
    \overline{\chi}\chi\frac{\overline{\psi}\psi}{2}\;,\label{eq:OF}
    \\
    \mathcal{O}_{\widetilde{F}} &= 
    \frac{1}{\Lambda^{2}}
    \overline{\chi}\gamma^\mu\chi\overline{\psi}\gamma_\mu\psi\;,\label{eq:OFtilde}
\end{align}
where $\Lambda$ is the cutoff scale, $\chi$ is a fermionic DM particle, while $\phi$ and $\psi$ denote the scalar and fermionic mediators, respectively.
To be more specific, we assume that $\phi$ is a real scalar in $\mathcal{O}_{S}$ and $\psi$ is a Majorana fermion in $\mathcal{O}_{F}$. The extension to the cases of complex scalar or Dirac fermion is trivial and only differs up to factors of 2. On the other hand, $\psi$ in $\mathcal{O}_{\widetilde{F}}$ can only be a Dirac fermion since $\overline{\psi}\gamma^\mu\psi\equiv 0$ for Majorana fermions. In all cases, $\chi$ is assumed to be a Dirac fermion. 

Since we are interested in the spin-independent part of the quantum forces in the non-relativistic limit, the bilinear form of the on-shell DM field can be simplified:
\begin{align}
\overline{u}_\chi u_\chi \approx 2m_\chi\;,\quad
\overline{u}_\chi\gamma^\mu u_\chi \approx 2m_\chi\left(1,\bf{0}\right),\label{eq:wavefunction}
\end{align}
where $u_\chi$ is the wavefunction part of $\chi$ that satisfies the momentum space Dirac equation and $m_\chi$ is the DM mass.

Note that our results for ${\cal O}_S$ and ${\cal O}_F$ can easily be mapped to the cases where the fermionic DM $\chi$ is replaced by a complex scalar DM $\Phi$:  
\begin{align}
{\cal O}_S' = \lambda' \left|\Phi\right|^2 \frac{\phi^2}{2}\;,\quad
{\cal O}_F' = \frac{1}{\Lambda'} \left|\Phi\right|^2 \frac{\overline{\psi}\psi}{2}\;.
\end{align}
To get the results for ${\cal O}_S'$ and ${\cal O}_F'$ from those for ${\cal O}_S$ and ${\cal O}_F$, one only needs to perform the replacement: $\Lambda \to 2m_\Phi/\lambda'$ and $\Lambda^2 \to 2m_\Phi \Lambda'$, where $m_\Phi$ is the mass of $\Phi$, $\lambda'$ is a dimensionless coupling, and $\Lambda'$ is the cutoff scale in ${\cal O}_F'$. 

In the remainder of this work, we will focus on the effective operators in Eqs.~(\ref{eq:OS})-(\ref{eq:OFtilde}).

\subsection{The modified propagator formalism}
\label{subsec:modifiedpropagator}
As shown in Eq.~(\ref{eq:Veff}), in the presence of a background of mediators, the quantum forces contributing to the Sommerfeld enhancement generally contain two terms: the vacuum potential $V_0$ and the background correction $V_{\rm bkg}$. Both $V_0$ and $V_{\rm bkg}$ induced from Eqs.~(\ref{eq:OS})-(\ref{eq:OFtilde}) have been calculated in the previous literature. $V_0$ with different spins of mediators were systematically studied in \cite{Fichet:2017bng,Brax:2017xho} (see also \cite{Ferrer:1998rw} for partial earlier results); for the calculation of $V_{\rm bkg}$, one can refer to e.g. \cite{Ghosh:2022nzo,VanTilburg:2024xib,Barbosa:2024pkl}. In the following, we quickly review the calculation of quantum forces using the formalism of modified propagators. Our goal is to provide a pedagogical derivation of both $V_0$ and $V_{\rm bkg}$ in a unified and field-theoretical framework. More details can be found in Appendix~\ref{app:quantum-force}.

The general method to derive the effective potential is first to calculate the $t$-channel elastic scattering amplitude $\chi(p_1) +\chi(p_2) \to \chi (p_1') + \chi(p_2')$, and then map it onto the potential in the non-relativistic limit via Fourier transform:
\begin{align}
V_{\rm eff}(r) = - \frac{1}{4m_\chi^2}\int \frac{{\rm d}^3\vecq}{\left(2\pi\right)^3}e^{i\vecq\cdot\vecr}{\cal M}_{\rm NR}\;,\label{eq:Veff-Fourier}    
\end{align}
where $q\equiv p_1'-p_1$ is the momentum transfer, and in the non-relativistic limit we have $q\approx (0,\vecq)$. ${\cal M}_{\rm NR}$ is the scattering amplitude in the non-relativistic limit, which is determined by the effective coupling between the DM and mediator particles and can be computed using quantum field theories. Note that the normalization factor $1/(4m_\chi^2)$ in Eq.~(\ref{eq:Veff-Fourier}) exactly cancels the contribution from the on-shell DM field (\ref{eq:wavefunction}) in the non-relativistic limit. 

For the effective operators in Eqs.~(\ref{eq:OS})-(\ref{eq:OFtilde}), the scattering amplitude has the following parametric form:
\begin{align}
{\cal M} \sim \int \frac{{\rm d}^4 k}{\left(2\pi\right)^4} D(k) D(k+q)\;, \label{eq:amplitude-parametric}   
\end{align}
where $D(k)$ denotes the propagator of the mediator particle with momentum $k$, and we have suppressed all Lorentz structures for simplicity. When scattering happens in the presence of a finite density of mediator particles, the propagator is modified from its form in vacuum:
\begin{align}
    D(k) = D_0(k) + D_{\rm bkg}(k)\;,
    \label{eq:mod-propagator}
\end{align}
where $D_0$ denotes the propagator in vacuum and $D_{\rm bkg}$ denotes the correction from the background, which is proportional to the number density of background particles. More specifically, we have
\begin{align}
\text{for boson}&: \; D_{\phi,0}(k) = 
    \frac{i}{k^{2}-m_{\phi}^{2}}\;,\quad\;\;
    D_{\phi,{\rm bkg}}(k) = + 2\pi \delta\left(k^{2}-m_{\phi}^{2}\right)
    f_{\phi}(\mathbf{k})\;,\label{eq:Dphi}\\
\text{for fermion}&: \; D_{\psi,0}(k) = 
    \frac{i\left(\slashed{k}+m_\psi\right)}{k^{2}-m_{\psi}^{2}}\;,\;\;
    D_{\psi,{\rm bkg}}(k) = -2\pi \left(\slashed{k}+m_\psi\right) \delta\left(k^{2}-m_{\psi}^{2}\right)
    f_{\psi}(\mathbf{k})\;,\label{eq:Dpsi}
\end{align}
where $f_\phi(\veck)$ and $f_\psi(\veck)$ denote the phase-space distribution function of the bosonic mediator $\phi$ and the fermionic mediator $\psi$, respectively. Note that the opposite sign in front of $D_{\phi,{\rm bkg}}$ and $D_{\psi,{\rm bkg}}$ is due to different statistics for bosons and fermions. Moreover, the $\delta$-function enforces the background particle to be on-shell. 

The formalism of modified propagators in Eqs.~(\ref{eq:mod-propagator})-(\ref{eq:Dpsi}) was first used in \cite{Horowitz:1993kw} to calculate the neutrino-mediated force in the cosmic neutrino background.\footnote{See  \cite{Weldon:1982aq,Weldon:1982bn} for the original development of this formalism to calculate the fermion and gauge-boson self-energies in the medium, and see \cite{Notzold:1987ik} for an early application of this formalism to calculating the neutrino MSW matter potential.} Later, it was widely used in the literature to study the quantum forces mediated by neutrinos and other light species beyond the SM in different kinds of backgrounds~\cite{Ferrer:1998ju,Ferrer:1998rw,Ferrer:1999ad,Ferrer:2000hm,Ghosh:2022nzo,VanTilburg:2024xib,Barbosa:2024pkl,Ghosh:2024qai,Grossman:2025cov,Cheng:2025fak,Gan:2025nlu}. Notably, it was shown in \cite{Ghosh:2022nzo,VanTilburg:2024xib,Grossman:2025cov} that this formalism can be equivalently derived using the coherent scattering between external particles and background particles. As a result, the validity of Eqs.~(\ref{eq:mod-propagator})-(\ref{eq:Dpsi}) does not require that the background particles be in thermal equilibrium.
 
The background effect is coherent because the phase-space distribution function enters at the amplitude level. Combining Eqs.~(\ref{eq:amplitude-parametric})
and (\ref{eq:mod-propagator}), the amplitude can also be split into two parts accordingly, ${\cal M}={\cal M}_0 + {\cal M}_{\rm bkg}$, where
\begin{align}
{\cal M}_0 &\sim  \int \frac{{\rm d}^4 k}{\left(2\pi\right)^4} D_0(k) D_0(k+q)\;,\label{eq:M0}\\
{\cal M}_{\rm bkg} &\sim  \int \frac{{\rm d}^4 k}{\left(2\pi\right)^4} \left[D_0(k) D_{\rm bkg}(k+q)+D_0(k+q) D_{\rm bkg}(k)\right]\label{eq:Mbkg}.
\end{align}

The term $D_{\rm bkg}(k)D_{\rm bkg}(k+q)$ contains two $\delta$-functions and only contributes to the \emph{imaginary} part of the amplitude. As a result, it does not contribute to the interacting potential between two $\chi$ particles and is therefore not relevant to the Sommerfeld enhancement.

Substituting ${\cal M}={\cal M}_0 + {\cal M}_{\rm bkg}$ back into Eq.~(\ref{eq:Veff-Fourier}) and taking the non-relativistic limit, we obtain the effective potential in Eq.~(\ref{eq:Veff}), where
\begin{align}
V_0(r) &= - \frac{1}{4m_\chi^2}\int \frac{{\rm d}^3\vecq}{\left(2\pi\right)^3}e^{i\vecq\cdot\vecr}{\cal M}_\text{0}\;,\label{eq:V0}\\
V_{\rm bkg}(r) &= - \frac{1}{4m_\chi^2}\int \frac{{\rm d}^3\vecq}{\left(2\pi\right)^3}e^{i\vecq\cdot\vecr}{\cal M}_\text{bkg}\;.\label{eq:Vbkg}
\end{align}
Note that there is no interference term between the vacuum part and the background part because all the calculations are at the amplitude level instead of the amplitude squared level. 

For the vacuum potential $V_0$, since both propagators are off-shell, it is a pure quantum effect. The loop amplitude ${\cal M}_0$ from the 4D integral in Eq.~(\ref{eq:M0}) is Lorentz invariant and is divergent in general. However, only the discontinuity part of ${\cal M}_0$ contributes to the Fourier transform, which must be finite according to the optical theorem. More specifically, using contour integration, it can be shown that Eq.~(\ref{eq:V0}) can be recast into~\cite{Feinberg:1989ps}:
\begin{align}
    V_0(r)=-\frac{1}{4m_\chi^2}\frac{1}{4\pi^{2}r}\int_{t_{\text{0}}}^{\infty}\text{d}t\, e^{-\sqrt{t}\,r}\, \text{Im}\left[\mathcal{M}_\text{0}\left(t\right)\right],\label{eq:disc}
\end{align}
where $t_0=(2m_i)^2$ (for $i=\phi$ or $i=\psi$) denotes the kinematic threshold of the decay to $ 2\phi$ or $2\psi$ particles.
Therefore, to get the vacuum potential, one only needs to extract the imaginary part of the loop amplitude instead of the whole amplitude.\footnote{Note that the vacuum amplitude ${\cal M}_0(t)$ is always real in the physical domain where $t\equiv q^2<0$. The imaginary part is nonzero only when $t>t_0\equiv 4m_{\phi,\psi}^2$.
The integral over $t$ in Eq.~(\ref{eq:disc}) is the result of analytic continuation --- see Appendix~\ref{app:quantum-force} for more details.} 

On the other hand, the background potential $V_{\rm bkg}$ corresponds to the process where only one of the propagators is on-shell. Therefore, it is essentially a classical effect. In particular, the amplitude ${\cal M}_{\rm bkg}$ in Eq.~(\ref{eq:Mbkg}) is not divergent due to the effective regularization from the background. Moreover, the Lorentz invariance of the amplitude is effectively broken by the background with a time-independent distribution function~\cite{Ghosh:2024qai}. As a result, one can always first integrate out the time component ${\rm d}k^0$ using the $\delta$-function in $D_{\rm bkg}$, leaving a 3D integral that depends on the phase-space distribution function $f(\veck)$. 

The quantum forces including both $V_0$ and $V_{\rm bkg}$ from each of the effective operators in Eqs.~(\ref{eq:OS})-(\ref{eq:OFtilde}) can be calculated using the above strategy. We leave the calculation details for Appendix~\ref{app:quantum-force}. In the following, we only collect the final results, which are directly relevant to our purpose of studying the Sommerfeld enhancement. 
For the scalar mediator ${\cal O}_S$ in Eq.~(\ref{eq:OS}), we have
\begin{align}
V_0^{S}(r)  &= -\frac{1}{32\pi^3 \Lambda^2} \frac{m_\phi}{r^2}K_1(2m_\phi r)\;,\label{eq:VS0}\\
V_{\text{bkg}}^{S} (r) &= 
    -\frac{1}{16\pi^{3}\Lambda^{2}r^{2}}
    \int_{0}^{\infty} {\rm d}\kappa 
    \,\frac{\kappa    f_{\phi}(\kappa)}{\sqrt{\kappa^2+m_\phi^2}}
 \sin(2\kappa r)\;.\label{eq:VSbkg}
\end{align}
For the fermionic mediator with scalar-scalar type coupling ${\cal O}_F$ in Eq.~(\ref{eq:OF}), we have
\begin{align}
V_0^{F}(r)  &= -\frac{3}{8\pi^3 \Lambda^4}\frac{m_\psi^2}{r^3}K_2(2m_\psi r)\;,\label{eq:VF0}\\
V_{\text{bkg}}^{F} (r) &=  
    -\frac{1}{4\pi^{3}\Lambda^{4}r^{4}}
    \int_{0}^{\infty} {\rm d}\kappa 
    \,
    \frac{\kappa f_{\psi}(\kappa)}{\sqrt{\kappa^2+m_\psi^2}}
    \left[
    \left(2r^{2}\left(\kappa^{2}+m_\psi^2\right)-1\right)\sin(2\kappa r) + 2\kappa r\cos(2\kappa r) 
    \right].\label{eq:VFbkg}
\end{align}
Finally, for the fermionic mediator with vector-vector type coupling ${\cal O}_{\widetilde{F}}$ in Eq.~(\ref{eq:OFtilde}), we have
\begin{align}
    V_{0}^{\widetilde{F}} (r) &= \frac{m_\psi^2}{2\pi^3 \Lambda^4 r^3}\left[K_2(2m_\psi r)+m_\psi r\,K_1(2m_\psi r)\right],\label{eq:VF0tilde}\\
V_{\rm bkg}^{\widetilde{F}} (r) &= -\frac{1}{2\pi^3\Lambda^4 r^4}\int_{0}^{\infty} {\rm d}\kappa 
    \,
    \frac{\kappa f_{\psi}(\kappa)}{\sqrt{\kappa^2+m_\psi^2}}\left[\left(1+2m_{\psi}^{2}r^{2}\right)\sin\left(2\kappa r\right)-2\kappa r \cos\left(2\kappa r\right)\right].\label{eq:VFbkgtilde}
\end{align}
In above expressions, $K_1$ and $K_2$ are the modified Bessel functions. For the background potential, we have assumed arbitrary isotropic backgrounds where $f(\veck) = f(\kappa)$ with $\kappa\equiv |\veck|$, which is a good approximation for thermal backgrounds. The case of non-thermal background will be discussed in Sec.~\ref{sec:non-thermal}. Also note that in Eq.~(\ref{eq:VFbkgtilde}) where $\psi$ is a Dirac fermion, we have neglected the asymmetry between particle and antiparticle distributions and assumed that they can be described by the same $f_\psi(\kappa)$. This is only done for the brevity of following discussions, and the inclusion of non-vanishing chemical potential is straightforward.

In the next three sections, we study the Sommerfeld enhancement with both scalar and fermionic mediators using Eqs.~(\ref{eq:VS0})-(\ref{eq:VFbkgtilde}). 

\subsection{Calculation of the Sommerfeld factor}
\label{subsec:algorithm}
Before diving into specific potentials, we briefly review the general numerical algorithm to calculate the Sommerfeld factor. We use the method developed in \cite{Iengo:2009ni,Iengo:2009xf} (see also \cite{Cassel:2009wt} for a similar method), which provides a convenient way to calculate the Sommerfeld factor for arbitrary partial waves simply by solving the Schr\"{o}dinger equation involving the relevant potential $V(r)$ --- see Appendix~\ref{app:S-calculation} for details.

Without loss of generality, we assume the incoming plane wave is along the $z$-axis. Then we can decompose the DM wavefunction as 
\begin{align}
 \psi(\vecr) = \sum_{\ell=0}^\infty P_\ell\left(\cos\theta\right)\frac{u_\ell(r)}{r}\;,  
\end{align}
where $P_\ell (\cos\theta)$ is the Legendre polynomial. Then the radial part of the Schr\"{o}dinger equation reads

\begin{align}
    u_{\ell}''(x) +\left[1- \mathcal{V}\left(x\right) 
    - \frac{\ell (\ell +1)}{x^{2}}
    \right]u_{\ell}(x)
    = 0\;,\label{eq:Schrodinger}
\end{align}
where $x\equiv pr$ is the normalized distance, $p\equiv Mv$ is the DM momentum, $M\equiv m_\chi/2$ is the reduced DM mass in the center of mass frame, and $v$ is the relative velocity of the two DM particles. Note that all derivatives in Eq.~(\ref{eq:Schrodinger}) are with respect to $x$. Moreover, ${\cal V}$ is the normalized potential, which is connected to the original potential $V(r)$ via 
\begin{align}
{\cal V}(x)\equiv \frac{2M}{p^2} V(x/p)\;.
\label{eq:Vnorm}
\end{align}
The boundary conditions are given by
\begin{align}
\lim_{x\to 0} u_\ell\left(x\right) = x^{\ell +1}\;,\qquad
\lim_{x\to 0}u_\ell'(x) =\left(\ell+1\right)x^\ell\;.
\label{eq:boundary-condition}
\end{align}
For any non-singular potential, the Schr\"{o}dinger equation (\ref{eq:Schrodinger}) can be numerically solved with the boundary conditions in Eq.~(\ref{eq:boundary-condition}) to get $u_\ell(x)$. The Sommerfeld factor $S_\ell$ for the $\ell$-th partial wave is determined by the value of $u_\ell$ in the asymptotic region~\cite{Iengo:2009ni,Iengo:2009xf,Bellazzini:2013foa} (see Appendix~\ref{app:S-calculation} for the derivation):
\begin{align}
S_\ell = \left[\frac{\left(2\ell +1\right)!!}{C_\ell}\right]^2\;,\qquad  C_\ell^2 = \lim_{x\to \infty} \left[u_\ell^2(x)+u_\ell^2(x-\pi/2)\right].\label{eq:Sl}
\end{align}

In this work, to be specific, we will focus on the case of the $s$-wave ($\ell =0$), though the calculation of higher partial waves is straightforward using Eq.~(\ref{eq:Sl}).

\section{Sommerfeld enhancement from quantum forces in vacuum}
\label{sec:vacuum}

\begin{figure}[t]
\centering
\includegraphics[scale=0.6]
{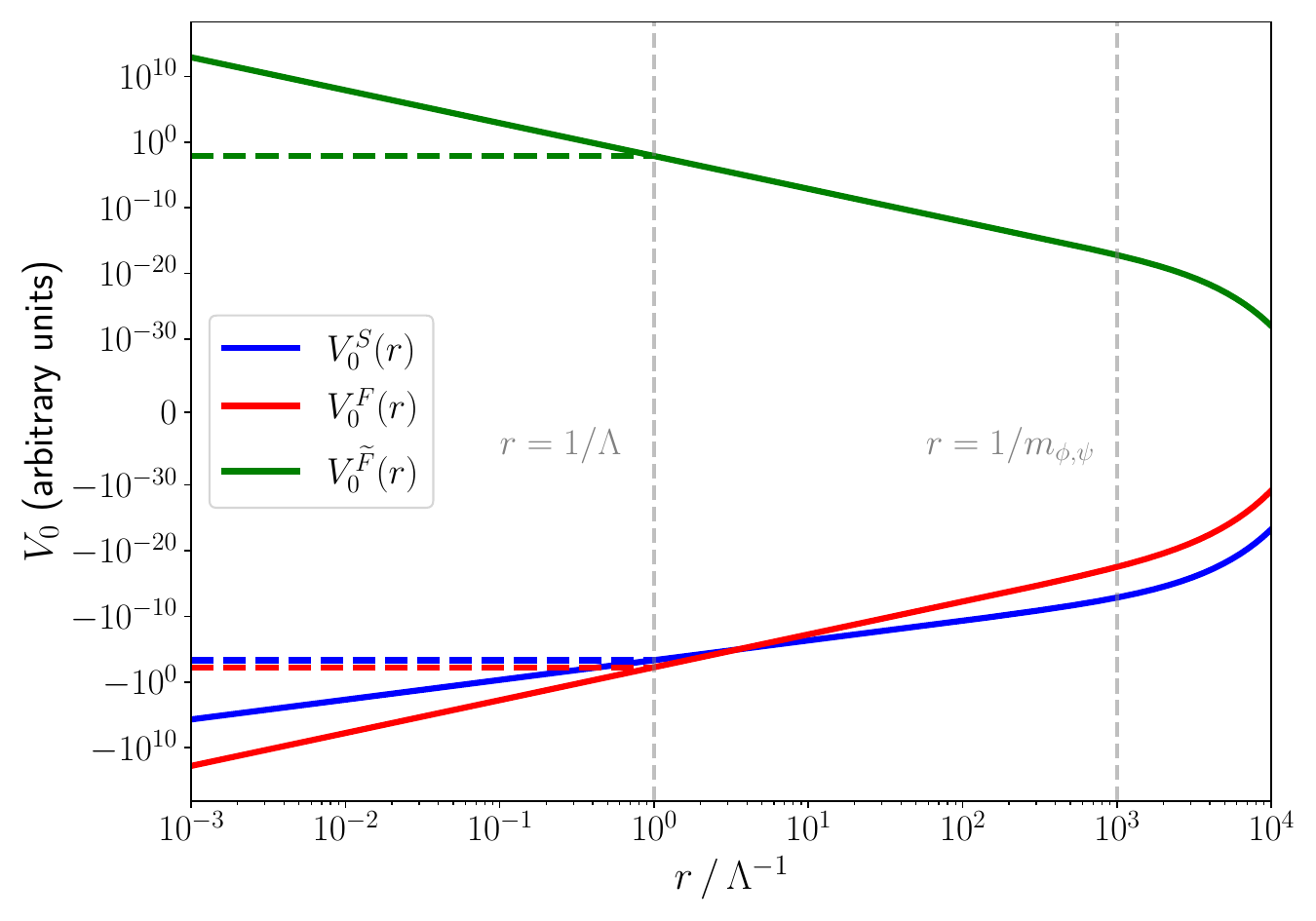}
\caption{\label{fig:vacuumpotentials} The vacuum potentials from the three effective operators $\mathcal{O}_{S}, \mathcal{O}_{F}$, and $\mathcal{O}_{\widetilde{F}}$. The solid lines correspond to the full expressions in Eqs.~(\ref{eq:VS0}), (\ref{eq:VF0}) and (\ref{eq:VF0tilde}). The dashed lines represent the regularization of the potential at distances below $r=1/\Lambda$ using Eq.~(\ref{eq:Vreg}). For definiteness, we have fixed the mediator mass to be $10^{-3} \Lambda$. As a result, all three vacuum potentials begin to exponentially decrease at $r\gtrsim 10^3/\Lambda$.}
\end{figure}

In this section, we study the Sommerfeld enhancement from quantum forces without including the background effect.
As explained in Sec.~\ref{sec:operators}, the vacuum potential is a pure quantum effect that always exists. For the effective operators in Eqs.~(\ref{eq:OS})-(\ref{eq:OFtilde}), the vacuum potentials induced from them are given by Eqs.~(\ref{eq:VS0}), (\ref{eq:VF0}) and (\ref{eq:VF0tilde}). They are plotted in Fig.~\ref{fig:vacuumpotentials}. 

Refs.~\cite{Iengo:2009ni,Iengo:2009xf} calculated the Sommerfeld enhancement with Yukawa-like potentials. The main difference between the Yukawa force and the quantum force is that the latter is more singular at short distances. To see this, we investigate the asymptotic behaviors of the vacuum potentials in Eqs.~(\ref{eq:VS0}), (\ref{eq:VF0}) and (\ref{eq:VF0tilde}), obtaining:
\begin{align}
V_0^S(r) &=  -\frac{1}{64\pi^3 \Lambda^2 r^3}\times
\begin{cases}
1 & \text{for  $r\ll m_\phi^{-1}$}\\
\sqrt{\pi}\left(m_\phi r\right)^{1/2}\, e^{-2m_\phi r} & \text{for $r\gg m_\phi^{-1}$}
\end{cases}\;,\label{eq:V0Sasym}\\
V_0^F(r) &= -\frac{3}{16\pi^3 \Lambda^4 r^5} \times
\begin{cases}
1 & \text{for $r\ll m_\psi^{-1}$}\\
\sqrt{\pi}\left(m_\psi r\right)^{3/2}\, e^{-2m_\psi r} & \text{for $r\gg m_\psi^{-1}$} 
\end{cases}\;,\label{eq:V0Fasym}\\
V_0^{\widetilde{F}}(r) &= +\frac{1}{4\pi^3 \Lambda^4 r^5}\;\,\,\times
\begin{cases}
1 & \text{for  $r\ll m_\psi^{-1}$}\\
\sqrt{\pi}\left(m_\psi r\right)^{5/2}\, e^{-2m_\psi r} & \text{for $r\gg m_\psi^{-1}$} 
\end{cases}\;.\label{eq:V0Ftildeasym}
\end{align}
Several comments are in order:
\begin{itemize}
    \item All three potentials are exponentially suppressed when the distance is greater than the inverse of twice the mediator mass. This is a generic feature of the vacuum potentials $V_0$. As we shall see later, the background potentials $V_{\rm bkg}$ are typically {\it not} exponentially suppressed at long distances.

    \item On the other hand, at short distances, the potentials from two-scalar (or two-fermion) exchange scale as $1/r^3$ (or $1/r^5$), which is more singular compared to the Yukawa potential.
    
    \item Both $V_0^S$ and $V_0^F$ are attractive at all distances, leading to Sommerfeld enhancement. However, $V_0^{\widetilde{F}}$ is repulsive due to the vector-vector Lorentz structure, which leads to Sommerfeld suppression that can decrease the DM annihilation cross section.
\end{itemize}

When there is no background, the Sommerfeld factor $S$ from quantum forces depends on three dimensionless parameters: 
\begin{align}
S = S\left(v,\eta_\chi,\eta_m\right),    
\end{align}
where $v$ is the relative velocity of the two DM particles, $\eta_\chi$ and $\eta_m$ are the reduced DM mass and the mediator mass normalized by the cutoff scale:
\begin{align}
\eta_\chi \equiv \frac{M}{\Lambda}\;,\quad
\eta_m \equiv \frac{m_{i}}{\Lambda}\;,\label{eq:eta}
\end{align}
where $i=\phi, \psi$ denotes the mediator.
\vspace{0.2cm}

The contents of this section are organized as follows. In Sec.~\ref{subsec:singularpotential}, before focusing on specific quantum forces, we discuss the general strategy for calculating the Sommerfeld enhancement/suppression with a singular potential. Then, in Sec.~\ref{subsec:V0S}-\ref{subsec:V0Ftilde}, we calculate the Sommerfeld factor separately from $V_0^S$, $V_0^F$, and $V_0^{\widetilde{F}}$ with the proper regularization procedure. It turns out that we get $S>1$ from $V_0^S$ and $V_0^F$, while $S<1$ from $V_0^{\widetilde{F}}$. We also provide two approximation methods, which are helpful for intuitive understanding of the behavior of $S$: the locations of resonance peaks for Sommerfeld enhancement can be analytically predicted using the box approximation in Sec.~\ref{subsec:box} (see another method in terms of bound state in Sec.~\ref{subsec:boundstate}); the exponentially decreasing behavior of $S$ for Sommerfeld suppression can be explained using the WKB approximation in Sec.~\ref{subsec:WKB}.

\subsection{General strategy to deal with singular potentials}
\label{subsec:singularpotential}
From Eqs.~(\ref{eq:V0Sasym})-(\ref{eq:V0Ftildeasym}), one finds that all three vacuum potentials are more divergent than the Yukawa potential at short distances. A potential more divergent than $1/r^2$ is known as a singular potential~\cite{Case:1950an,Frank:1971xx}, which does not lead to the regular series solutions of the Schr\"{o}dinger equation around the origin. In fact, for a singular potential $V(r)$, using the WKB approximation, one can show that the radial wavefunction $u(r)$ for any state around $r=0$ scales as~\cite{Case:1950an,Frank:1971xx}
\begin{align}
u(r) \sim \left|V(r)\right|^{-1/4}\exp\left(\pm i \int^r  {\rm d}r'\sqrt{-V(r')}\right).
\end{align}
If $V(r)$ is attractive, the wavefunction goes to zero with infinitely rapid oscillations as $r \to 0$, which does not lead to a physical solution. On the other hand, if $V(r)$ is a repulsive singular potential, $V(r) \sim +1/r^d$ (with $d>2$), the wavefunction in the neighborhood of the origin is found to be~\cite{Case:1950an,Frank:1971xx}
\begin{align}
u(r) \sim r^{d/4} \exp\left(-\frac{2}{d-2}\,r^{-\frac{d-2}{2}}\right), 
\label{eq:solution-repulsive}
\end{align}
which still corresponds to a unique and physical solution that vanishes at the origin.

For our purpose of calculating the Sommerfeld factor, we need to solve the Schr\"{o}dinger equation including quantum forces to find the DM wavefunctions. Among the three singular potentials considered here, $V_0^S$ and $V_0^F$ are attractive, while $V_0^{\widetilde{F}}$ is repulsive. According to the above discussions, $V_0^S$ and $V_0^F$ cannot be used to calculate $S$ directly since they lead to unphysical solutions of the Schr\"{o}dinger equation around the origin, and a regularization of the potential at short distances is necessary; $V_0^{\widetilde{F}}$, on the other hand, can be directly inserted into the Schr\"{o}dinger equation, and it can provide a physical result of $S$ in the region where the effective theory is valid. In addition to that, since we start from the non-renormalizable operators in Eqs.~(\ref{eq:OS})-(\ref{eq:OFtilde}), the effective theory breaks down at $r<1/\Lambda$. In  momentum space, this corresponds to the region where the momentum transfer $|\vecq|\sim Mv$ between two DM particles exceeds the cutoff scale $\Lambda$. In this region, the expressions in Eqs.~(\ref{eq:VS0}), (\ref{eq:VF0}) and (\ref{eq:VF0tilde}) no longer hold. 

The Schr\"{o}dinger equation  with singular potentials can be renormalized with a finite number of counterterms as long as $|\vecq|\ll \Lambda$ is satisfied~\cite{Lepage:1997cs}. In this region, the separation of scales is possible, and the potential can be split into the long-range part, and a finite number of local operators which encode the information of UV theory. The Wilson coefficients of these local operators are determined by comparing theoretical predictions using regularized potentials with low-energy observables --- a procedure similar to the renormalization in quantum field theories. Moreover, it was shown~\cite{Beane:2000wh} that any singular potential at short distances can be renormalized by a finite well with fixed width and varying height (i.e., a one-parameter counterterm), regardless of what the UV completion is. For a fixed value of the width $r_0$, the height of the well can be  determined by matching the solution of Schr\"{o}dinger equation with a finite-well potential to the solution with the full singular potential at the boundary $r=r_0$~\cite{Beane:2000wh}. In practice, it is usually convenient to express the height of the well as a function of a low-energy observable (such as the phase shift from DM scattering) that is determined by experiments~\cite{Bellazzini:2013foa}. Once the height of the well is fixed by the low-energy observable, one can use it to make predictions independent of the UV physics.
This provides a way to describe low-energy phenomena resulting from singular potentials without specifying the UV completion of effective operators.

Based on the above consideration, we regularize our vacuum potentials at short distances with the following procedure:
\begin{align}
    V_\text{0,reg}^i (r) = 
    \begin{cases}
    V_0^i(r) & \text{for $r>1/\Lambda$}\\
    cV_0^i (r=1/\Lambda) & \text{for $r\leq 1/\Lambda$}
    \end{cases}\;,\quad
    \text{for $i=S,F,\widetilde{F}$}\;.
    \label{eq:Vreg}
\end{align}
That is, we take the full expression of the potential below the cutoff scale where the effective theory is valid, and we use a finite well to approximate the potential above the cutoff scale. The dimensionless parameter $c$ controls the height of the well and is determined by comparing the theoretical prediction using Eq.~(\ref{eq:Vreg}) with any low-energy observable of DM scattering/annihilation.

In general, the behavior of quantum forces above the cutoff scale depends on UV completion. In specific UV models, the quantum forces are no longer singular at short distances because the effective operators are renormalized above the cutoff scale. For example, in the SM, the $1/r^5$ two-neutrino exchange potential derived from the four-Fermi effective theory needs to be replaced at short distances by the potential obtained from the renormalizable electroweak theory. The resulting potential is found to scale as $1/r$ when $r$ is smaller than the inverse of the electroweak scale~\cite{Ghosh:2024ctv}. For examples of UV completions of quantum forces beyond the SM and the resulting Sommerfeld enhancement, see e.g. \cite{Beneke:2019qaa,Coy:2022cpt,Xu:2021daf}. Throughout this section, to make our result as model independent as possible, we use quantum forces derived from effective operators plus the regularization procedure in Eq.~(\ref{eq:Vreg}) to calculate the Sommerfeld factor. Our results derived in this way are independent of UV completion in the parameter space where the momentum transfer between two DM particles is small compared to the cutoff scale. From the point of view of the effective theory, the height of the well should be determined by using a low-energy observable (such as the phase shift) as an input. However, in this work, since we are mainly interested in the theoretical aspects of the structure of the Sommerfeld factor caused by quantum forces, we will always take $c=1$ in Eq.~(\ref{eq:Vreg}) for the following calculations.\footnote{This choice is not necessary and is just ``theoretically natural'' since the regularized potential in Eq.~(\ref{eq:Vreg}) is continuous at the cutoff scale if $c=1$. For different values of $c$, we have verified that the numerical values of the Sommerfeld factor will be changed, but the qualitative properties of the resonance structure are not affected. In particular, there can still be significant Sommerfeld enhancement for vanishingly small $c$ (see Appendix~\ref{app:regular-dependence}). This fact indicates that the long-range part of the singular potential has non-negligible contributions to the Sommerfeld effect.} We show the results with $c\neq 1$ in Appendix~\ref{app:regular-dependence}.

\subsection{Scalar mediator}
\label{subsec:V0S}
\begin{figure}[t]
\centering
\includegraphics[scale=0.4]
{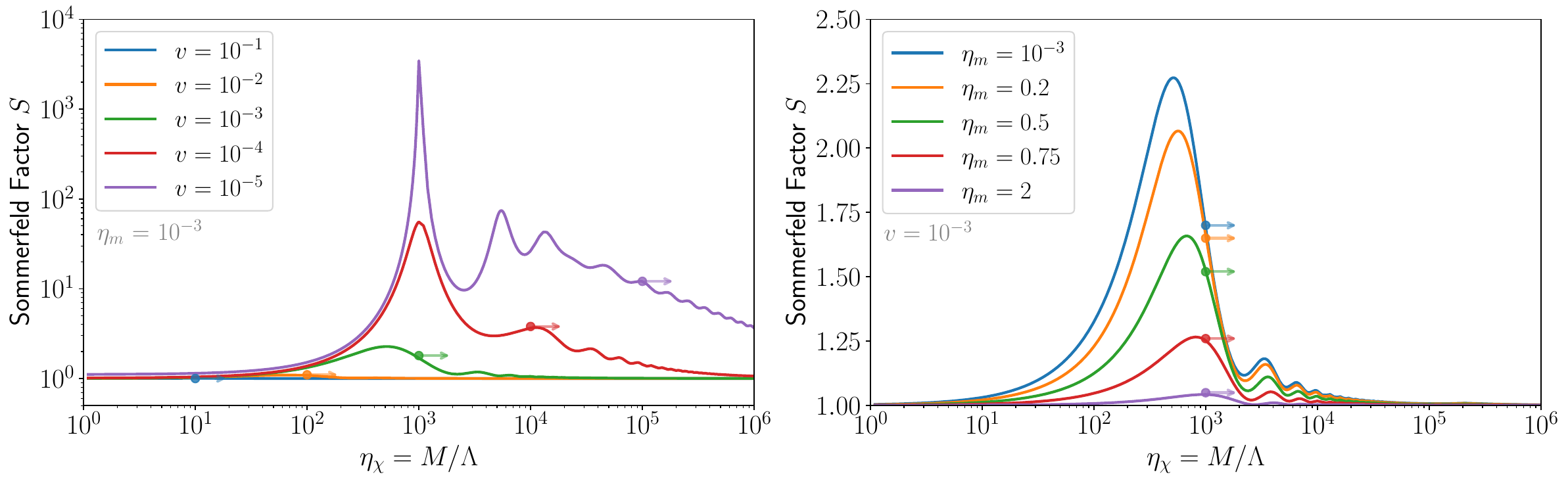}
\caption{\label{fig:scalarvacuum} Sommerfeld enhancement from the two-scalar potential $V_0^S$ in Eq.~(\ref{eq:VS0}), where the regularization procedure in Eq.~(\ref{eq:Vreg}) is used to get a physical result. We plot the Sommerfeld factor $S$ as a function of the normalized reduced DM mass $\eta_\chi\equiv M/\Lambda$. \emph{Left:} We fix the mediator mass to be $\eta_m\equiv m_\phi/\Lambda=10^{-3}$ and change the velocity. \emph{Right:} We fix the velocity to be $v=10^{-3}$ and change the mediator mass . The dots with arrows on each curve represent the masses for which the momentum transfer exceeds the cutoff scale, i.e., $|\vecq|\sim Mv \geq \Lambda$; the exact value of $S$ to the right of these dots depends on the UV completion of the effective operator. To the left of these dots, the value of $S$ is determined by a single parameter controlling the UV physics; see the text around Eq.~(\ref{eq:Vreg}) for details.}
\end{figure}

We first study the case of the scalar mediator. The Sommerfeld factor $S$ is calculated with the vacuum two-scalar potential in Eq.~(\ref{eq:VS0}) regularized by Eq.~(\ref{eq:Vreg})  and using the numerical algorithm in Sec.~\ref{subsec:algorithm}. The result is shown in Fig.~\ref{fig:scalarvacuum}. In the left panel, we fix the mediator mass and plot $S$ as a function of the reduced DM mass for different velocities; in the right panel, we fix the velocity to be $v=10^{-3}$ and plot $S$ with different values of the mediator mass. It can be seen that a larger Sommerfeld enhancement corresponds to a smaller velocity or a lighter mediator. We have also verified that for a fixed velocity, the Sommerfeld factor is almost constant for any value of $\eta_m \lesssim 10^{-3}$.

In Fig.~\ref{fig:scalarvacuum}, we find that $S \to 1$  in the limit of $M\to \infty$. This can be understood in the following way. We expand the normalized potential (\ref{eq:Vnorm}) in the large $M$ limit: 
\begin{align}
\mathcal{V}_0^S(x)\equiv \frac{2M}{p^2}V_0^S(x/p) \approx - \frac{M^2 v}{32\pi^3 \Lambda^2 x^3}\;.    
\end{align}
The cutoff scale is $x_c\equiv Mv/\Lambda$, where for $x<x_c$, the potential is regularized by a constant $\mathcal{V}_c\equiv \mathcal{V}_0^S(Mv/\Lambda)=- \Lambda/(32\pi^3 M v^2)$. In the limit of $M\gg \Lambda$, we have $x_c \to \infty$ and $\mathcal{V}_c \to 0$; in this case, the effective potential is negligible in the whole domain of $x$ and therefore $S\to 1$.

One should keep in mind that when $M$ is sufficiently large, the momentum transfer $|\vecq|\sim Mv$ becomes greater than the cutoff scale $\Lambda$, and the effective theory breaks down. In this region, the finite-well approximation (\ref{eq:Vreg}) fails and the exact value of $S$ depends on the UV completion of the effective operator ${\cal O}_S$. We define the critical mass as a function of the velocity:
\begin{align}
M_c(v) \equiv \Lambda/v\;.    
\end{align}
The locations of $M_c$ are represented by the dots with arrows on each curve in Fig.~\ref{fig:scalarvacuum}. For $M<M_c$, our result for $S$ is not sensitive to UV physics. For $M>M_c$, the value of $S$ depends on the UV completion of ${\cal O}_S$ and may change if one uses a different regularization procedure from Eq.~(\ref{eq:Vreg}).

While the Sommerfeld factor plotted in Fig.~\ref{fig:scalarvacuum} is calculated numerically, the approximate location of the peaks can be predicted analytically using the box approximation. This will be discussed in Sec.~\ref{subsec:box}.

\subsection{Fermionic mediator: attractive potential}
\label{subsec:V0F}
\begin{figure}[t]
\centering
\includegraphics[scale=0.4]
{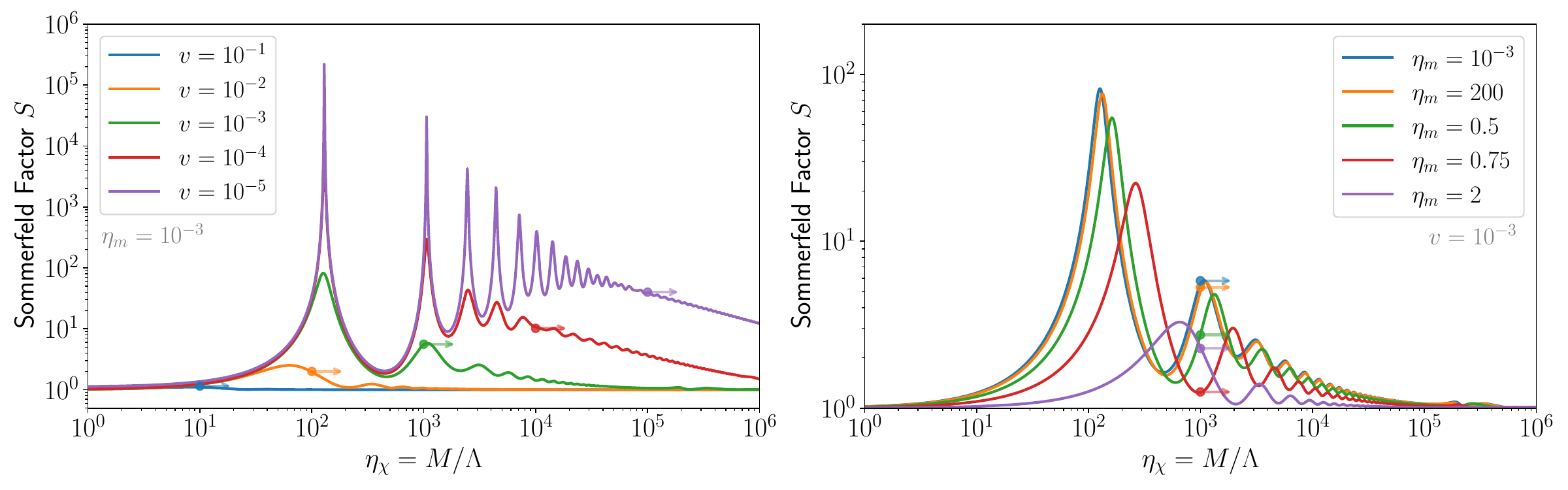}
\caption{\label{fig:fermionvacuum} Same conventions as Fig.~\ref{fig:scalarvacuum}, but with the attractive two-fermion potential $V_0^F$ in Eq.~(\ref{eq:VF0}).}
\end{figure}
Now we turn to the case of the fermionic mediator. We first look at the potential $V_0^F$ in Eq.~(\ref{eq:VF0}). Similarly to the two-scalar potential, $V_0^F$ is attractive at all distances, so it will lead to Sommerfeld enhancement. 

We use the same regularization procedure in Eq.~(\ref{eq:Vreg}) for $V_0^F$ to calculate the Sommerfeld factor, and the result is shown in Fig.~\ref{fig:fermionvacuum}. The peaks in Fig.~\ref{fig:fermionvacuum} are generically higher than those in Fig.~\ref{fig:scalarvacuum}, indicating a stronger Sommerfeld enhancement. This is because for a fixed cutoff scale, the two-fermion potential $V_0^F \sim 1/r^5$ is more singular than the two-scalar potential $V_0^S \sim 1/r^3$ at short distances, which can also be seen from Fig.~\ref{fig:vacuumpotentials}. As a result, the distortion of the DM wavefunction caused by $V_0^F$ is stronger compared to that from $V_0^S$. Similarly to the case of a scalar mediator, for heavy DM with $M \gtrsim \Lambda/v$, the exact value of $S$ depends on the UV completion of ${\cal O}_F$. This is indicated by the dots with arrows in Fig.~\ref{fig:fermionvacuum}.

\subsection{Fermionic mediator: repulsive potential}
\label{subsec:V0Ftilde}
\begin{figure}[t]
\centering
\includegraphics[scale=0.4]
{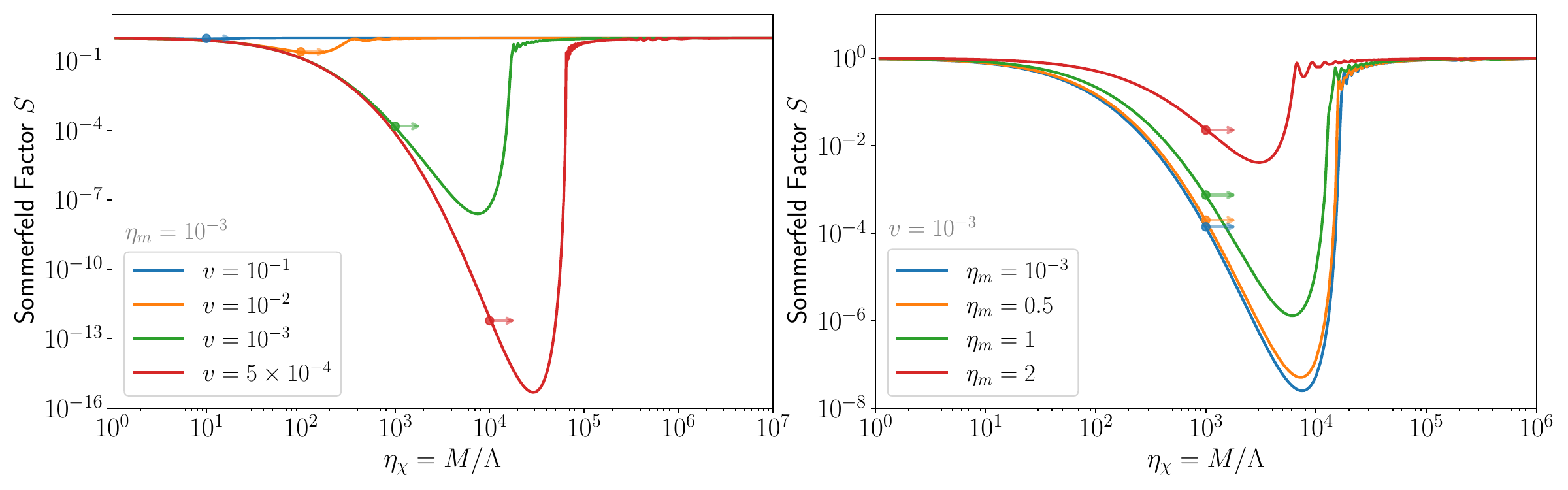}
\caption{\label{fig:fermionvacuum-Repulsive} Sommerfeld suppression from the repulsive two-fermion potential $V_{0}^{\widetilde{F}}$ in Eq.~(\ref{eq:VF0tilde}), where the regularization procedure in Eq.~(\ref{eq:Vreg}) is used. 
Conventions are the same as Fig.~\ref{fig:scalarvacuum}.}
\end{figure}

As the third example, we study the potential $V_0^{\widetilde{F}}$ in Eq.~(\ref{eq:VF0tilde}). Unlike $V_0^S$ and $V_0^F$, this time, we have a repulsive potential. Therefore, we expect that it will lead to Sommerfeld suppression with $S<1$. 

Note that for a singular repulsive potential, the Sommerfeld factor can be computed by solving the Schr\"{o}dinger equation including the full potential (\ref{eq:VF0tilde}) without regularization. This is because in this case the solution of the wavefunction around the origin is unique and well-defined, as shown in Eq.~(\ref{eq:solution-repulsive}). However, the effective interaction described by $\widetilde{\cal O}_F$ still becomes invalid when the momentum transfer exceeds the cutoff scale. As a result, the value of $S$ calculated from the unregularized potential (\ref{eq:VF0tilde}) is not reliable when $M \gtrsim \Lambda/v$. In order to have a physical result for $S$ for any value of the DM mass and velocity, we also regularize the potential $V_0^{\widetilde{F}}$ using the procedure in Eq.~(\ref{eq:Vreg}). The Sommerfeld factor calculated from the regularized $V_0^{\widetilde{F}}$ is shown in Fig.~\ref{fig:fermionvacuum-Repulsive}. Similarly to Figs.~\ref{fig:scalarvacuum} and \ref{fig:fermionvacuum}, the dots with arrows on each curve indicate the region where the effective theory description breaks down.

The main difference is that in Fig.~\ref{fig:fermionvacuum-Repulsive}, there is a region where $S \ll 1$. This can be clearly seen even in the region where effective theory is valid (i.e., to the left of the dots). Moreover, a stronger suppression occurs when the velocity decreases or when the mediator mass is reduced. This suppression is a typical consequence of a repulsive long-range force: as two DM particles approach each other with small velocity, the repulsion between them strongly disfavors the occurrence of annihilation and strongly reduces the effective cross section. In fact, the suppression is exponential --- a typical feature of quantum tunneling, which can be quantitatively understood using the WKB approximation. We leave the detailed analysis for Sec.~\ref{subsec:WKB}.

\subsection{Locations of peaks: box approximation}
\label{subsec:box}

Our calculations of the Sommerfeld factor $S$ in Figs.~\ref{fig:scalarvacuum} and \ref{fig:fermionvacuum} are numerical, because it is not possible to get analytical solutions of the Schr\"{o}dinger equation with quantum forces. However, the resonance peaks in Figs.~\ref{fig:scalarvacuum} and \ref{fig:fermionvacuum} have a clear physical meaning: they occur when the potential admits a bound state with energy approaching zero. In this case, the (positive, but small) kinetic energy of the colliding non-relativistic DM particles approximately matches the (negative, but also small) bound-state energy, producing a resonance. In this subsection, we use the box approximation to derive the locations of peaks analytically (a different method is provided in Sec.~\ref{subsec:boundstate}). The box approximation method was previously used to derive the peaks for the Yukawa potential in \cite{Lattanzi:2008qa} and for the $1/r^3$ potential from one pseudo-scalar exchange in \cite{Bellazzini:2013foa}.

The basic idea is that if an attractive potential decreases fast after some transition point at $r>r_0$ while changing slowly at $r<r_0$, then it can be approximated by a finite well:
\begin{align}
V_{\rm box}(r) = -U_0\, \Theta \left(r_0-r\right), \label{eq:Vbox}
\end{align}
where $\Theta$ is the step function, $r_0$ is the approximate range of the potential, and $U_0$ is the height of well fixed by matching $V_{\rm box}$ to the exact potential at $r=r_0$. 

For a finite-well potential, the Schr\"{o}dinger equation of a scattering problem can be analytically solved, so the Sommerfeld factor can be computed analytically~\cite{Lattanzi:2008qa,Bellazzini:2013foa}: 
\begin{align}
S^{-1} &= \cos^2\left(r_0\sqrt{p^2+2MU_0}\right) + \frac{p^2}{p^2+2MU_0}\sin^2\left(r_0\sqrt{p^2+2MU_0}\right) \nonumber\\
& \approx \cos^2\left(r_0 \sqrt{2MU_0}\right) + \frac{p^2}{2M U_0}\sin^2\left(r_0 \sqrt{2MU_0}\right),\label{eq:Sbox}
\end{align}
where in the second line we have used non-relativistic approximation, $p\ll \sqrt{2MU_0}$. To maximize the value of $S$ in Eq.~(\ref{eq:Sbox}), the DM mass should satisfy the discrete relation:
\begin{align}
r_0\sqrt{2MU_0} = \left(n+\frac{1}{2}\right) \pi \quad \implies \quad
M_{n} = 
    \frac{\pi^{2}}{2r_{0}^{2}U_0^{}}
    \left(n+\frac{1}{2}\right)^{2}, \quad
n = 0,1,2,...\label{eq:Mnbox}
\end{align}
At these discrete values, $S$ is maximized and is proportional to $1/v^2$:
\begin{align}
S_n =   \frac{2U_0}{M_n v^2} = \frac{4 r_0^2 U_0^2}{(n+1/2)^2 \pi^2 v^2}\;.\label{eq:Smax}
\end{align}
Note, however, that $S\sim 1/v^2$ violates the partial-wave unitarity bound if the velocity is sufficiently small~\cite{Hisano:2002fk,Hisano:2003ec,Hisano:2004ds}. Therefore, it is necessary to regularize $S$ in the small velocity limit~\cite{Blum:2016nrz} (see Appendix~\ref{app:unitarity}).

Physically, the condition~(\ref{eq:Mnbox}) can be understood as follows. The potential~(\ref{eq:Vbox}) supports bound states, with energies $E_n<0$. A peak in the Sommerfeld factor occurs when the (positive) energy of the scattering state nearly matches the (negative) energy of one of the  bound states, leading to a resonant enhancement of the scattering amplitude. The scattering state energy is always small (compared to the characteristic energy scale of the interaction, $U_0$), so the matching condition is that the  
(absolute value of) one of the bound state energies be small as well. The bound state energies are determined by solving a transcendental equation
\begin{align}
\cot \left(\sqrt{2M(U_0+E_n)}\,r_0\right)   = - \sqrt{\frac{-E_n}{U_0+E_n}}\;. 
\label{eq:BS}
\end{align}
Resonance occurs when one of the bound states approaches zero energy, $E_n\to 0$ for some $n$. Substituting this into Eq.~(\ref{eq:BS}) reproduces the condition~(\ref{eq:Mnbox}). 

For the attractive quantum potentials $V_0^S$ and $V_0^F$  with the regularization procedure in Eq.~(\ref{eq:Vreg}), the transition
between slowly-changing and rapidly-dropping behaviors occurs 
around $r=1/\Lambda$. To optimize the parameters of the box approximation, we write
\begin{align}
r_0^i = b_i/\Lambda\;,\qquad
-U_0^i = V_0^i\left(r=b_i/\Lambda\right),\qquad
\text{for $i=S,F$}\;.
\end{align}
Here $b_i$ (for $i=S,F$) are expected to be ${\cal O}(1)$ numbers, which can be fixed by comparing Eq.~(\ref{eq:Mnbox}) with the numerical result for the first peak. The locations of the subsequent peaks can then be predicted.

\renewcommand\arraystretch{1.3}
\begin{table}[t]
	\centering
	\begin{tabular}{c|c|c||c|c}
		\hline\hline
		& Numeric ($V_0^S$) & Box Approx ($b_S = 0.41$) & Numeric ($V_0^F$) & Box Approx ($b_F = 0.86$) \\
		\hline
		$n=0$ & $1.0\times 10^3$ & $1.0 \times 10^3$ & $1.3 \times 10^2$ & $1.3 \times 10^2$ \\
		\hline
		$n=1$ & $5.5 \times 10^3$ & $9.0 \times 10^3$ & $1.1 \times 10^3$ & $1.2 \times 10^3$\\
		\hline
		$n=2$ & $1.4 \times 10^4$ & $2.5 \times 10^4$ & $2.5 \times 10^3$ & $3.2 \times 10^3$\\
		\hline\hline
	\end{tabular}
	\caption{\label{tab:peaklocationsbox}Comparison of the locations of the first three peaks (in units of $\eta_\chi \equiv M/\Lambda$) between numerical results and box approximation results for attractive quantum forces $V_0^S$ (first two columns) and $V_0^F$ (last two columns). We fix $\eta_m \equiv m_{\phi,\psi}/\Lambda = 10^{-3}$ as a benchmark. The dimensionless coefficients 
    $b_i$ (for $i=S,F$) are determined by minimizing the difference between the numerical result and the box approximation result for the location of the first peak $(n=0)$.}
\end{table}
\renewcommand\arraystretch{1}

Applying the box approximation to $V_0^S$ and $V_0^F$, we arrive at the prediction of the locations of peaks:
\begin{align}
    M^{S}_{n} &= 
    \frac{16\pi^{5}\Lambda^{2}}
    {m_{\phi}K_{1}(2m_{\phi}b_S/\Lambda)}
    \left(n+\frac{1}{2}\right)^{2}, 
    \\
    M^{F}_{n}  &= 
    \frac{4\pi^{5}\Lambda^{3}b_F}
    {3m_{\psi}^{2}K_{2}(2m_{\psi}b_F/\Lambda)}
    \left(n+\frac{1}{2}\right)^{2}.
\end{align}
In terms of the dimensionless parameters in Eq.~(\ref{eq:eta}), we have
\begin{align}
\eta_{\chi,n}^S = \frac{16 \pi^5}{\eta_m K_1(2  \eta_m b_S)}\left(n+\frac{1}{2}\right)^{2},\label{eq:etaSbox}\\
\eta_{\chi,n}^F = \frac{4 \pi^5 b_F}{3\eta_m^2 K_2(2  \eta_m b_F)}\left(n+\frac{1}{2}\right)^{2}.\label{eq:etaFbox}
\end{align}

A comparison between the analytical expression in Eqs.~(\ref{eq:etaSbox})-(\ref{eq:etaFbox}) and the numerical result in Figs.~\ref{fig:scalarvacuum}-\ref{fig:fermionvacuum} is presented in Table~\ref{tab:peaklocationsbox}. It can be seen that the box approximation works better for $V_0^F$ than $V_0^S$. This is because $V_F \sim 1/r^5$ decreases faster compared to $V_S \sim 1/r^3$ at $r > 1/\Lambda$. As a result, the shape of $V_F$ is closer to the finite well (\ref{eq:Vbox}) than that of $V_0^S$.

\subsection{Sommerfeld suppression and WKB approximation}
\label{subsec:WKB}

\begin{figure}[t]
\centering
\includegraphics[scale=0.6]
{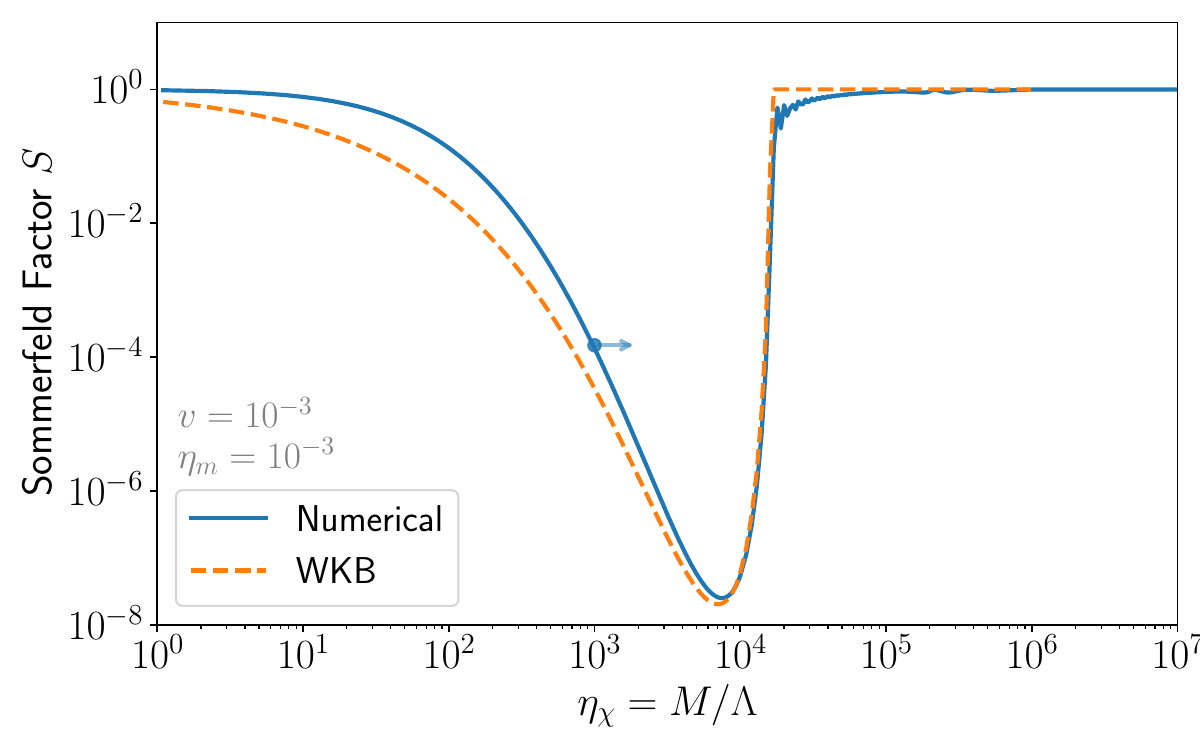}
\caption{\label{fig:vacuumWKB} Comparison between the numerical calculation of the Sommerfeld factor from the repulsive potential $V_0^{\widetilde{F}}$ (solid line) and the prediction using WKB approximation (dashed line). As a benchmark, the mediator mass $\eta_m \equiv m_\psi/\Lambda=10^{-3}$ and the relative velocity $v=10^{-3}$ are assumed in both cases. The dot and arrow again label where the effective theory becomes invalid, as in Fig.~\ref{fig:scalarvacuum}.}
\end{figure}

We see in Fig.~\ref{fig:fermionvacuum-Repulsive} that a long-range repulsive potential can exponentially reduce the effective cross section of DM annihilation. This can be understood as follows. For annihilation to occur, the two DM particles need to come within a distance of order $M^{-1}$ of each other. To do so, the particles must pass through a potential barrier, whose height is greater than their kinetic energy. Tunneling through the barrier is classically forbidden but is allowed by quantum mechanics. According to the WKB approximation, the transmission coefficient ${\cal T}$ (tunneling rate) is given by
\begin{align}
    \mathcal{T} 
    =
    e^{-2\gamma}\;,\qquad {\rm with}\;\;
    \gamma = 
    \int_{r_{1}}^{r_{2}} {\rm d}r \sqrt{2M\left|E-V(r)\right|}\;, \label{eq:transmission}
\end{align}
where $E = M v^{2}/2$ is the kinetic energy, while $r_1$ and $r_2$ are the classical turning points of the potential barrier determined by $V(r_{1,2})=E$. (For a monotonically decreasing potential, we can take $r_1 = 0$.)

As a simple example, for a repulsive Coulomb potential $V(r) = \alpha/r$, using Eq.~(\ref{eq:transmission}) one obtains ${\cal T} = e^{-2\pi \alpha/v}$, which gives exactly the same exponential suppression behavior as the known result of the Sommerfeld factor from a repulsive Coulomb potential. Therefore, we expect that the tunneling rate can be used to estimate the Sommerfeld suppression from a general repulsive potential:
\begin{align}
S \sim e^{-2\gamma}\;,\qquad {\rm with}\;\;
    \gamma = 
    \int_{r_{1}}^{r_{2}} {\rm d}r \sqrt{2M\left|E-V(r)\right|}\;. \label{eq:WKB} 
\end{align}
The Sommerfeld factor predicted by this formula for the repulsive two-fermion potential $V_0^{\widetilde{F}}$ (regularized by Eq.~(\ref{eq:Vreg})) is shown by the dashed line 
in Fig.~\ref{fig:vacuumWKB}, while the numerical calculation of the Sommerfeld factor from the same potential is shown by the solid line. The two results agree very well.

\section{Correction from thermal background: scalar mediator}
\label{sec:thermal-scalar}

Starting from this section, we study the Sommerfeld enhancement from quantum forces in the presence of a background of mediator particles. In this section and Sec.~\ref{sec:thermal-fermion}, we assume that the background is thermal\footnote{By thermal background we refer to the scenario where the phase-space distribution function of the mediator particles can be described by a universal temperature. This temperature may or may not be equal to the temperature of the SM bath.} and isotropic and focus on scalar and fermionic mediators, respectively. The case of a non-thermal background will be discussed in Sec.~\ref{sec:non-thermal}.

As explained in Sec.~\ref{sec:intro}  and Sec.~\ref{subsec:modifiedpropagator}, in a background of finite number density of mediator particles, the propagators are modified, leading to a finite-density correction to the above quantum forces and the Sommerfeld factor (see the right panel of Fig.~\ref{fig:Schematic} for a schematic Feynman diagram).  
For the scalar mediator $\phi$, the propagator in the background is modified to be
\begin{align}
    \frac{i}{k^{2}-m_{\phi}^{2}}
    ~~\rightarrow~~
     \frac{i}{k^{2}-m_{\phi}^{2}}
     + 2\pi\delta\left(k^{2}-m_{\phi}^{2}\right)f_{\phi}(\veck)\;,
     \label{eq:ModifiedScalarProp}
\end{align}
where $f_{\phi}(\veck)$ is the phase-space distribution functon, which is normalized such that $\int {\rm d}^3 \veck \, f_\phi(\veck)/(2\pi)^3$ gives the number density of $\phi$ particles in the background.

In Appendix \ref{app:quantum-force}, we show how to calculate the background potential with an arbitrary $f_\phi(\veck)$. Throughout this section, we assume an isotropic background, i.e., $f_\phi(\veck)=f_\phi(\kappa)$ with $\kappa\equiv |\veck|$. Then the background potential is given by Eq.~(\ref{eq:VSbkg}). In addition, we consider the scenario where the mediator mass $m_\phi$ is negligible compared to the cutoff scale $\Lambda$ and the typical energy scale of the background (such as the temperature $T$). This assumption can greatly simplify the calculation and isolate the finite-temperature effect on the Sommerfeld enhancement. Our results in this section are valid as long as $m_\phi \ll \Lambda,T$ is satisfied. 

Taking the limit of the massless mediator, the background potential is simplified to
\begin{align}
	\label{eq:VSbkgmassless}
    V_{\text{bkg}}^{S} (r) = -
    \frac{1}{16\pi^{3}\Lambda^{2}r^{2}}
    \int_{0}^{\infty} {\rm d}\kappa 
    \,
    f_{\phi}(\kappa)\sin(2\kappa r)\;.
\end{align}
In this case, there are four typical scales:  the DM mass $M$, the temperature of the background $T$, the cutoff scale $\Lambda$, and the scale of momentum transfer $|\vecq| \sim Mv$. The Sommerfeld factor depends on three dimensionless parameters:
\begin{align}
S=S\left(v,\eta_\chi,\eta_T\right),
\end{align}
where $\eta_\chi\equiv M/\Lambda$ and $\eta_T \equiv T/\Lambda$. The validity of the effective operator formalism requires $|\vecq| <\Lambda$ (or $\eta_\chi<1/v$). In addition, the kinetic energy of a massless mediator in thermal equilibrium is of order $T$, so $T<\Lambda$ (or $\eta_T<1$) is also required. Note that if $\phi$ and $\chi$ are in the same thermal bath, they share the same temperature and therefore $v$ and $T$ are not independent, $v\sim 
\sqrt{T/M}$, which is the case of thermal freeze-out to be discussed in Sec.~\ref{subsec:freezeout}. In this section, we consider the most general case in which $T$ and $v$ are independent. The readers should keep in mind that $T$ refers to the temperature of the mediator, but not to the temperatue of $\chi$.

Below we compute the Sommerfeld factor $S$ from Eq.~(\ref{eq:VSbkgmassless}) with two distributions: Maxwell-Boltzmann (MB) and Bose-Einstein (BE).

\subsection{Maxwell-Boltzmann}
\begin{figure}[t]
	\centering
	\includegraphics[width=4in]
	{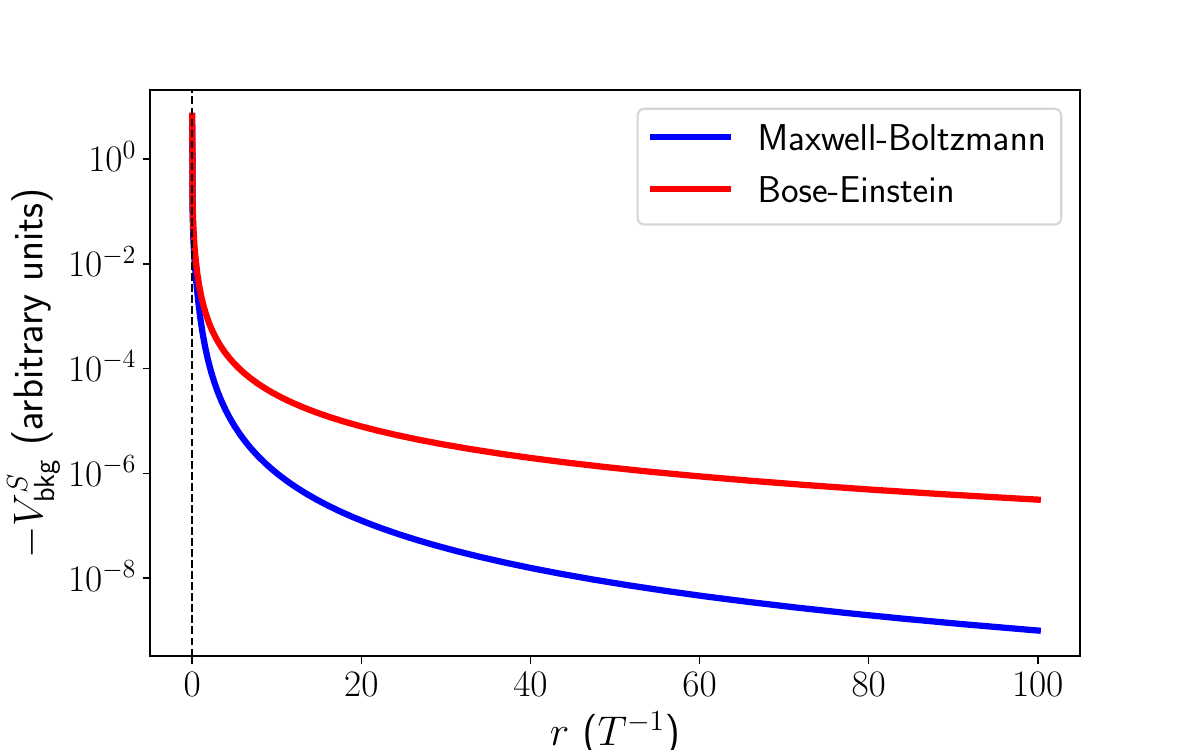}
	\caption{The two-scalar mediated background potentials (arbitrary units) as a function of distances (in units of the inverse temperature) for Maxwell-Boltzmann (blue curve) and Bose-Einstein  (red curve) distributions. }
	\label{fig:ScalarPotentials}
\end{figure}

First we study the MB distribution, where $f_{\phi}(\kappa) = e^{-\kappa/T}$. The integral in Eq.~(\ref{eq:VSbkgmassless}) can be analytically worked out:
\begin{align}
    V^{S}_{\text{bkg,MB}}(r) 
    &= 
    -\frac{1}{8\pi^{3}\Lambda^{2}r}
    \frac{T^{2}}{4r^{2}T^{2}+1} \label{eq:VSMB}\\
    &= -\frac{T^2}{8\pi^3 \Lambda^2}\times
    \begin{cases}  
    1/r & \text{for $r\ll T^{-1}$}\\
    1/(4T^2 r^3) & \text{for $r\gg T^{-1}$}
    \end{cases}.
\end{align} 

This potential is  plotted as the blue curve in Fig.~\ref{fig:ScalarPotentials}. It is attractive at all distances. In the long-distance limit, it scales as $1/r^3$ and does not depend on temperature. In the short-distance limit, where $1/\Lambda<r < 1/T$, the temperature behaves as a regulator, so the potential is no longer singular and scales as $1/r$. For $r<1/\Lambda$, the effective theory breaks down and the potential still scales as $1/r$, but the specific form depends on the UV completion.

\begin{figure}[t]
	\centering
	\includegraphics[width=6in]
	{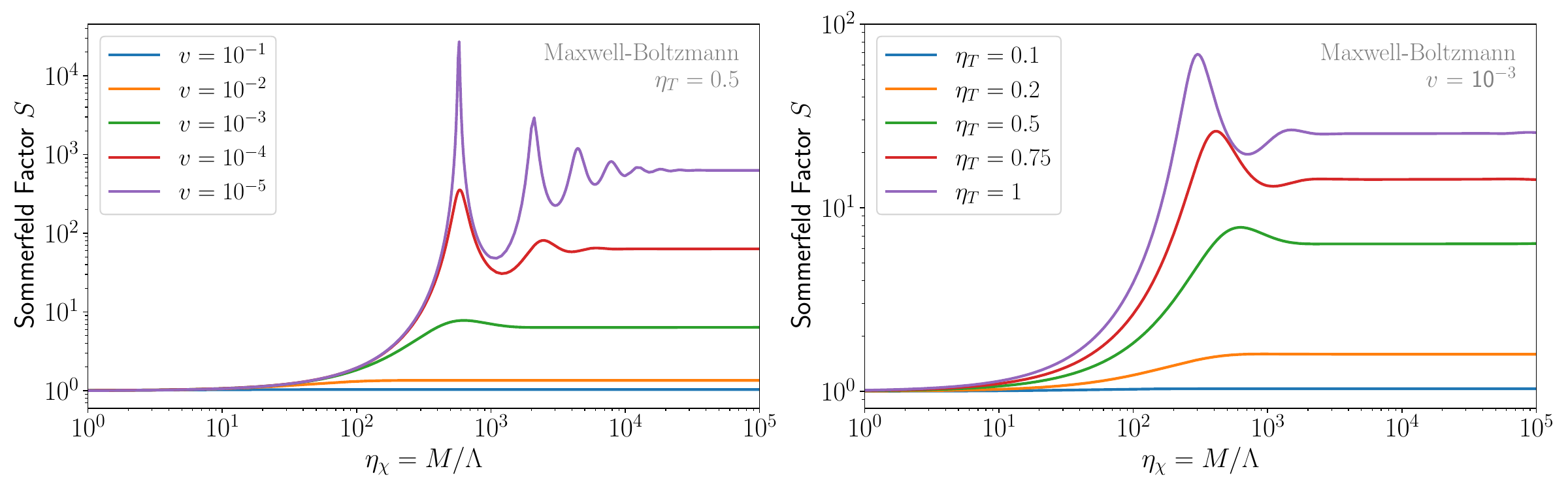}
	\caption{\label{fig:Scalar_MB_Sommerfeld}Sommerfeld enhancement from the two-scalar potential in a thermal background with Maxwell-Boltzmann distribution (\ref{eq:VSMB}). \emph{Left}: we fix the temperature to be $\eta_T \equiv T/\Lambda =0.5$ and vary the velocity. \emph{Right}: we fix the velocity to be $10^{-3}$ and vary the temperature. }
\end{figure}

Since the potential is not singular at short distances, it does  not need to be regularized. One can insert Eq.~(\ref{eq:VSMB}) into the Schr\"{o}dinger equation directly to get the physical solutions of wave functions. Using the strategy in Sec.~\ref{subsec:algorithm}, we compute the Sommerfeld factor induced by Eq.~(\ref{eq:VSMB}) and the result is shown in  Fig.~\ref{fig:Scalar_MB_Sommerfeld}. The typical resonance peaks also appear, which are purely induced by the finite-temperature effect in this case. In Sec.~\ref{subsec:boundstate}, we will use the bound-state approximation method to predict the locations of peaks analytically. 

The Sommerfeld factor in Fig.~\ref{fig:Scalar_MB_Sommerfeld} tends to a constant in the large $M$ limit. This is because in the limit of $Mv\gg T$, the normalized potential (\ref{eq:Vnorm}) scales as 
 \begin{align}
  \mathcal{V}^{S}_{\text{bkg,MB}}(x) &=  \frac{2M}{p^2}V^{S}_{\text{bkg,MB}}(x/p) =-\frac{M^2 T^2 v}{4 \pi ^3 \Lambda ^2 \left(M^2 v^2 x+4 T^2 x^3\right)}  \overset{Mv \gg T}{\approx} -\frac{T^2}{4\pi^3 \Lambda^2 v}\frac{1}{x}\;.
 \end{align}
This is an attractive Coulomb potential, where the Sommerfeld factor is predicted to be a constant:
\begin{align}
S_{\rm MB} = \frac{T^2}{4\pi^2  \Lambda^2 v} = \frac{\eta_T^2}{4\pi^2 v} \;,\qquad \text{for $Mv\gg T$}\;,\label{eq:SMB-largeM}
\end{align}
which agrees very well with the numerical results in Fig.~\ref{fig:Scalar_MB_Sommerfeld}.

Note that the effective theory breaks down when $M v > \Lambda$, so the exact value of the Sommerfeld factor in Fig.~\ref{fig:Scalar_MB_Sommerfeld} at $\eta_\chi>1/v$ depends on the UV theory. However, since the background potential already scales as $1/r$ at $r<1/T$, it does not change significantly as $r$ decreases further below $1/\Lambda$. Therefore, we expect that UV completion does not cause a qualitative difference for the Sommerfeld factor compared to the result extrapolated from the effective theory.


\subsection{Bose-Einstein}
For the case of BE distribution, $f_{\phi}(\kappa) = 1/(e^{\kappa/T}-1)$, the integral in Eq.~(\ref{eq:VSbkgmassless}) gives
\begin{align}
    V^{S}_{\text{bkg,BE}}(r) 
    &= 
    -\frac{1}{64\pi^{3}\Lambda^{2}r^{3}}
     \left[2\pi r T \coth(2\pi r T) - 1\right]\label{eq:VSBE}\\
    & = -\frac{T^2}{32\pi^2 \Lambda^2}\times
    \begin{cases}
     2\pi/(3r) & \text{for $r\ll T^{-1}$}\\
     1/(T r^2) & \text{for $r\gg T^{-1}$}
    \end{cases}.
\end{align}
The potential for BE distribution is also plotted in Fig.~\ref{fig:ScalarPotentials} (red curve) and compared with that for MB distribution.

We define the following quantity to characterize the difference between BE and MB statistics:
\begin{align}
{\cal Q}_S (y) &\equiv \frac{V^S_\text{bkg,BE}(r)}{V^S_\text{bkg,MB}(r)} = \frac{1+4y^2}{8y^2}\left[-1+2\pi y \coth\left(2\pi y\right)\right],\quad \text{with $y\equiv r T$}\;.\label{eq:QS}
\end{align}
The asymptotic behavior of ${\cal Q}_S$ turns out to be
\begin{align}
{\cal Q}_S(y) = 
\begin{cases}
\pi^2/6 & \text{for $y\ll 1$}\\
\pi y & \text{for $y\gg 1$}
\end{cases}\;.
\end{align}
So the effect of quantum statistics becomes significant in the low-energy limit ($r\gg 1/T$). This can be understood by the difference in the distribution functions between $f_{\rm MB}=e^{-\kappa/T}$ and $f_{\rm BE}=1/(e^{\kappa/T}-1)$: when $r\sim 1/\kappa \gg 1/T$, we have $f_{\rm MB}\approx 1$ and $f_{\rm BE}\approx T/\kappa \sim Tr \gg 1$, that is, there is a large number of bosons sitting in the low-energy state, similar to the Bose-Einstein condensate.

\begin{figure}[t]
	\centering
	\includegraphics[width=6in]
	{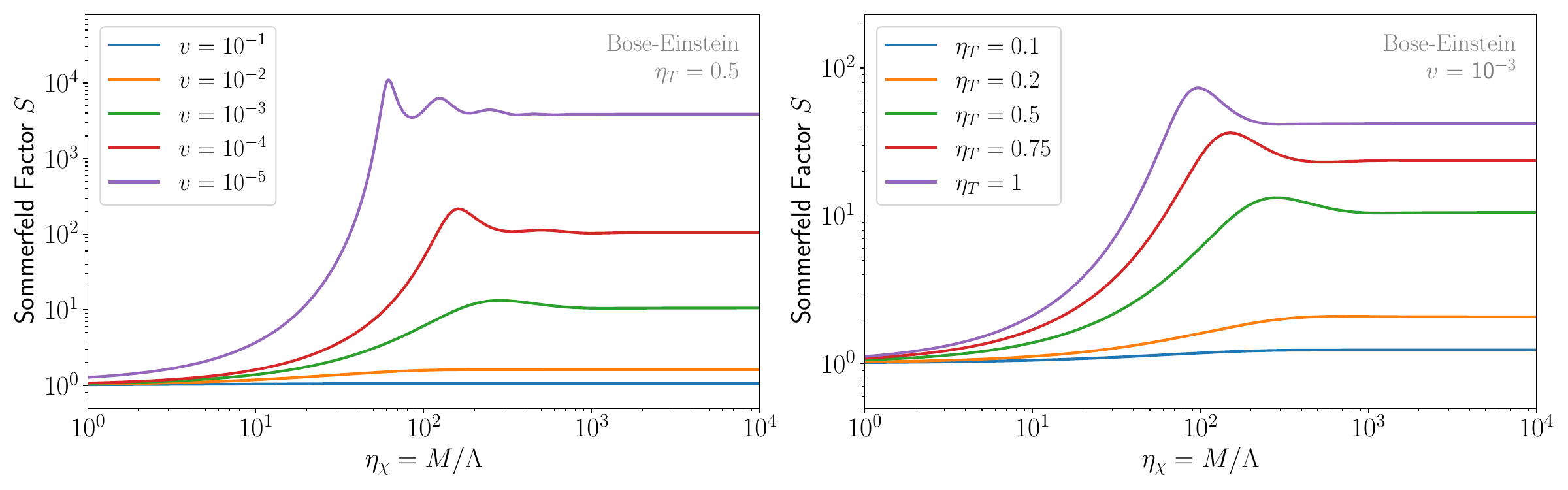}
	\caption{Same as Fig.~\ref{fig:Scalar_MB_Sommerfeld}, but for the background potential with Bose-Einstein distribution (\ref{eq:VSBE}). }
	\label{fig:Scalar_BE_Sommerfeld}
\end{figure}

The Sommerfeld factor calculated from Eq.~(\ref{eq:VSBE}) is plotted in Fig.~\ref{fig:Scalar_BE_Sommerfeld}.  Compared to the MB case, the peaks from BE distribution are less sharp. This is because the transition behavior of $V_\text{bkg,BE}^S$ around $1/T$ (from $1/r^2$ to $1/r$) is smoother than that of $V_\text{bkg,MB}^S$ (from $1/r^3$ to $1/r$). 

In the large $M$ limit, the normalized potential from BE distribution scales as 
\begin{align}
  \mathcal{V}^{S}_{\text{bkg,BE}}(x) \overset{M\gg T}{\approx} -\frac{T^2}{24\pi \Lambda^2 v}\frac{1}{x}\;,
\end{align}
which again leads to a constant Sommerfeld factor:
\begin{align}
S_{\rm BE} = \frac{\eta_T^2}{24 v}	 \;,\qquad \text{for $Mv\gg T$}\;. \label{eq:SBE-largeM}
\end{align}
Comparing with Eq.~(\ref{eq:SMB-largeM}), for a fixed temperature, we have $S_{\rm BE}/S_{\rm MB}=\pi^2/6 \approx 1.6$ in the large $M$ limit, which is exactly the value of ${\cal Q}_S$ (\ref{eq:QS}) in the small distance limit.

\subsection{Locations of peaks: bound-state approximation}
\label{subsec:boundstate}

As discussed in Sec.~\ref{subsec:box}, the resonance peaks in the Sommerfeld enhancement as a function of the DM mass $M$ correspond to the case where two $\chi$ can form a bound state with near-zero binding energy. Understanding the structure of the resonance peaks analytically is then equivalent to solving for the bound-state spectrum of a particular potential. For most potentials, a closed-form analytic solution does not exist. However, one may still gain insight by replacing the actual interaction potential with an approximate potential which has a similar structure, but admits an analytic solution. A well-known example is the box approximation discussed in Sec.~\ref{subsec:box}, which works well when the potential changes rapidly around a transition point. For the vacuum two-scalar (or two-fermion) potential, the transition around $r=1/\Lambda$ is from $1/r^3$ (or $1/r^5$) to a constant. This can be well captured by the box approximation. For background potentials, the transition around $r=1/T$ is from $1/r^3$ (for MB) or $1/r^2$ (for BE) to $1/r$, which is smoother than the transition in the vacuum case. As a result, we expect that the box approximation does not fit the background potentials as well as the vacuum case. In this subsection, we use a different approximation --- the bound-state approximation --- to predict the location of peaks of the Sommerfeld enhancement from background potentials.

Consider an attractive potential $V(r)$ which has Coulomb-like behavior, $V(r) \sim \alpha/r$, at short distances, and drops off faster than Coulomb at large distances. Such a potential can be approximated as follows: 
\begin{align}
V_{\rm app}(r) = 
-\frac{\alpha}{r} +\frac{\alpha}{r_0}\;,
\label{eq:Vapp}
\end{align}
where $r_0$ is a constant. Physically, $r_0$ models the characteristic distance that separates the Coulomb and non-Coulomb behaviors of the original potential. Note that $V_{\rm app}$ does not have the same asymptotic behavior at large distance as $V$, which tends to zero. However, this turns out to be qualitatively unimportant for finding the bound states, whose wavefunctions decay exponentially at large $r$. The bound state spectrum of the potential~(\ref{eq:Vapp}) is given by the familiar hydrogen-atom spectrum shifted by a positive additive constant $C=\alpha/r_0$. Demanding that the potential supports a zero-energy bound state yields   
\begin{align}
-\frac{\alpha^2 M_n}{2n^2}\,+\,C=0\quad \implies \quad M_n = \frac{2n^2 C}{\alpha^2}\;,\quad n=1,2,3\cdots
\label{eq:hydrogenspectrum}
\end{align}
As long as the spectrum of $V_{\rm app}$ is a good approximation to that of the original potential, we expect the Sommerfeld enhancement to exhibit a peak whenever the condition~(\ref{eq:hydrogenspectrum}) is satisfied. Note that for a pure Coulomb potential, $C=0$ and we predict that there are no peaks, consistent with the well-known analytic result.

\subsubsection*{Yukawa force}
For the Yukawa potential $V_Y(r) = -\alpha e^{-m_\phi r}/r$, the transition point is roughly determined by the inverse of the mass of the mediator. We evaluate the potential at $r = 1/m_\phi$:
\begin{align}
    V_Y \left(r=1/m_\phi\right) = -\frac{\alpha}{r} + \alpha m_\phi\left(1-e^{-1}\right).
\end{align}
In this case, we obtain $C = \alpha m_\phi (1-e^{-1}) \approx 0.6\,\alpha m_\phi$, so the condition to form a bound state is $\alpha m_\phi \gtrsim M v^2/2$. Using Eq.~(\ref{eq:hydrogenspectrum}) and including the non-relativistic kinetic energy, the locations of the peaks are predicted to be
\begin{align}
\frac{M_n}{m_\phi} = \frac{2\alpha\left(1-e^{-1}\right)}{\alpha^2/n^2 +v^2} \overset{v\to 0}{=} 2\left(1-e^{-1}\right)\frac{n^2}{\alpha}\;. \label{eq:MnYukawa}
\end{align}
This gives the known distribution of the peaks for the Sommerfeld enhancement from the Yukawa potential~\cite{Lattanzi:2008qa}.

\subsubsection*{Quantum force}
Now we apply this method to the potential due to a background of Maxwell-Boltzmann or Bose-Einstein scalars, given in Eqs.~(\ref{eq:VSMB}) and (\ref{eq:VSBE}). Evaluating the potential at $r=b_{i}/T$ (where $b_{i}\sim {\cal O}(1)$ is a dimensionless positive constant to be determined separately for $i$=MB and $i$=BE), one obtains
\begin{align}
    V^{S}_{\text{bkg}, i}(r=b_{i}/T) = -\frac{\alphaeff^{i}}{r} + T\alphaeff^{i}g_{i}(b_{i})\;,
\end{align}
where 
\begin{align}
    \alphaeff^{i} = \frac{T^{2}}{\pi^{3}\Lambda^{2}}
    \begin{cases}
        1/8  & i = \text{MB} \\
        1/64 & i = \text{BE} 
    \end{cases}\;,
\end{align}
is the effective coupling and
\begin{align}
    g_{i}(b_i) = 
    \begin{cases}
        \frac{4b_{i}}{4b_{i}^{2}+1} &  i = \text{MB} \\
        \frac{1-2\pi b_{i} \coth(2\pi b_i) + b_{i}^{2}}{b_{i}^{3}} & i = \text{BE} 
    \end{cases}\;.
\end{align}
The condition of having bound states is given by
\begin{align}
T \alphaeff^{i}\,g_{i}(b_{i}) > \frac{1}{2}Mv^2\;, \quad \text{$i=$MB, BE}\;.
\end{align}

According to Eq.~(\ref{eq:hydrogenspectrum}), the estimations for the mass spectra of bound states are then
\begin{align}
    M_{n}^{i} (v) = \frac{2n^2\alphaeff^{i} g_{i}(b_{i})}{(\alphaeff^{i})^{2} + v^{2}n^{2}}\,T \approx \frac{2n^2 g_i(b_i)}{\alpha_{\rm eff}^i}\,T\;, \qquad \text{$i=$MB, BE}\;,
    \label{eq:Mn}
\end{align}
where the approximation holds if $v\ll \alpha_{\rm eff}^i$.
We fix the value of $b_{i}$ by matching exactly for the first peak, $n=1$, with the numerical result. In Table \ref{tab:peaklocationsboundstate}, we compare the bound-state approximation in Eq.~(\ref{eq:Mn}) and the numerical result in Figs.~\ref{fig:Scalar_MB_Sommerfeld}-\ref{fig:Scalar_BE_Sommerfeld} for the first three peaks.

\renewcommand\arraystretch{1.3}
\begin{table}[t]
	\centering
	\begin{tabular}{c|c|c||c|c}
		\hline\hline
		& Numerical (MB) & Bound State ($b_{\rm MB} = 1.62$) & Numerical (BE) & Bound State ($b_{\rm BE} =6.85$) \\
		\hline
		$n=1$ & 560 & 560 & 120 & 120 \\
		\hline
		$n=2$ & 2100 & 2200 & 280 & 480\\
		\hline
		$n=3$ & 4500 & 5000 & 600 & 1100\\
		\hline\hline
	\end{tabular}
	\caption{\label{tab:peaklocationsboundstate}Comparison of the locations of the first three peaks (in units of $\eta_\chi \equiv M/\Lambda$) between numerical results and bound-state approximation results for Maxwell-Boltzmann (first two columns) and Bose-Einstein (last two columns). We fix $\eta_T \equiv T/\Lambda = 0.5$ and $v=10^{-5}$ as a benchmark. The dimensionless coefficients $b_i$ (for $i={\rm MB},{\rm BE}$) are determined by minimizing the difference between the numerical result and the bound-state approximation result for the location of the first peak $(n=1)$.}
\end{table}
\renewcommand\arraystretch{1}

It is worth pointing out that this result can be understood as a generalization of the standard result for Sommerfeld enhancement in the context of a massive mediator at the tree level. The attractive Yukawa potential $V_Y=-\alpha e^{-m_\phi r}/r$ is well known to have resonant Sommerfeld enhancement for certain values of the scattered particle's mass. These resonances are located approximately at $M_n\sim n^{2}m_\phi/\alpha$~\cite{Lattanzi:2008qa}. In our context of background potentials, the mediator mass is negligible, while the locations of resonances are found to be at $M_n \sim n^2 \Lambda^2/T$ in Figs.~\ref{fig:Scalar_MB_Sommerfeld} and \ref{fig:Scalar_BE_Sommerfeld}.
This can be understood by noting that the mediator will acquire an effective mass in the thermal bath that scales with the temperature of the bath.\footnote{This should not be confused with the thermal mass of $\phi$ in a bath of $\chi$ (with temperature $T_\chi$) obtained through its coupling to $\chi$, which is given by $[m(T_\chi)]^2\sim \langle \bar\chi \chi\rangle/\Lambda \sim T^3_\chi/\Lambda$. On the other hand, the effective mass $m_{\rm eff}\sim T_\phi$ considered here should be understood as the typical energy scale of $\phi$ in a background of relativistic $\phi$ particles (with temperature $T_\phi$). In general, $T_\chi \neq T_\phi$.} The bath also gives an effective coupling depending on the temperature. Identifying $m_{\rm eff}\sim T$ and $\alpha_{\rm eff}\sim T^{2}/\Lambda^{2}$, we find that the locations of the resonance peaks for background potentials have a similar structure to that for the Yukawa potential: $M_n \sim n^2 m_{\rm eff}/\alpha_{\rm eff}$.

\subsection{Combination of vacuum and background effects}
\begin{figure}[t]
\centering
\includegraphics[scale=0.5]
{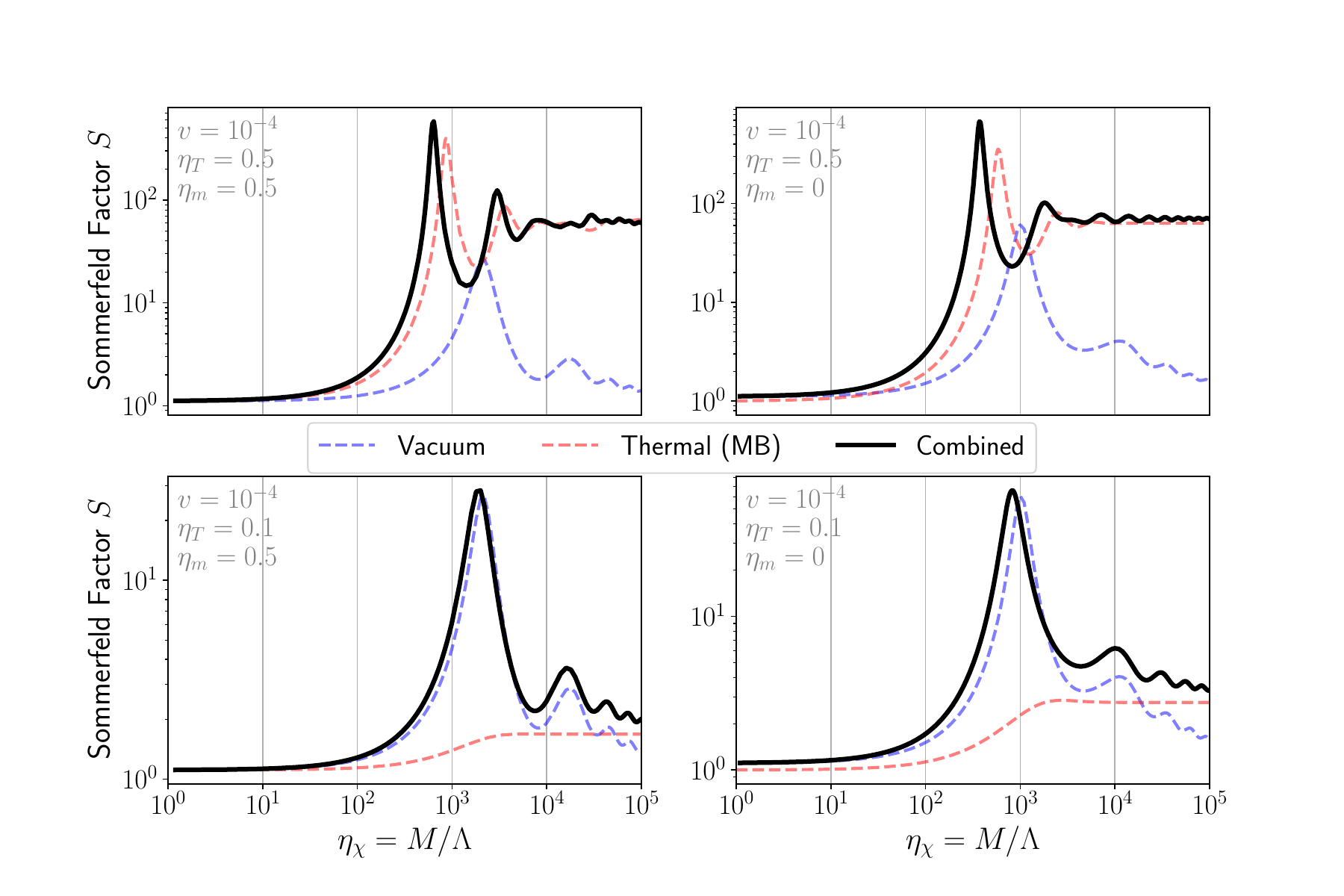}
\caption{\label{fig:CombinedScalarMB}Sommerfeld enhancement from two-scalar exchange potentials. The blue and red dashed lines denote the contribution from the vacuum and background (with MB distribution) potentials, respectively, while the black solid lines represent the combination of both effects. We fix the temperature to be $\eta_T\equiv T/\Lambda=0.5$ (0.1) in the top (bottom) two panels; we fix the mediator mass to be $\eta_m\equiv m_\phi/\Lambda=0.5$ (0) in the left (right) two panels; we fix the relative velocity to be $v=10^{-4}$ in all panels.}
\end{figure}

In the presence of a background of mediators, the total potential is given by the sum of the vacuum and the background potentials:
\begin{align}
V_{\text{combined}}^S = V_{0}^S + V_{\text{bkg}}^S\;.    
\end{align}
We obtain the combined contribution to the Sommerfeld enhancement by solving the Schr\"{o}dinger equation with $V_{\text{combined}}^S$.

In Fig.~\ref{fig:CombinedScalarMB}, we show the contribution to the Sommerfeld enhancement from $V_0^S$ and $V_{\rm bkg}^S$, as well as the total effect from $V_{\rm combined}^S$. For illustration, we assume the MB distribution for the background. In the large $M$ limit (which corresponds to a large momentum transfer or small distances), the background contribution dominates over the vacuum one. This is because the background potential scales as $1/r$ at short distances, while we regularize the vacuum potential as a constant at $r<1/\Lambda$. Note that the comparison at $M>\Lambda/v$ is not model-independent since it goes beyond the effective theory region, and the result may be changed with a different regulator for the vacuum potential.
For $M<\Lambda/v$, where the effective theory is valid and both potentials are insensitive to the UV completion, we observe that both vacuum and background potentials can dominate in certain regions of parameter space. The background effect is more significant at higher temperatures, as one can see in the top two panels. The left two panels show that 
a larger mediator mass leads to a smaller $S$, which is expected from previous calculations of $S$ with different $\eta_{m}$.

\section{Correction from thermal background: fermionic mediator}
\label{sec:thermal-fermion}

In this section, we compute the background contribution to the Sommerfeld enhancement with a fermionic mediator $\psi$.  

Similarly to the case of the scalar mediator, the fermion propagator is modified in a background of finite number density of $\psi$ particles, giving a correction to the quantum forces mediated by two fermions. The fermionic propagator is modified as
\begin{align}
    \frac{i(\slashed{k} + m_{\psi})}{k^{2}-m_{\psi}^{2}}
    ~~\rightarrow~~
    \left(\slashed{k} + m_{\psi}\right)
    \bigg( 
     \frac{i}{k^{2}-m_{\psi}^{2}}
     - 2\pi\delta\left(k^{2}-m_{\psi}^{2}\right)f_{\psi}(\veck) 
     \bigg)\;,
     \label{eq:ModifiedFermionProp}
\end{align}
where again $f_{\psi}(\veck)$ is the phase-space distribution of the fermionic mediator background, such that $\int {\rm d}^3 \veck \, f_\psi(\veck)/(2\pi)^3$ gives the number density of $\psi$ in the background.

For the two-fermion mediated potentials, we focus on the contributions from the effective operator with scalar-scalar coupling in Eq.~(\ref{eq:OF}) and with vector-vector coupling in Eq.~(\ref{eq:OFtilde}), where the corresponding background potentials are denoted by $V_{\rm bkg}^F$ and $V_{\rm bkg}^{\widetilde{F}}$, respectively. In Appendix~\ref{app:quantum-force}, we show how to calculate $V_{\rm bkg}^F$ and $V_{\rm bkg}^{\widetilde{F}}$ with an arbitrary $f_{\psi}(\veck)$. Throughout this section, again, we assume that the background is isotropic and the mediator mass is negligible for the same reason mentioned in the last section. In this case, the background potentials are given by
\begin{align}
    V_{\text{bkg}}^{F} (r) &= 
    -\frac{1}{4\pi^{3}\Lambda^{4}r^{4}}
    \int_{0}^{\infty} {\rm d}\kappa 
    \,
    f_{\psi}(\kappa)
   \left[
    \left(2\kappa^{2}r^{2}-1\right)\sin\left(2\kappa r\right) + 2\kappa r\cos\left(2\kappa r\right) 
    \right],\label{eq:VFbkgmasless}\\
   V_{\rm bkg}^{\widetilde{F}} (r) &= -\frac{1}{2\pi^3\Lambda^4 r^4}\int_{0}^{\infty} {\rm d}\kappa 
    \,
    f_\psi(\kappa)\left[\sin\left(2\kappa r\right)-2\kappa r \cos\left(2\kappa r\right)\right].\label{eq:VFbkgtildemassless} 
\end{align}
In the following we study the Sommerfeld factors from Eqs.~(\ref{eq:VFbkgmasless}) and (\ref{eq:VFbkgtildemassless}) separately for specific phase-space distributions. 

\subsection{Sommerfeld factor from scalar-scalar coupling}
\label{subsec:fermionscalar}

We consider two thermal distributions for the fermionic mediator: the Maxwell-Boltzmann (MB) distribution and the Fermi-Dirac (FD) distribution. For both cases, we can obtain an analytic expression for the integral in Eq.~(\ref{eq:VFbkgmasless}). We then compute the Sommerfeld factor due to these potentials, and discuss its interpretation in terms of quasi-bound states. 

\subsubsection*{Maxwell-Boltzmann potential}

\begin{figure}[t]
\centering
\includegraphics[width=4in]
{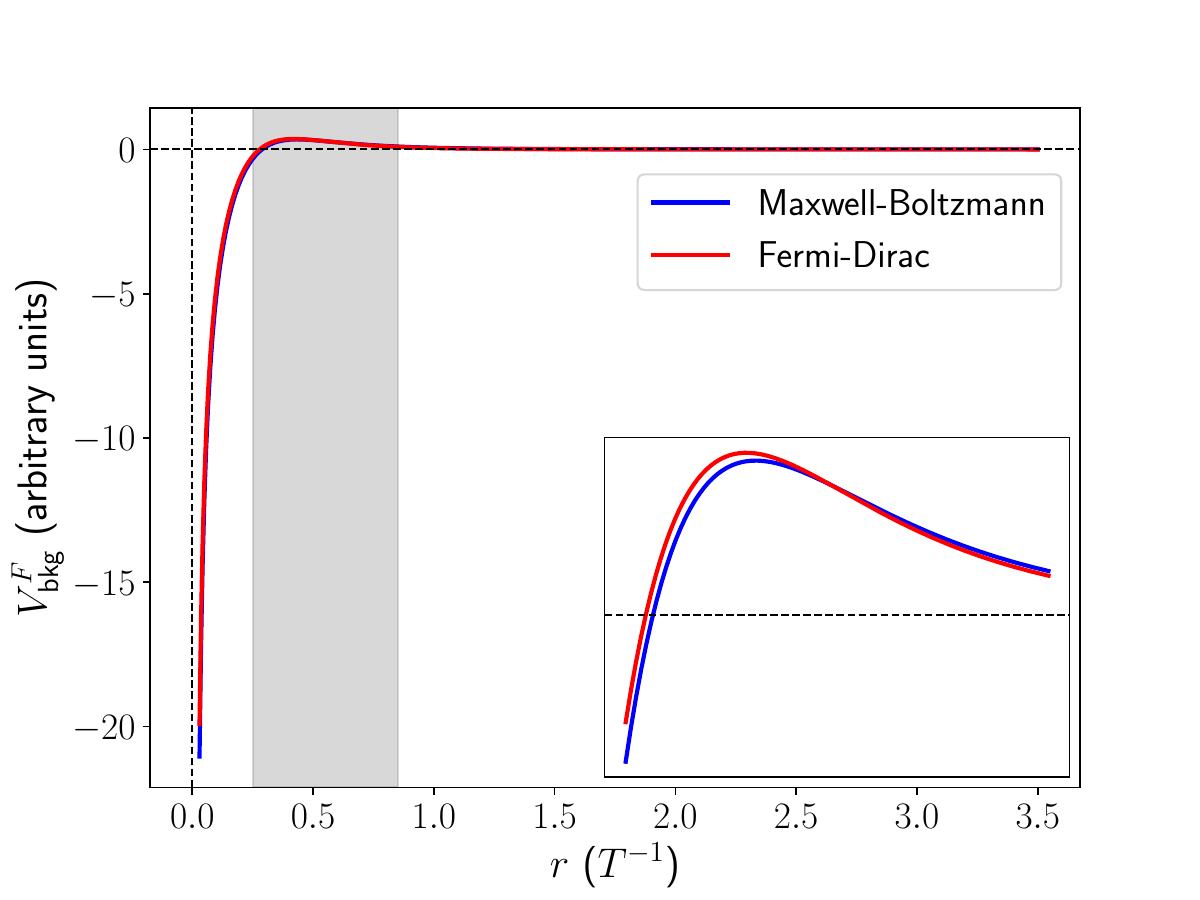}
\caption{Two-fermion exchange background potentials (arbitrary units) with scalar-scalar coupling as a function of distances (in units of the inverse temperature) for Maxwell-Boltzmann (blue) and Fermi-Dirac (red) distributions.  }
\label{fig:FermionPotentials}
\end{figure}

First, we study the MB distribution, where $f_{\psi}(\kappa) = e^{-\kappa/T}$. The background potential (\ref{eq:VFbkgmasless}) simplifies to
\begin{align}
    V^{F}_{\text{bkg,MB}}(r) 
    = 
    -\frac{2T^4}{\pi^{3}\Lambda^{4}r}
    \frac{1 - 12r^{2}T^{2}}
         {\left(1+4r^{2}T^{2}\right)^{3}}\;.
\label{eq:FermionMBPotential}        
\end{align}
This potential is plotted in Fig.~\ref{fig:FermionPotentials} with the blue curve, and its asymptotic behavior is given by
\begin{align}
V^F_\text{bkg,MB}(r)= \frac{2T^4}{\pi^3 \Lambda^4}\times
\begin{cases}
-1/r& \text{for $r\ll T^{-1}$}\\
3/(16T^4 r^5) & \text{for $r\gg T^{-1}$}
\end{cases}\;.\label{eq:FermionMBPotentialasy}  
\end{align}
Interestingly, the two-fermion potential is attractive at short distances but becomes repulsive at large distances, which will lead to a very different behavior of the Sommerfeld effect compared to the previous cases, as we show below. 

\subsubsection*{Fermi-Dirac potential}
For the FD distribution $f_{\psi}(\kappa)=1/\left(e^{\kappa/T}+1\right)$, the background potential (\ref{eq:VFbkgmasless}) reduces to
\begin{align}
    V^{F}_{\text{bkg,FD}}(r) 
    =
    \frac{1}{16\pi^{3}\Lambda^{4}r^{5}}
    \left\{3+ \xi\,\text{csch}^{3}(2\xi)
    \left[
    1 - 6\xi^2 - \left(1+2\xi^2\right)\cosh\left(4\xi\right) - 2\xi\sinh\left(4\xi\right)
    \right]
    \right\},
    \label{eq:FermionFDPotential} 
\end{align}
where $\xi\equiv \pi r T$.
This potential is plotted in Fig.~\ref{fig:FermionPotentials} with the red curve and compared to the previous MB case. 
Its asymptotic behavior is given by
\begin{align}
V^F_\text{bkg,FD}(r)= \frac{T^4}{8\pi^3 \Lambda^4}\times
\begin{cases}
-7\pi^4/(45 r)& \text{for $r\ll T^{-1}$}\\
3/(2T^4r^5) & \text{for $r\gg T^{-1}$}
\end{cases}\;.    \label{eq:FermionFDPotentialasy} 
\end{align}
The potentials for the MB and FD distributions are very similar. Comparing the two-fermion background potentials in Fig.~\ref{fig:FermionPotentials} with the two-boson background potentials in Fig.~\ref{fig:ScalarPotentials}, an important difference is that for fermions, the classical and quantum statistics lead to the same scaling behaviors at both long and short distances, while for bosons, they only agree at short distances, but the quantum correction dominates at long distances. 

\subsubsection*{Quasi-bound state and the WKB approximation}

\begin{figure}[t]
\centering
\includegraphics[width=6in]
{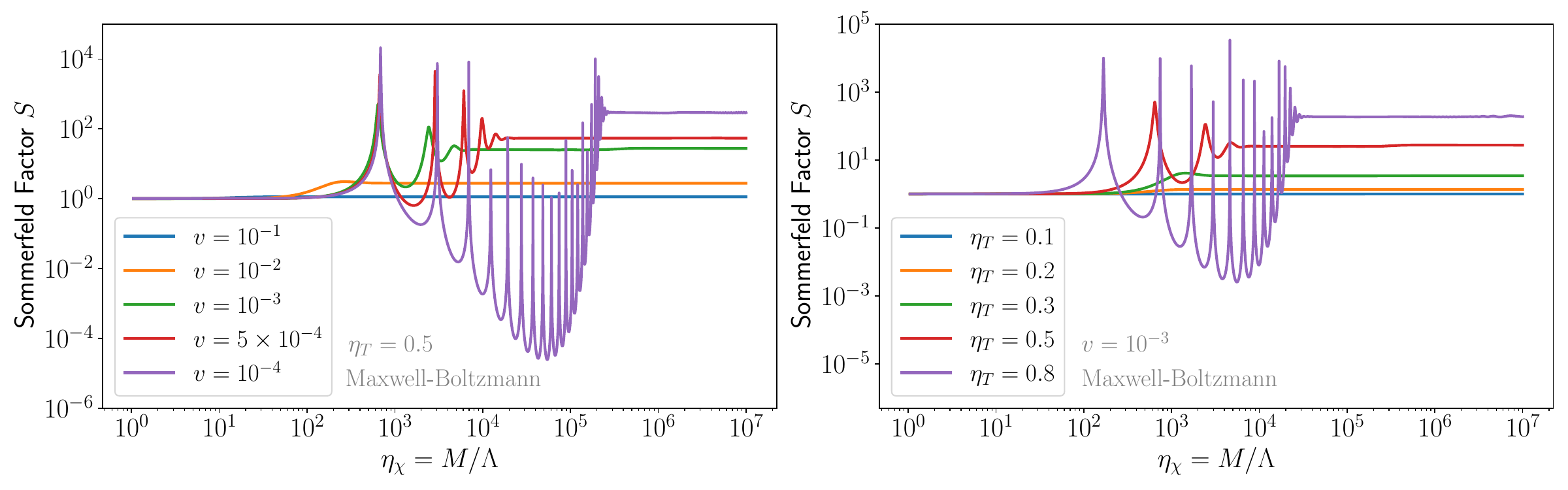}
\caption{Sommerfeld factor from the two-fermion potential with scalar-scalar coupling in a thermal background with Maxwell-Boltzmann distribution (\ref{eq:FermionMBPotential}). Same conventions as Fig.~\ref{fig:Scalar_MB_Sommerfeld}. We have numerically verified that the potential with Fermi-Dirac distribution (\ref{eq:FermionFDPotential}) leads to almost the same Sommerfeld factor as Maxwell-Boltzmann distribution.}
\label{fig:Fermion_MB_Sommerfeld}
\end{figure}

Next, we turn to studying the Sommerfeld enhancement from the two-fermion background potential. Numerically, the MB and FD distributions only lead to small differences for potentials, as shown in Fig.~\ref{fig:FermionPotentials}. As a result, the Sommerfeld factor calculated from $V_\text{bkg,MB}^F$ and $V_\text{bkg,FD}^F$ are almost the same. (We have numerically verified this point.) Therefore, in the following, we only discuss the result from the MB distribution. 
The Sommerfeld factor is plotted in Fig.~\ref{fig:Fermion_MB_Sommerfeld}.

From Fig.~\ref{fig:Fermion_MB_Sommerfeld}, we find that there is Sommerfeld suppression for certain ranges of velocity, temperature, and DM mass. This can be understood in the following way. First, in the limit of $Mv \gg T$ (which corresponds to $r\ll 1/T$), the background potential is reduced to an attractive Coulomb potential, as shown in Eq.~(\ref{eq:FermionMBPotentialasy}), so we get a constant value of $S>1$ in the large $M$ limit:
\begin{align}
S = \frac{4 T^4}{\pi^2 \Lambda^4 v} = \frac{4 \eta_T^4}{\pi^2 v}\;,\qquad \text{for $Mv\gg T$}\;.
\end{align}
Next, we look for the condition to get $S<1$. From Fig.~\ref{fig:FermionPotentials}, we see that the Sommerfeld suppression occurs if the kinetic energy is smaller than the height of the energy barrier, in which case the Sommerfeld factor describes the tunneling rate. To obtain the height of the barrier, we maximize the background potential (\ref{eq:FermionMBPotential}) and get
\begin{align}
 r_* = \frac{1}{T} \sqrt{\frac{1}{12}+\frac{1}{3\sqrt{10}}} \approx \frac{0.43}{T}  \implies \left(V_\text{bkg,MB}^F\right)_{\rm max} = V_\text{bkg,MB}^F(r=r_*) \approx 0.035 \eta_T^4 T\;.
\end{align}
On the other hand, the kinetic energy at $r=r_*$ is given by $Mv^2/2 \sim v/(2 r_*)$, where we have used the fact that the momentum around $r_*$ is of order $1/r_*$. Therefore, the condition to have the Sommerfeld suppression is given by
\begin{align}
\frac{v}{2r_*} \lesssim  V_\text{bkg,MB}^F(r=r_*) \quad  \implies \quad v \lesssim  0.03\,\eta_T^4\;.
\end{align}
Note that this condition does not depend on the mass of the DM. For example, for $\eta_T=T/\Lambda=0.5$, the above condition reads $v\lesssim 2 \times 10^{-3}$.

\begin{figure}[t]
\centering
\includegraphics[width=3.9in]
{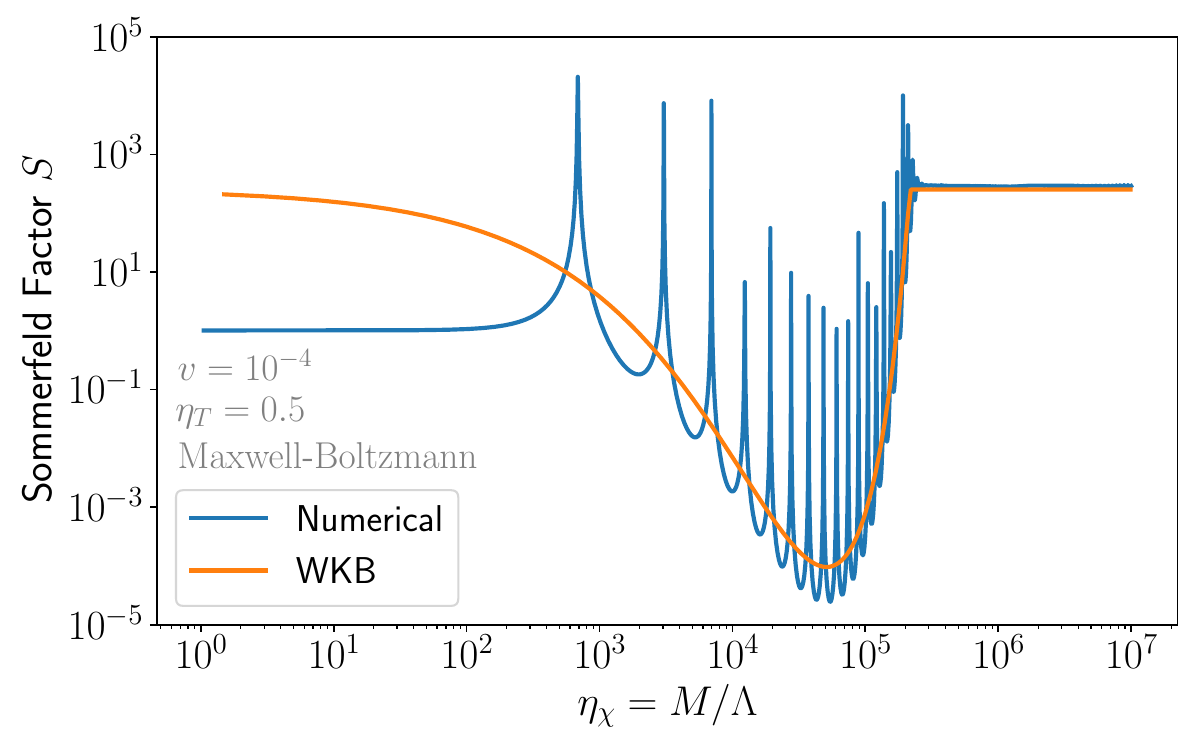}
\caption{Comparison between the numerical calculation of the Sommerfeld factor from the potential $V_\text{bkg,MB}^{F}$ in Eq.~(\ref{eq:FermionMBPotential}) (blue curve) and the prediction using WKB approximation (orange curve). As a benchmark, the temperature $\eta_T \equiv T/\Lambda=0.5$ and the relative velocity $v=10^{-4}$ are assumed in both cases.}
\label{fig:WKB}
\end{figure}

Using the WKB method, we can calculate the tunneling rate through the potential barrier shown in Fig.~\ref{fig:FermionPotentials}. The transmission coefficient $\mathcal{T}$ can be calculated using Eq.~(\ref{eq:transmission}) with the potential given in Eq.~(\ref{eq:FermionMBPotential}).
As a function of mass, we find that $\mathcal{T}$ follows the trend of the Sommerfeld suppression curves. This is shown as the orange curve in Fig.~\ref{fig:WKB}, where $\mathcal{T}$ is rescaled to match the numerical result for large $M$.

In addition to the tunneling suppression, the Sommerfeld factor in Figs.~\ref{fig:Fermion_MB_Sommerfeld} and~\ref{fig:WKB} also exhibits a resonant peak structure. The potential in Fig.~\ref{fig:FermionPotentials} can support quasi-bound states with $V_{\rm max}>E>0$. Presumably, the peaks in $S$ correspond to parameters for which the energy of the colliding DM pair matches one of these quasi-bound states. It would be an interesting exercise to fully work out the quasi-bound state spectrum from the potentials in Fig.~\ref{fig:FermionPotentials} and compare it with the location of the peaks in $S$, although we will not pursue it here.

\subsubsection*{Combination of vacuum and background effects}
\begin{figure}[t]
\centering
\includegraphics[scale=0.5]
{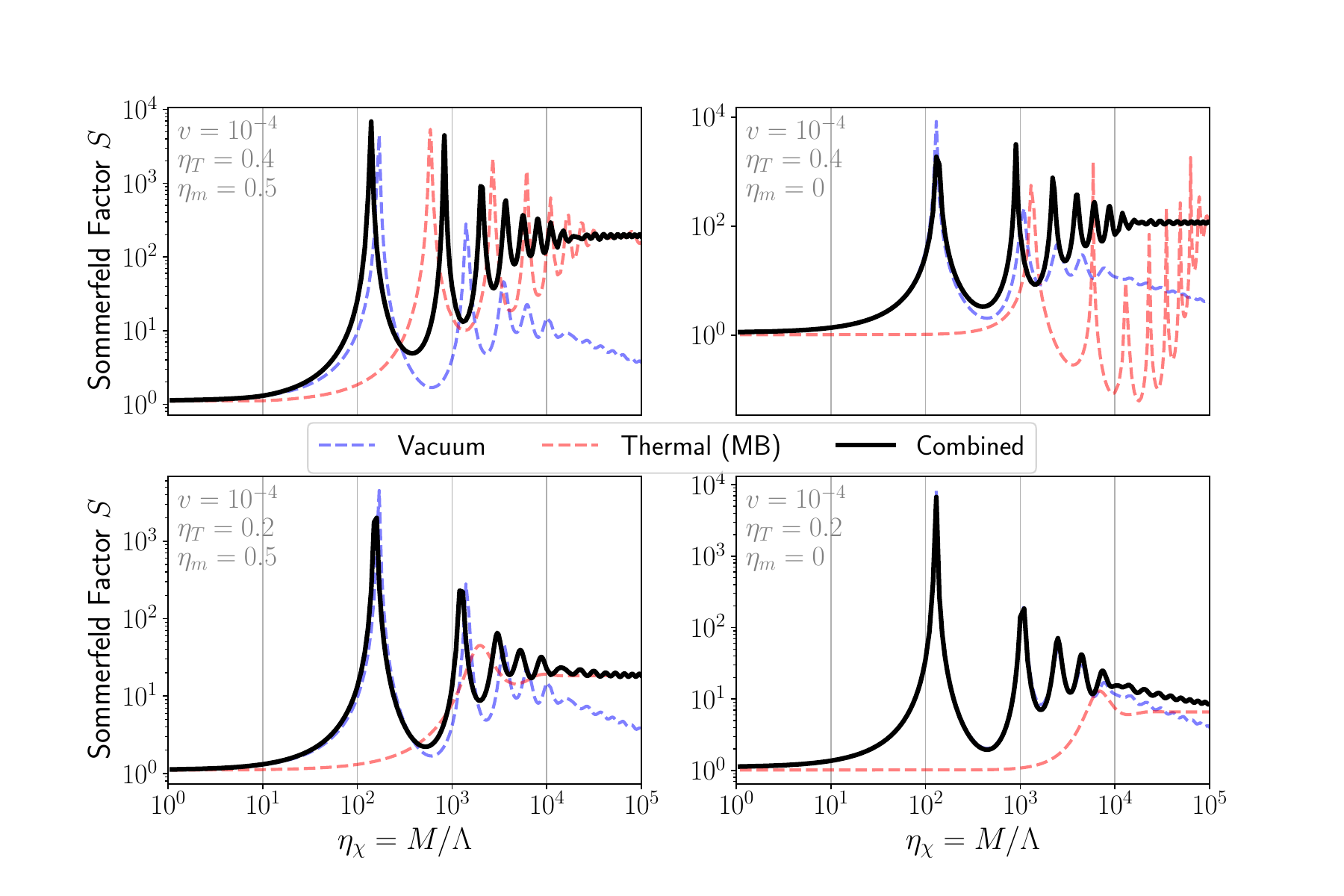}
\caption{\label{fig:CombinedFermionMB}Sommerfeld enhancement from two-fermion exchange potentials with scalar-scalar coupling, including both vacuum and background effects. Same conventions as Fig.~\ref{fig:CombinedScalarMB}.}
\end{figure}

In the presence of the fermionic background, the total potential relevant to the Sommerfeld enhancement is given by 
\begin{align}
V_{\text{combined}}^F = V_{0}^F + V_{\text{bkg}}^F\;. 
\end{align}
The Sommerfeld factor computed from $V_{\rm combined}^F$ as a function of the DM mass is shown in Fig.~\ref{fig:CombinedFermionMB} with different temperatures and mediator masses. For illustration, we assume the MB distribution for the background and fix the relative velocity to be $v=10^{-4}$.  Again we see that 
a larger temperature leads to a larger $S$. We observe again that for smaller $\eta_{\chi}$ and $\eta_{T}$, the vacuum contribution tends to dominate, whereas the background is dominant for large $\eta_{\chi}$ regardless of the temperature. 

\subsection{Sommerfeld factor from vector-vector coupling}
\label{subsec:fermionvector}

\begin{figure}[t]
\centering
\includegraphics[width=3.5in]
{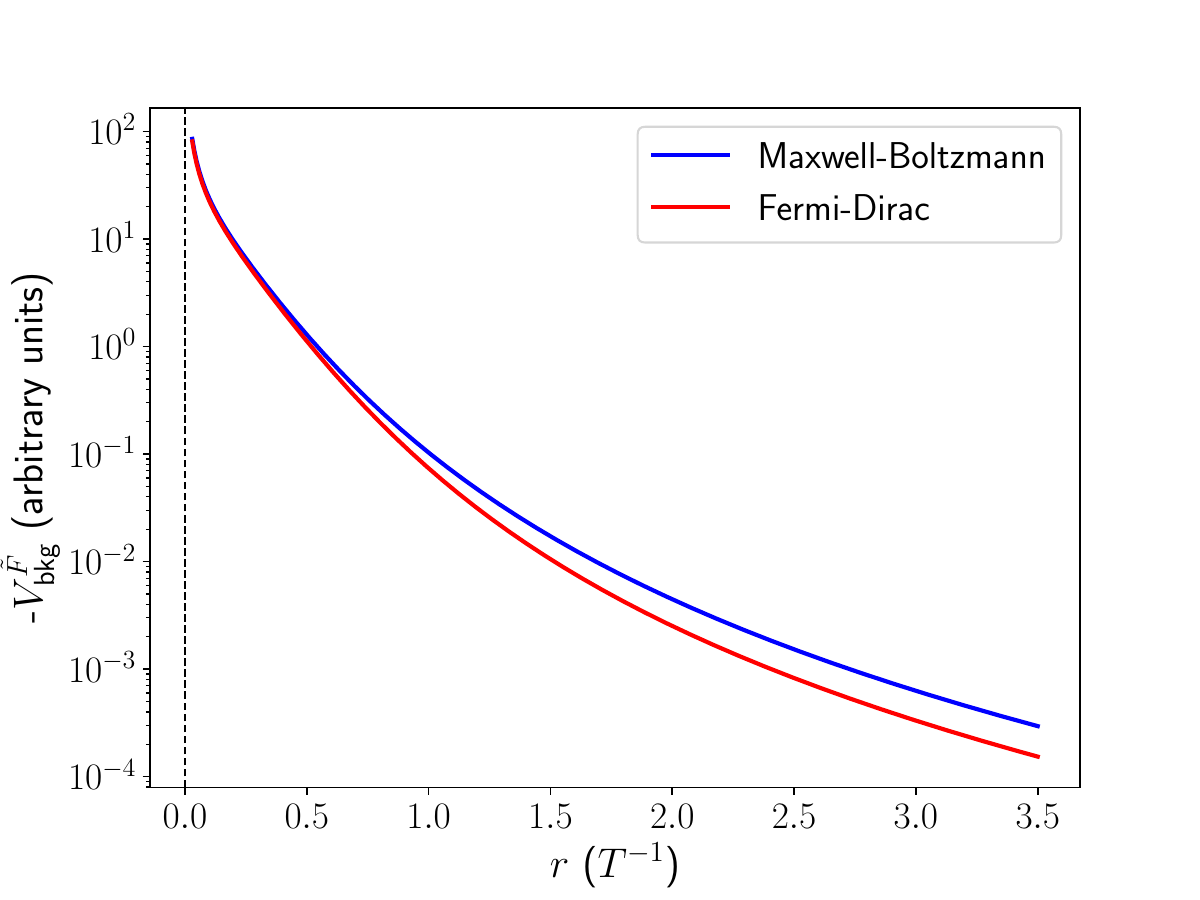}
\caption{Two-fermion exchange background potentials (arbitrary units) with vector-vector coupling as a function of distance (in units of inverse temperature), shown for Maxwell-Boltzmann (blue) and Fermi-Dirac (red) distributions.}
\label{fig:VFtilde}
\end{figure}

\begin{figure}
\centering
\includegraphics[width=6in]
{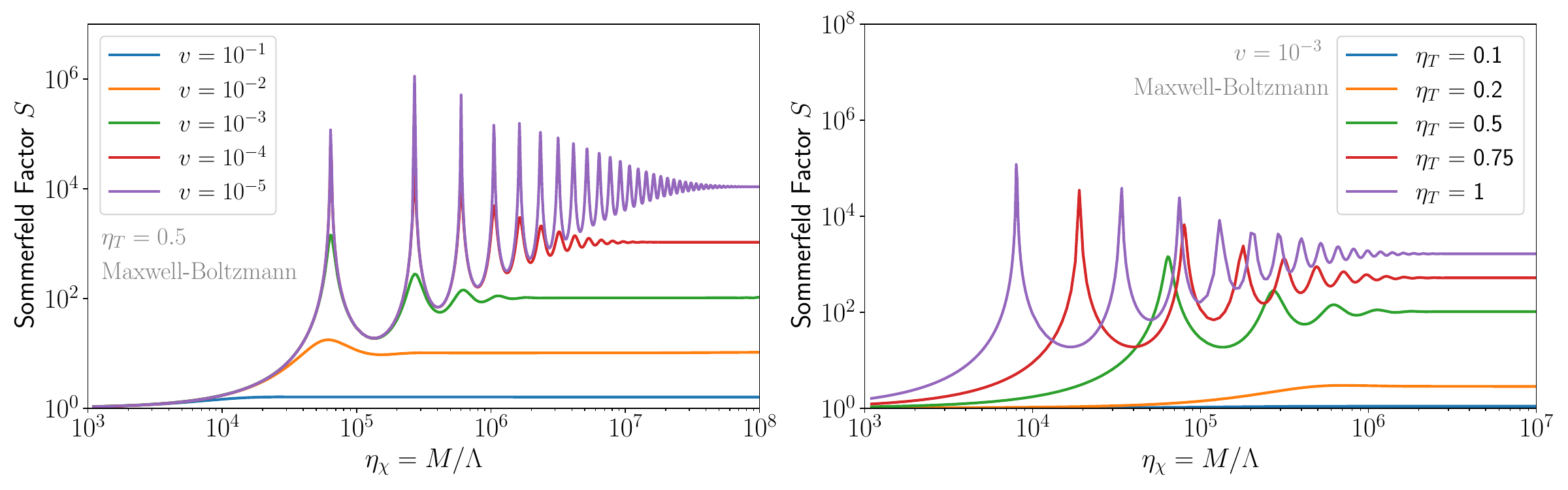}
\caption{Sommerfeld enhancement from the two-fermion potential with vector-vector coupling in a thermal background with Maxwell-Boltzmann distribution (\ref{eq:FermionTildeMBPotential}). Same conventions as Fig. \ref{fig:Scalar_MB_Sommerfeld}.} 
\label{fig:FermionTilde_MB_Sommerfeld}
\end{figure}

Next, we consider the two-fermion potential with vector-vector coupling (\ref{eq:VFbkgtildemassless}) and the resulting Sommerfeld enhancement. We show the result in a manner parallel to that in Sec.~\ref{subsec:fermionscalar}.

The integral in Eq.~(\ref{eq:VFbkgtildemassless}) can also be worked out analytically for standard distributions of the fermionic background. For Maxwell-Boltzmann distribution, we obtain
\begin{align}
    V_{\rm bkg, \text{MB}}^{\widetilde{F}} (r) 
    &= -\frac{8T^4}{\pi^{3}\Lambda^{4}r}  
    \frac{1}{\left(1+4r^{2}T^{2}
    \right)^{2}}\label{eq:FermionTildeMBPotential}\\
    &= -\frac{8 T^4}{\pi^3 \Lambda^4} \times
    \begin{cases}
1/r& \text{for $r\ll T^{-1}$}\\
1/(16 T^4r^5) & \text{for $r\gg T^{-1}$}
\end{cases}\;.
\end{align}
For the Fermi-Dirac distribution, we obtain
\begin{align}
    V_{\rm bkg, \text{FD}}^{\widetilde{F}} (r) 
    &= -\frac{1}{8\pi^{3}\Lambda^{4}r^{5}}  
    \left[2-2 \xi\, 
   \text{csch}(2 \xi ) -\xi^2\text{csch}^2(\xi)-\xi^2\text{sech}^2(\xi)
    \right] \quad (\text{with $\xi\equiv \pi rT$}) \label{eq:FermionTildeFDPotential}\\
    & = -\frac{T^4}{4\pi^{3}\Lambda^{4}}\times
    \begin{cases}
     14\pi^4/(45 r)& \text{for $r\ll T^{-1}$}\\
1/(T^4r^5) & \text{for $r\gg T^{-1}$}
    \end{cases}\;.
\end{align}

A plot of $V_{\rm bkg}^{\widetilde{F}}$ for the two distributions is shown in Fig.~\ref{fig:VFtilde}. Contrary to the case in Sec.~\ref{subsec:fermionscalar}, the potential $V_{\rm bkg}^{\widetilde{F}}$ does not change sign and remains attractive throughout the range. As a result, it can only lead to $S>1$ instead of $S<1$. The Sommerfeld factor calculated from $V_{\rm bkg}^{\widetilde{F}}$ is shown in Fig.~\ref{fig:FermionTilde_MB_Sommerfeld} with the 
MB distribution. We have also numerically verified that the case of FD distribution does not lead to a significantly different $S$ factor.

\begin{figure}[t]
\centering
\includegraphics[scale=0.5]
{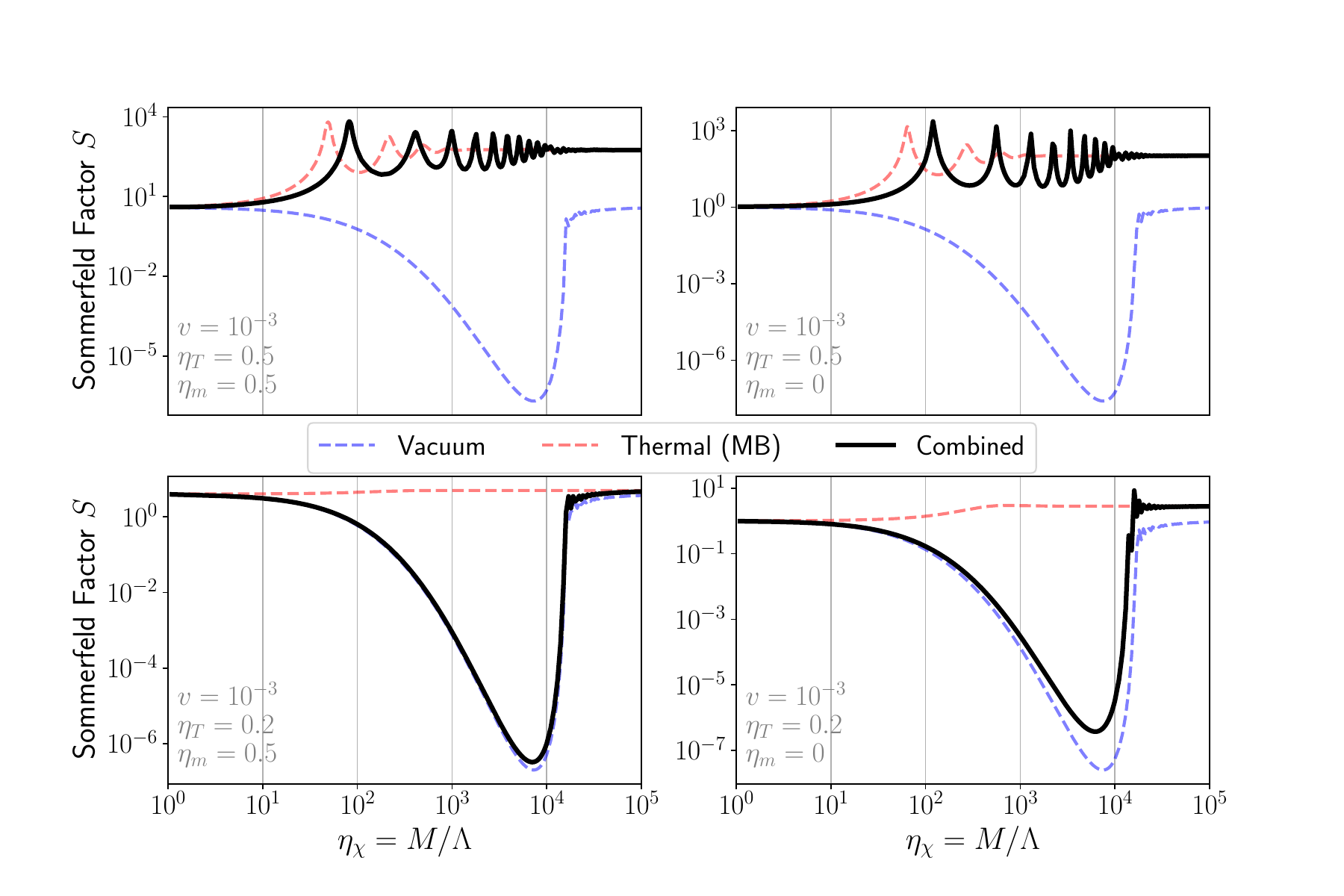}
\caption{\label{fig:CombinedFermion2MB}Sommerfeld enhancement from two-fermion exchange potentials with vector-vector coupling, including both vacuum and background effects. Same conventions as Fig.~\ref{fig:CombinedScalarMB}.}
\end{figure}

Finally, the Sommerfeld factor calculated from the combined potential $V_{\rm combined}^{\widetilde{F}}=V_{0}^{\widetilde{F}}+V_{\rm bkg}^{\widetilde{F}}$ is shown in Fig.~\ref{fig:CombinedFermion2MB}. As expected, the parameter regions where $S$ gets its contributions mainly from the vacuum or background are the same as in the previous two cases.

\section{Sommerfeld enhancement in non-thermal background}
\label{sec:non-thermal}
In the previous two sections, we studied the effect on the Sommerfeld enhancement from thermal backgrounds. In these scenarios, we find that the background correction becomes more significant as the temperature increases. This is because the background effect is coherent (i.e., proportional to the number density of background particles), while in thermal equilibrium systems, the number density is related to the temperature: $n\sim T^3$. However, if the DM particles and mediators are in the same thermal bath, the momentum of DM is also related to the temperature: $p\sim \sqrt{MT}$. Then the temperature is bounded by the requirement that DM particles should be non-relativistic: $T\lesssim M$. As a result, the background effect on the Sommerfeld enhancement is also bounded.

One way of breaking the above constraint is that the background particles are condensed in a low-energy state. This can only happen for bosonic mediators. In this case, the number density is not bounded by the DM mass, and the background effect on Sommerfeld enhancement keeps increasing as one increases the occupation number. 

In this section, we study this intriguing possibility and calculate the Sommerfeld factor with the potential $V_{\rm bkg}$ induced from a non-thermal background.\footnote{More precisely, by non-thermal background we mean that background particles are not in \emph{chemical} equilibrium within themselves or with other species, i.e., the rates of all number-violating processes of background particles are smaller than the Hubble expansion rate. As a result, the large number density of background particles is not destroyed via the self-interaction or interactions with other species.
Background particles may still be in kinetic equilibrium among themselves or with other species.} Although the calculation of $V_{\rm bkg}$ in thermal backgrounds was done a long time ago~\cite{Horowitz:1993kw,Ferrer:1998ju,Ferrer:1998rw,Ferrer:1999ad,Ferrer:2000hm}, the significant effect from non-thermal backgrounds on quantum forces was realized very recently~\cite{Ghosh:2022nzo,VanTilburg:2024xib,Barbosa:2024pkl,Ghosh:2024qai,Grossman:2025cov,Cheng:2025fak,Gan:2025nlu}. The two-neutrino exchange potential in a non-thermal and directional neutrino background was first calculated in \cite{Ghosh:2022nzo}, in which it was found that $V_{\rm bkg}\sim 1/r$ along the direction of the background momentum. This is a large enhancement compared to the vacuum potential which scales as $1/r^5$. Later, the two-boson exchange potential in non-thermal backgrounds was studied in great detail, first in \cite{VanTilburg:2024xib} for the quadratically coupled mediators and then in \cite{Grossman:2025cov} for axions. Again, it was found that $V_{\rm bkg}\sim 1/r$ as long as $r$ is small compared to the wavelength of background particles. In addition, for bosonic mediators, $V_{\rm bkg}$ can be significantly enhanced by the large occupation number of background particles~\cite{VanTilburg:2024xib,Grossman:2025cov,Gan:2025nlu}. 

We are interested in the non-thermal background effect on the Sommerfeld enhancement. As explained above, the effect is more significant for bosonic mediators due to the condensate. Therefore, we focus on the operator ${\cal O}_S$ in Eq.~(\ref{eq:OS}) throughout this section. In a background of $\phi$ with an arbitrary phase-space distribution function $f_\phi(\veck)$, the background potential between two $\chi$ particles is found to be:
\begin{align}
V_{\rm bkg}^{S}(\vecr) =  -\frac{1}{4\pi r \Lambda^2} \int \frac{{\rm d}^3\veck}{\left(2\pi\right)^3}\frac{f_\phi(\veck)}{2E_\veck} \left[\cos\left(\kappa r-\veck\cdot\vecr\right)+\cos\left(\kappa r+\veck\cdot\vecr\right)\right],\label{eq:Vbkg-general}   
\end{align}
where $E_\veck=(\veck^2+m_\phi^2)^{1/2}$ is the energy of the background particles. This result was first derived in \cite{VanTilburg:2024xib} with a slightly different form using the retarded propagator and then in \cite{Grossman:2025cov} using the Feynman progagator. In Appendix~\ref{app:quantum-force}, we provide details to derive Eq.~(\ref{eq:Vbkg-general}) without resorting to propagator ambiguity (see Eq.~(\ref{eq:Fourierbkg})).
In the limit where the background is isotropic, $f(\veck)=f(\kappa)$ with $\kappa \equiv |\veck|$, the general expression in Eq.~(\ref{eq:Vbkg-general}) is reduced to Eq.~(\ref{eq:VSbkg}). 

The cosine terms in Eq.~(\ref{eq:Vbkg-general}) correspond to the decoherence effect from phase space at long distances~\cite{Ghosh:2022nzo,VanTilburg:2024xib,Grossman:2025cov,Gan:2025nlu}: at $r\gg \lambda_{\rm dB}$, where $\lambda_{\rm dB}$ is the typical de Broglie wavelength of background particles, the background particles are no longer coherent and the oscillation becomes rapid, killing the leading $1/r$ term. As a result, $V_{\rm bkg}$ is more suppressed than $1/r$ at $r\gg \lambda_{\rm dB}$. On the other hand, all the background particles within the volume of $\lambda_{\rm dB}^3$ are coherent, and the cosine terms reduce to 1 in the limit of $r \ll \lambda_{\rm dB}$, making $V_{\rm bkg}\sim 1/r$ in the coherent region.

For our purpose of calculating the Sommerfeld factor, in terms of the normalized Schr\"{o}dinger equation (\ref{eq:Schrodinger}), the radial distance $r$ is normalized by the DM momentum: $r \equiv x/(Mv)$. The coherent length of $\phi$ particles in a condensate background is of order $\lambda_{\rm dB}\sim 1/(m_\phi v_\phi)$, where $v_\phi$ is the velocity of $\phi$. We are interested in the scenario of light mediators that satisfies $m_\phi v_\phi \ll M v$. In this case, it is a good approximation to take the coherent limit of Eq.~(\ref{eq:Vbkg-general}) for the calculation of the Sommerfeld factor. Then the background potential is reduced to 
\begin{align}
V_{\rm bkg}^{S}(r) &=  -\frac{1}{4\pi r \Lambda^2} \int \frac{{\rm d}^3\veck}{\left(2\pi\right)^3}\frac{f_\phi(\veck)}{E_\veck} = -\frac{\tilde{n}_\phi}{4\pi r \Lambda^2 m_\phi}\;,\label{eq:Vbkg-coherent}
\qquad
\text{for $m_\phi v_\phi \ll M v$}\;,
\end{align}
where $\tilde{n}_\phi$ is the effective number density defined as 
\begin{align}
\tilde{n}_\phi \equiv \int \frac{{\rm d}^3\veck}{\left(2\pi\right)^3}\frac{f_\phi(\veck)}{E_\veck/m_\phi}\;.   
\end{align}
If background particles are non-relativistic, $E_\veck \approx m_\phi$, then $\tilde{n}_\phi$ is reduced to the standard number density $n_\phi$, and the result in  Eq.~(\ref{eq:Vbkg-coherent}) is independent of the specific distribution function. On the other hand, in the relativistic limit, $E_\veck \gg m_\phi$, we have $\tilde{n}_\phi = c\,m_\phi^{} n_\phi^{2/3}$, where $c$ is a dimensionless parameter depending on the distribution function; we find $c= 1/ (2\pi^{2/3})\approx 0.23$ for the MB distribution  and $c= (\pi^2/\zeta(3))^{2/3}/12 \approx 0.34$ for the BE distribution with zero chemical potential.

The potential induced from non-thermal background in Eq.~(\ref{eq:Vbkg-coherent}) is an attractive Coulomb-like potential, from which the Sommerfeld factor can be computed analytically (for $s$-wave):
\begin{align}
 S &= \frac{\alpha_{\phi}/v}{1-e^{-\alpha_{\phi}/v}} \qquad \text{with $\alpha_{\phi}\equiv \frac{\tilde{n}_\phi}{2\Lambda^2 m_\phi}$}\label{eq:S-nonthermal}\\
 &= 
 \begin{cases}
  \alpha_\phi/v & \text{if $\alpha_\phi \gg v$}\\
  1 & \text{if $\alpha_\phi \ll v$}
 \end{cases}\;.
\end{align}

Compared to the Sommerfeld enhancement studied in Sec.~\ref{sec:thermal-scalar} and \ref{sec:thermal-fermion}, the crucial difference is that the number density in a non-thermal background is not related to the temperature, and thus the magnitude of $S$ is not bounded by the DM mass or the temperature.
We will discuss the implications of Eq.~(\ref{eq:S-nonthermal}) for the DM indirect detection in Sec.~\ref{subsec:indirect}.

\section{Phenomenology}
\label{sec:pheno}
In this section, we discuss how the phenomenology of DM cosmic evolution and detection is affected by the Sommerfeld enhancement/suppression from quantum forces. We focus on three effects: (i) thermal freeze-out of DM in the early universe, (ii) Cosmic Microwave Background (CMB) spectral distortion caused by DM annihilation at the redshift $10^3\lesssim z\lesssim 10^{6}$, and (iii) indirect detection of DM at present. In all phenomenological applications, we restrict to $s$-wave DM. We also assume that the mediator is light compared to the cutoff scale, $\eta_m = m_{\phi,\psi}/\Lambda \lesssim 10^{-3}$; in this case, the mediator is effectively massless and the specific value of $m_{\phi,\psi}$ does not affect the Sommerfeld factor induced by vacuum and thermal potentials.

\subsection{Thermal freeze-out}
\label{subsec:freezeout}
As a first application, we investigate how the Sommerfeld enhancement/suppression affects the DM dynamics during thermal freeze-out. For concreteness, we consider the freeze-out process in which two $\chi$ particles annihilate into any two other particles. 

\subsubsection*{Perturbative limit}
We first briefly review the calculation of the relic abundance without including the Sommerfeld effect. The Boltzmann equation governing this process is given by (see \cite{Kolb:1990vq,Gondolo:1990dk,Cirelli:2024ssz} for reviews)
\begin{align}
\frac{{\rm d}n_\chi}{{\rm d}t} + 3 H n_\chi = -\langle \sigma v\rangle\left(n_\chi^2-n_{\chi,{\rm eq}}^2\right),
\end{align}
where $n_\chi$ is the DM number density, $H$ is the Hubble parameter, $n_{\chi,{\rm eq}}$ is the DM number density in thermal equilibrium, and $\langle \sigma v\rangle$ denotes the thermal average of DM annihilation cross section times the relative velocity. Defining the yield $Y\equiv n_\chi/s$ (with $s$ the entropy density) and the inverse temperature $x\equiv m_\chi/T$ (with $T$ the temperature of photon), the Boltzmann equation can be simplified to 
\begin{align}
    Y'(x) 
    = 
    -\frac{\lambda(x)}{x^{2}}
   \left[ 
    Y^{2}(x) - Y^{2}_{\text{eq}}(x)
    \right],
    \label{eq:BoltzmannEq}
\end{align}
where prime denotes the derivative with respect to $x$ and
\begin{align}
 \lambda(x) = \sqrt{\frac{\pi}{45}} \frac{g_{*S}}{\sqrt{g_*}} \left(1+\frac{T}{3} \frac{{\rm d} \log g_{*S}}{{\rm d}T} \right) m_\chi m_{\rm Pl} \langle \sigma v\rangle\;,  \label{eq:lambdax}
\end{align}
with $m_{\rm Pl}\approx 1.2 \times 10^{19}~{\rm GeV}$ the Planck mass, while $g_{*}$ and $g_{*S}$ denote the number of relativistic degrees of freedom relevant to the energy density and entropy, respectively (we adopt the same notations as \cite{Kolb:1990vq}). By assuming a Maxwell-Boltzmann distribution during the thermal equilibrium, the thermally averaged cross section has an explicit Lorentz-invariant form~\cite{Gondolo:1990dk}:
\begin{align}
\langle \sigma v\rangle =  \frac{1}{8m_\chi^4 T K_2^2(m_\chi/T)} \int_{4m_\chi^2}^\infty {\rm d}s\,\sqrt{s}\left(s-4m_\chi^2\right)K_1(\sqrt{s}/m_\chi) \,\sigma(s)\;,
\end{align}
where $\sigma(s)$ is the annihilation cross section as a function of the center-of-mass squared energy, and $K_i$ are modified Bessel functions. In addition, we have $Y_{\rm eq}(x)=45 g_\chi x^2 K_2(x)/(4\pi^4 g_{*S})$, where $g_\chi$ is the number of internal degrees of freedom of $\chi$. (In our case, $g_\chi=4$ if $\chi$ is a Dirac fermion.) 

All expressions are exact so far. Since $\chi$ is non-relativistic during the freeze-out process, we can take the  approximation of $x\gg1$, obtaining $Y_{\rm eq}(x)\approx 45 g_\chi x^{3/2}e^{-x}/(4\sqrt{2}\,\pi^{7/2}g_{*S})\equiv a x^{3/2} e^{-x}$. In addition, the cross section can be expanded as a series of velocity: $\sigma v = \sigma_0 + \sigma_1 v^2 + \cdots$, leading to
\begin{align}
\langle \sigma v\rangle \approx \frac{x^{3/2}}{2\sqrt{\pi}}\int_0^\infty {\rm d}v\, v^2 e^{-xv^2/4}\left(\sigma_0+\sigma_1 v^2+\cdots\right) = \sigma_0 + \sigma_1 \frac{6}{x} + \cdots\;,
\end{align}
where $\cdots$ denotes terms suppressed by higher powers of the velocity. Throughout this section, we focus on the $s$-wave cross section and take $\langle \sigma v\rangle \approx \sigma_0$ as the leading-order approximation.

The Boltzmann equation (\ref{eq:BoltzmannEq}) can be analytically solved using the approximation method in \cite{Kolb:1990vq}. Before freezing out, the yield tracks closely the equilibrium value, $Y\approx Y_{\rm eq}$, so Eq.~(\ref{eq:BoltzmannEq}) is reduced to: $Y'_{\rm eq} \approx -2 \lambda Y_{\rm eq}(Y-Y_{\rm eq})/x^2$, which leads to $Y-Y_{\rm eq}\approx -x^2 Y'_{\rm eq}/(2\lambda Y_{\rm eq}) \approx x^2/(2\lambda)$. We define the freeze-out temperature $x_{\rm fo}$ as the point when $Y$ fails to track $Y_{\rm eq}$: $Y(x_{\rm fo})-Y_{\rm eq}(x_{\rm fo})\equiv Y_{\rm eq}(x_{\rm fo})$. Then $x_{\rm fo}$ can be solved iteratively:
\begin{align}
x_{\rm fo} &= \log\left(2a\lambda/\sqrt{x_{\rm fo}}\right)\approx \log\left(2a\lambda\right) - \frac{1}{2}\log\left[\log\left(2a\lambda\right)\right], \quad \text{with $a\lambda\approx 0.038 g_\chi m_\chi m_{\rm Pl} \sigma_0/\sqrt{g_*}$}\;,\label{eq:xfo}
\end{align}
where we have neglected the weak dependence of $g_{*S}$ on the temperature to get the numerical value. 

When $x>x_{\rm fo}$, $Y_{\rm eq}$ is negligible compared to $Y$ as it is exponentially suppressed. Then Eq.~(\ref{eq:BoltzmannEq}) is reduced to $Y' \approx -\lambda Y^2/x^2$. Integrating both sides, we obtain the yield at arbitrary $x=x_{\rm IR}>x_{\rm fo}$:
\begin{align}
\frac{1}{Y(x_{\rm IR})} -\frac{1}{Y(x_{\rm fo})} = \int_{x_{\rm fo}}^{x_{\rm IR}} {\rm d}x \frac{\lambda}{x^2} \quad  \implies \quad Y_\infty \approx \frac{x_{\rm fo}}{\lambda} \approx \frac{3.78\,x_{\rm fo}}{\left(g_{*S}/\sqrt{g_*}\right) m_\chi m_{\rm Pl} \sigma_0}\;, \label{eq:yieldfinal}
\end{align}
where $Y_\infty$ denotes the yield at $x\gg x_{\rm fo}$, and we have used $Y_\infty /Y(x_{\rm fo}) \approx 1/x_{\rm fo}\ll 1$. Note that in the absence of Sommerfeld enhancement, the integral in Eq.~(\ref{eq:yieldfinal}) is insensitive to IR physics, so one can effectively take $x_{\rm IR}$ to be infinity and identify $Y_\infty$ as the yield at present. This approximation, however, does not hold in the presence of long-range interactions, as we show below.

The final yield $Y_\infty$ is connected to the relic abundance via $\Omega_{\chi} = Y_\infty s_0 m_\chi/\rho_c$, where $\rho_c \approx 1.054 \times 10^{-5}h^2~{\rm GeV}/{\rm cm^3}$ is the critical density with $h$ the reduced Hubble parameter, and $s_0 \approx 2891/{\rm cm}^3$ is the present entropy density~\cite{ParticleDataGroup:2024cfk}. As a benchmark value, we take $\sigma_0 = 2.2 \times 10^{-26}{\rm cm^3/s} \approx 1/(23~{\rm TeV})^2$ and $m_\chi = 1~{\rm TeV}$, then Eq.~(\ref{eq:xfo}) gives the freeze-out temperature $x_{\rm fo}\approx 26$, which leads to the correct DM relic abundance from Eq.~(\ref{eq:yieldfinal}):
\begin{align}
\frac{\Omega_{\chi} h^2}{0.12} \approx \left(\frac{x_{\rm fo}}{26}\right)\left(\frac{2.2\times 10^{-26}{\rm cm^3/s}}{\sigma_0}\right).\label{eq:relicabundance}   
\end{align}
This is known as the WIMP miracle, because the above benchmark value of $\sigma_0$ happens to roughly agree with the typical cross section of the weak interaction: $\sigma_0\sim \alpha_{W}^2/m_{W}^2$, where $\alpha_W\sim 10^{-2}$ denotes the coupling strength of the weak interaction and $m_W\sim 100~{\rm GeV}$ is the mass of the weak gauge boson.
Note that for a fixed $\sigma_0$, the relic abundance only has a logarithmic dependence on the DM mass through $x_{\rm fo}$.

\subsubsection*{Correction from Sommerfeld effect}
Next, we investigate how the above results are affected by the non-perturbative Sommerfeld correction from quantum forces. The effect from an attractive Yukawa potential was discussed in \cite{Kamionkowski:2008gj,Dent:2009bv,Zavala:2009mi,Feng:2010zp}.

In the presence of long-range attractive/repulsive potentials, the DM annihilation cross section is enhanced/suppressed. However, we see from Eq.~(\ref{eq:xfo}) that the freeze-out temperature $x_{\rm fo}$ only depends logarithmically on the cross section. In addition, the relative velocity before freeze-out is not very small, because 
$v\sim \sqrt{T/m_\chi}\sim1/\sqrt{x} \gtrsim 0.2$ for $x \lesssim x_{\rm fo} \sim {\cal O}(25)$. Therefore, we expect that the non-perturbative effect is not significant at $x\lesssim x_{\rm fo}$, and the freeze-out temperature can still be approximated using Eq.~(\ref{eq:xfo}) even in the presence of the Sommerfeld correction.

At $x>x_{\rm fo}$, the Sommerfeld effect becomes more significant as the velocity continues to decrease. The yield after freeze-out in Eq.~(\ref{eq:yieldfinal}) is modified to be
\begin{align}
\frac{1}{Y(x_{\rm IR})} = \frac{1}{Y(x_{\rm fo})} + \sqrt{\frac{\pi}{45}}\frac{g_{*S}}{\sqrt{g_*}} m_\chi m_{\rm Pl}\sigma_0
\int_{x_{\rm fo}}^{x_{\rm IR}} {\rm d}x \frac{S_{\rm eff}(x_\chi)}{x^2}\;,\label{eq:YIR}    
\end{align}
where we have used Eq.~(\ref{eq:lambdax}) and neglected the dependence of $g_{*}$ and $g_{*S}$ on the temperature, and $S_{\rm eff}$ is the Sommerfeld factor averaged over velocity:\footnote{In numerical calculations, we find that it is a good approximation to use $S_{\rm eff}(x_\chi) \approx S(\bar{v},T)$, where $\bar{v}=4/\sqrt{\pi x_\chi}$ is the average velocity weighted by the Maxwell–Boltzmann distribution.}
\begin{align}
 S_{\rm eff}(x_\chi) = \frac{x_\chi^{3/2}}{2\sqrt{\pi}} \int_0^\infty {\rm d}v\,v^2 e^{-x_\chi v^2/4}S(v,T)\;.\label{eq:Seff}   
\end{align}
Here, $S(v,T)$ is the Sommerfeld factor calculated in the previous sections, which is a function of the velocity and the temperature of the mediator in general when both $V_0$ and $V_{\rm bkg}$ are included.\footnote{Throughout this subsection, we assume that the mediator $\phi$ or $\psi$ are in thermal equilibrium with the SM bath, so they share the same temperature as the SM bath: $T_{\phi,\psi}=T$. A different scenario of non-thermal background is discussed in Sec.~\ref{sec:non-thermal} and Sec.~\ref{subsec:indirect}.}
Moreover, $x_\chi$ is defined as $x_\chi \equiv m_\chi/T_\chi$, where $T_\chi$ is the temperature of DM. Note that $T_\chi$ is not necessarily equal to $T$ (the temperature of the SM bath) after freeze-out, so $x_\chi \neq x$ in general.

More specifically, freeze-out implies that $\chi$ is no longer in \emph{chemical} equilibrium with the SM bath, i.e., the $\chi$-number-violating process becomes ineffective. However, $\chi$ can still be in \emph{kinetic} equilibrium with the SM bath through elastic scattering with other species. When $\chi$ is in kinetic equilibrium but not in chemical equilibrium, its comoving number density is fixed, but its temperature can still track the temperature of the SM bath, so that $T_\chi=T$. As the temperature further decreases, $\chi$ does not maintain kinetic equilibrium with the SM bath, and $T_\chi$ starts to deviate from $T$, because $T\sim 1/a$ while $T_\chi \sim 1/a^2$ for non-relativistic $\chi$ particles (with $a$ the scale factor of the universe). We define $x_{\rm kd}\equiv m_\chi/T_{\rm kd}$, where $T_{\rm kd}$ is the temperature below which $\chi$ cannot maintain kinetic equilibrium. The exact value of $x_{\rm kd}$ depends on the particle physics model that describes the interaction between $\chi$ and SM species, but $x_{\rm kd}\geq x_{\rm fo}$ typically holds for the case we are considering (see \cite{profumo} for a more detailed and comprehensive discussion of the ordering). In the following analysis, we take $x_{\rm kd}$ as a free parameter.

Therefore, $x_\chi$ in Eq.~(\ref{eq:Seff}) can be connected to $x$ through a piecewise function:
\begin{align}
x_\chi (x) =
\begin{cases}
x & \text{for $x_{\rm fo} \leq x < x_{\rm kd}$}\\
x^2/x_{\rm kd} & \text{for $x\geq x_{\rm kd}$}
\end{cases}\;.\label{eq:xpiecewise}
\end{align}
As a result, the integral in Eq.~(\ref{eq:YIR}) should be split into two regions (assuming $x_{\rm IR}>x_{\rm kd}$):
\begin{align}
    \int_{x_{\rm fo}}^{x_{\rm IR}} {\rm d}x \frac{S_{\rm eff}(x_\chi)}{x^2} = \int_{x_{\rm fo}}^{x_{\rm kd}} {\rm d}x \frac{S_{\rm eff}(x)}{x^2} + \int_{x_{\rm kd}}^{x_{\rm IR}} {\rm d}x \frac{S_{\rm eff}(x^2/x_{\rm kd})}{x^2}\;. \label{eq:Seffpiecewise}
\end{align}
The velocity decreases as $v\sim 1/\sqrt{x_\chi}\sim1/\sqrt{x}$ in the first region, similar to the freeze-out process, while it decreases faster in the second region as $v\sim 1/\sqrt{x_\chi}\sim 1/x$. So, we can expect a more significant Sommerfeld effect from the second term in Eq.~(\ref{eq:Seffpiecewise}).

We want to investigate the correction on the final yield from the Sommerfeld effect, which is obtained by comparing Eq.~(\ref{eq:yieldfinal}) and Eq.~(\ref{eq:YIR}). For the case of attractive potential, we obtain
\begin{align}
\frac{Y_\infty(S=1)}{Y_\infty(S>1)} \approx x_{\rm fo}\left[\int_{x_{\rm fo}}^{x_{\rm kd}} {\rm d}x \frac{S_{\rm eff}(x)}{x^2} + \int_{x_{\rm kd}}^{x_{\rm IR}} {\rm d}x \frac{S_{\rm eff}(x^2/x_{\rm kd})}{x^2}\right],\label{eq:yieldratio}    
\end{align}
where $Y_\infty (S>1)$ and $Y_\infty(S=1)$ denote the final yield with and without Sommerfeld enhancement, respectively. For example, in the Coulomb limit, where the Sommerfeld factor is given by $S(v)=\alpha/v$ (with $\alpha \gg v$), we obtain $S_{\rm eff}(x_\chi)=\alpha \sqrt{x_\chi/\pi}$ in Eq.~(\ref{eq:Seff}), and the ratio in Eq.~(\ref{eq:yieldratio}) can be calculated analytically
\begin{align}
\frac{Y_\infty(S=1)}{Y_\infty(S>1)} \overset{\text{Coulomb}}{=} \frac{2}{\sqrt{\pi}}\alpha \sqrt{x_{\rm fo}} \left[1-\sqrt{\frac{x_{\rm fo}}{x_{\rm kd}}}+\frac{1}{2}\sqrt{\frac{x_{\rm fo}}{x_{\rm kd}}}\log\left(\frac{x_{\rm IR}}{x_{\rm kd}}\right)\right].\label{eq:yieldratio-Coulomb}
\end{align}
In the limit of late kinetic decoupling ($x_{\rm fo}\ll x_{\rm kd}\approx x_{\rm IR}$), the Sommerfeld enhancement in Eq.~(\ref{eq:yieldratio-Coulomb}) decreases the final yield by a factor of $\alpha \sqrt{x_{\rm fo}}$; in the limit of early kinetic decoupling ($x_{\rm fo}\approx x_{\rm kd}\ll x_{\rm IR}$), the Sommerfeld correction is more significant, and the final yield in Eq.~(\ref{eq:yieldratio-Coulomb}) is decreased by a factor of $\alpha \sqrt{x_{\rm fo}}\log(x_{\rm IR}/x_{\rm fo})$. For the Sommerfeld enhancement from quantum forces, as calculated in previous sections, there is no analytical expression for $S$, so we use Eq.~(\ref{eq:yieldratio}) for numerical calculation.

 \begin{figure}[t]
 \centering
 \includegraphics[width=6in]
 {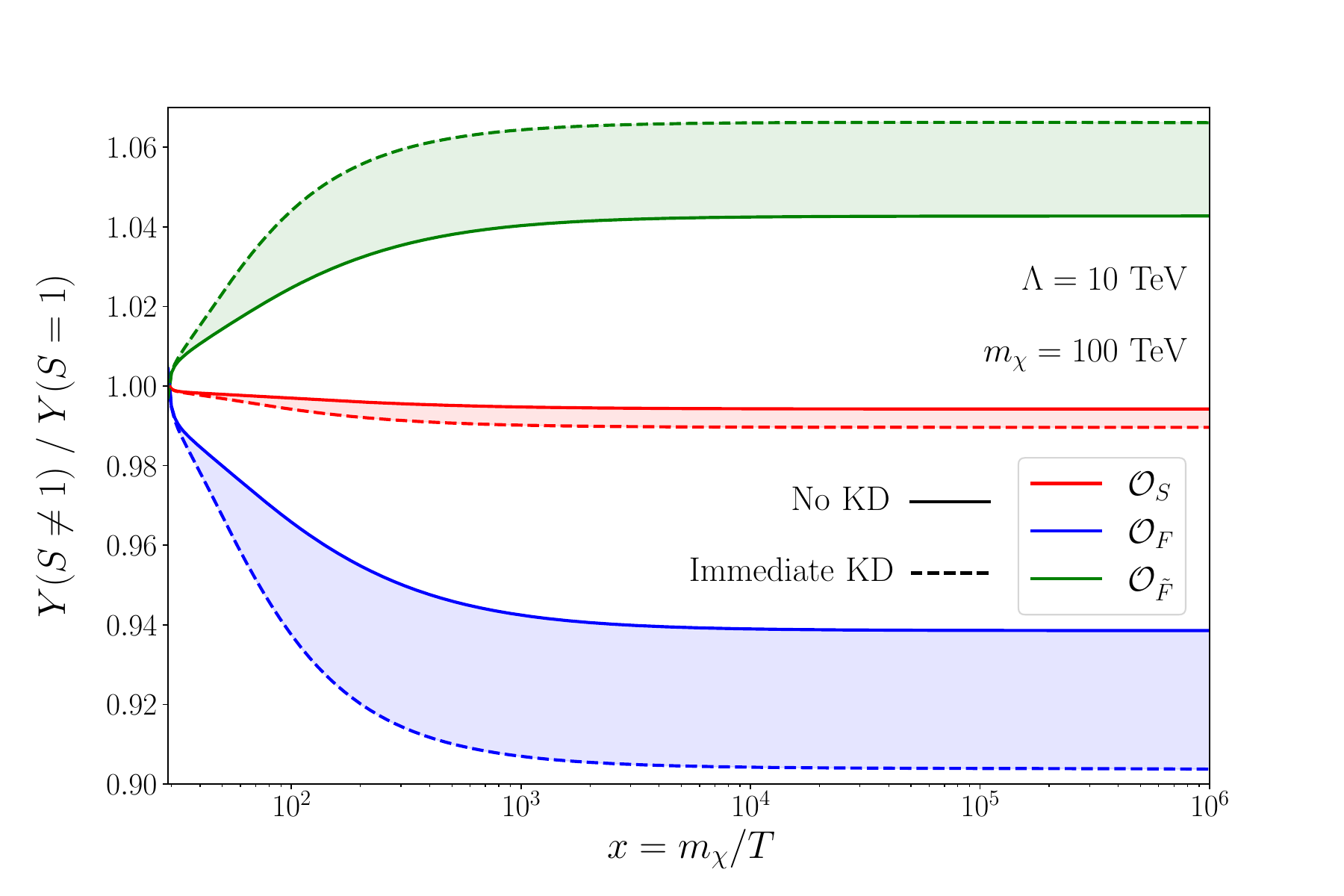}
 \caption{Evolution of the DM yield as the inverse temperature with the Sommerfeld effect turned on versus turned off for the three operators considered in this work, each shown in the two extreme scenarios of no kinetic decoupling $x_{\rm kd}=x_{\rm IR}$ (solid line) and immediate kinetic decoupling $x_{\rm kd}=x_{\rm fo}$ (dashed line). For any realistic value of $x_{\rm kd} \in  (x_{\rm fo},x_{\rm IR})$, this ratio will lie in the shaded regions. For $x<x_{\rm fo}$, we have verified that the ratio is very close to 1, so we show the result starting from $x_{\rm fo}$.
 The DM mass is fixed to $m_\chi = 100$ TeV, and the cutoff scale is chosen as $\Lambda = 10$ TeV.}
 \label{fig:SommerfeldFreezeout1}
 \end{figure} 

 \begin{figure}[t]
 \centering
 \includegraphics[width=6in]
 {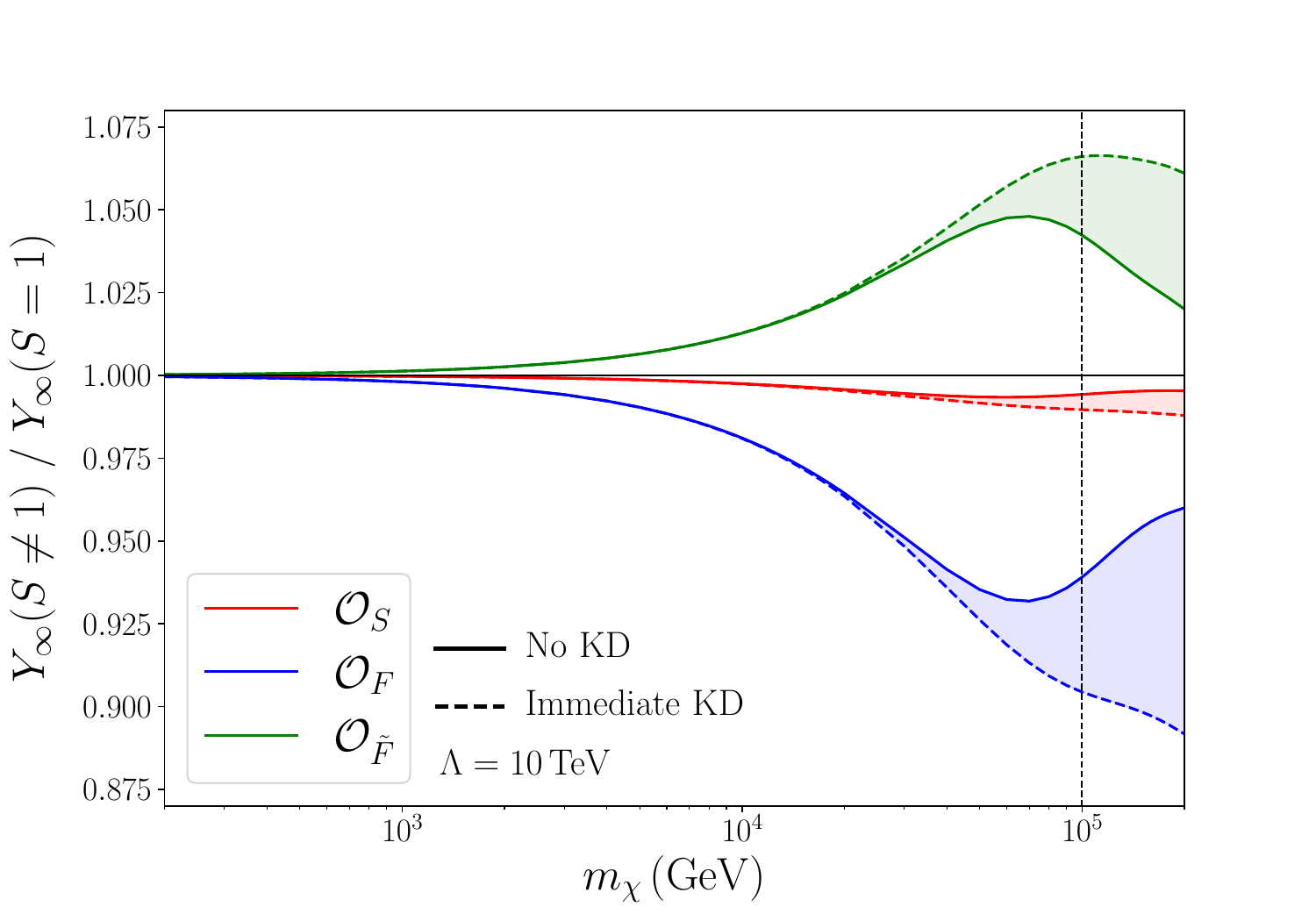}
 \caption{Ratio of the DM final yield as a function of DM mass with the Sommerfeld effect turned on versus turned off for the three operators considered in this work. The solid and dashed lines have the same meaning as Fig.~\ref{fig:SommerfeldFreezeout1}.
 The cutoff scale is chosen as $\Lambda = 10$ TeV. The vertical dashed line corresponds to the unitary bound for thermal freeze-out DM: $m_\chi \lesssim 100~{\rm TeV}$.}
 \label{fig:SommerfeldFreezeout2}
 \end{figure} 

For the case of repulsive potential, the Sommerfeld suppression tends to prevent the annihilation that occurs at $x>x_{\rm fo}$, which corresponds to the second term on the right-hand side of Eq.~(\ref{eq:YIR}). 
In order to estimate the maximal possible effect,
we take the strong Sommerfeld suppression limit, in which the annihilation after $x_{\rm fo}$ is negligible
\begin{align}
\frac{Y_\infty(S=1)}{Y_\infty(S\ll 1)} \approx \frac{x_{\rm fo}/\lambda}{Y(x_{\rm fo})} \approx x_{\rm fo}^{-1}\;.  
\end{align}
Therefore, strong Sommerfeld suppression can increase the final yield by a factor of up to $x_{\rm fo}\approx 25$. 
In practice, the Sommerfeld suppression is not strong enough to neglect the annihilation after $x_{\rm fo}$, and it is necessary to use Eq.~(\ref{eq:YIR}) to calculate the yield.

We are interested in the effects on the DM yield from the quantum forces considered in this work (${\cal O}_S$, ${\cal O}_F$, ${\cal O}_{\widetilde{F}}$).
In Fig.~\ref{fig:SommerfeldFreezeout1}, we plot the freeze-out yield that includes the Sommerfeld effect induced by quantum forces ($S\neq 1$) relative to that in the perturbative limit ($S=1$) as a function of the inverse temperature. For illustration, we fix $m_\chi =100~{\rm TeV}$ and $\Lambda = 10~{\rm TeV}$. (If $\Lambda\ll {\cal O}(10)~{\rm TeV}$, the annihilation channel $\chi\chi \to \phi\phi$ or $\chi\chi \to \psi\psi$ will dominate over the WIMP-scale interaction during freeze-out and lead to a DM yield too small to explain the relic abundance.)
The yield with the Sommerfeld correction is computed from Eq.~(\ref{eq:YIR}). For each effective operator, we include the combined effect from vacuum and background potentials. The freeze-out temperature is nearly unaffected by the Sommerfeld correction and is calculated using Eq.~(\ref{eq:xfo}), with $\sigma_0$ determined by Eq.~(\ref{eq:relicabundance}) to obtain the correct relic abundance in the perturbative limit. In Fig.~\ref{fig:SommerfeldFreezeout1}, the solid lines correspond to no kinetic decoupling ($x_{\rm kd}=x_{\rm IR}$), while the dashed lines correspond to immediate kinetic decoupling ($x_{\rm kd}=x_{\rm fo}$); for any realistic case where $x_{\rm kd} \in (x_{\rm fo}, x_{\rm IR})$, the results lie in the shaded regions. At $x < x_{\rm fo}$, we have checked that the Sommerfeld effect is negligible because the velocity is large. At $x>x_{\rm fo}$, the Sommerfeld effect from ${\cal O}_S$ and ${\cal O}_F$ tends to preserve DM annihilation and decrease the final yield compared to the perturbative limit, while that from ${\cal O}_{\widetilde{F}}$ contributes in the opposite direction and tends to increase the final yield. From Fig.~\ref{fig:SommerfeldFreezeout1}, we see that the Sommerfeld correction from three operators on the DM yield is of order a few percent.

In Fig.~\ref{fig:SommerfeldFreezeout2}, we plot a similar ratio of the final yields  as a function of the DM mass $m_{\chi}$. We identify the final yield $Y_\infty$ as the value of $Y$ at $x\gg 1$ that does not  further change as one increases $x$. 
The conventions are similar as Fig.~\ref{fig:SommerfeldFreezeout1} and we again fix the cutoff scale $\Lambda = 10$ TeV.  The single peak structure we observe here comes from the different features we have discussed previously in Sommerfeld enhancement/suppression plots. However, here the peaks are smoothed out due to the integration over $x_{\chi}$. The Sommerfeld effect becomes more significant as one increases $m_\chi$. However, the partial-wave unitarity bound is violated for $m_\chi \gtrsim 100~{\rm TeV}$~\cite{Griest:1989wd}, so we do not plot the result with higher masses. This is why we only see a single broadened peak. 

In conclusion, we find that compared to the perturbative limit, the Sommerfeld effect from ${\cal O}_S$ and ${\cal O}_F$ can decrease the DM yield from thermal freeze-out by up to $1\%$ and $10\%$, respectively, while that from ${\cal O}_{\widetilde{F}}$ can increase the final yield by up to 7\%.

\subsection{CMB spectral distortions}
\label{subsec:CMB}

In the very early universe, any physics that injects energy into the CMB blackbody spectrum is balanced by photon-number-changing processes (such as double Compton scattering and Bremsstrahlung) and equilibrating processes (such as Compton scattering). When these balancing processes become inefficient, the CMB spectrum can be distorted away from a perfect blackbody: $f(x_\gamma)=1/(e^{x_\gamma}-1)$, where $x_\gamma \equiv E/T$ with $E$ and $T$ the energy and temperature of CMB photons. The annihilation of DM particles into SM species during this epoch will then leave imprints on the CMB spectral distortion. In this subsection, we investigate how this spectral distortion is affected by the Sommerfeld effect from quantum forces.

\subsubsection*{$\mu$-distortion}
Double Compton scattering freezes out at the redshift $z_{\text{DC}} \approx 2 \times 10^{6}$ and Compton scattering freezes out at $z_{\text{C}} \approx 5\times 10^{4}$. When the CMB is a perfect blackbody, the chemical potential $\mu$ is 0, but once the number of photons is fixed (after $z_{\text{DC}}$), the spectrum develops a nonzero chemical potential $\mu$ from the energy injection, known as the $\mu$-distortion of the CMB~\cite{Hu:1993gc,McDonald:2000bk}. Below we show how to connect the $\mu$-distortion to the energy injection from DM annihilation. 

At $z_{\rm C}\lesssim z\lesssim z_{\rm DC}$, the injected energy distorts the CMB spectrum to $f(x_\gamma) \to f(\tilde{x}_\gamma,\mu)=1/(e^{\tilde{x}_\gamma+\mu}-1)$, where $\tilde{x}_\gamma\equiv E/\widetilde{T}$. Note that the temperature before and after injection is different in general $T\neq \widetilde{T}$. Requiring the conservation of the photon number, one obtains $\zeta(3)T^3 = {\rm Li_3}(e^{-\mu})\widetilde{T}^3$, where $\zeta(n)$ is the Riemann zeta function with $\zeta(3)\approx 1.202$ and ${\rm Li}_n$ is the polylogarithm function. The photon energy density before injection is $\rho_\gamma = \pi^2 T^4/15$, while after injection it becomes
\begin{align}
\rho_\gamma + \delta\rho_\gamma &= \frac{1}{\pi^2}\int_0^\infty {\rm d}E  \frac{E^3}{e^{\tilde{x}_\gamma +\mu}-1} = \frac{6}{\pi^2} {\rm Li}_4(e^{-\mu})\left[\frac{\zeta(3)}{{\rm Li_3}(e^{-\mu})}\right]^{\frac{4}{3}}T^4\nonumber\\
&\approx \frac{\pi^2 T^4}{15}  + \mu\frac{\pi^2 T^4}{15}\left(\frac{2\pi^6-810\,\zeta^2(3)}{9\pi^4 \zeta(3)}\right) + {\cal O}(\mu^2)\;,
\end{align}
where in the second equality we used the relation between $T$ and $\widetilde{T}$ from photon number conservation, and in the last equality we have assumed $\mu\ll1$. Therefore, the $\mu$-distortion parameter is given by
\begin{align}
\mu = \left(\frac{9\pi^4\zeta(3)}{2\pi^6-810\,\zeta^2(3)}\right)\frac{\delta \rho_\gamma}{\rho_\gamma} \approx 1.4\,  \frac{\delta \rho_\gamma}{\rho_\gamma}
= 1.4\int_{t_{\text{DC}}}^{t_{\text{C}}}
    \frac{
     \dot{\rho}_\gamma
     }{\rho_\gamma}\,{\rm d}t\;,\label{eq:mudistortion}
\end{align}
where the lower and upper limits of the integral correspond to the time when double Compton scattering and Compton scattering freeze out, respectively.

The energy injection rate $\dot{\rho}_\gamma$ can be computed given the specific particle physics model. In our case, it comes from DM annihilation into species that can interact with photons. Following the derivation in \cite{McDonald:2000bk}, we have $\dot{\rho}_\gamma= 2 m_\chi f_{\rm EM} \langle \sigma v\rangle n_\chi^2$, where $\langle \sigma v\rangle$ denotes the thermal average of the annihilation cross section of two non-relativistic $\chi$ particles times the relative velocity and $n_\chi$ is the number density of $\chi$. The factor of 2 comes from the fact that the annihilation particle $\chi$ is not its own antiparticle, which is consistent with our assumption that $\chi$ is a Dirac fermion. Here $f_{\rm EM}\leq 1$ is the fraction of the annihilation energy that interacts electromagnetically. The exact value of $f_{\rm EM}$ depends on the specific annihilation channel; 
for $z\gtrsim 10^3$ (which is satisfied in our case), it was found in \cite{Slatyer:2009yq} that $f_{\rm EM}\gtrsim 0.25$ and is independent of the redshift for all relevant channels. 
It is more convenient to change the variable of integration in Eq.~(\ref{eq:mudistortion}) from time to redshift. Using $\rho_\gamma \propto (1+z)^4$, $n_\chi\propto (1+z)^3$ and the relation $t\propto (1+z)^{-2}$ during the radiation-dominated era, we have
\begin{align}
    \mu=1.4\,\frac{\delta\rho_\gamma}{\rho_\gamma} 
    &= 1.4\,
    \frac{4 m_{\chi} f_{\rm EM} n_{\chi,0}^{2} t_{*}}{\rho_{\gamma,0}}
    \int_{z_{\text{C}}}^{z_{\text{DC}}}
    \frac{\braket{\sigma v}}{1+z}\, {\rm d}z = 1.4\,\frac{4 f_{\rm EM} \rho_c^2 \Omega_\chi^2 t_{*}}{\rho_{\gamma,0} m_\chi}\int_{z_{\text{C}}}^{z_{\text{DC}}}
    \frac{\braket{\sigma v}}{1+z}\,{\rm d}z\;,\label{eq:mudistortion2}
\end{align}
where $t_{*}\equiv m_{\rm Pl}\sqrt{3 /(16\pi g_{*,0}\rho_{\gamma,0})} \approx 2.4 \times 10^{19}~{\rm s}$ is the reference time, $g_{*,0}\approx3.36$ is the number of relativistic degrees of freedom at present relevant to the energy density, $\rho_{\gamma,0}\approx 2.6 \times 10^{-10}~{\rm GeV/cm^3}$ and $n_{\chi,0}$ denote the energy density of CMB and the number density of $\chi$ at present, respectively. In the second equality, we have replaced $n_{\chi,0}$ with the relic abundance via $\Omega_\chi = m_\chi n_{\chi,0}/\rho_c$.

\subsubsection*{$y$-distortion}
Between $z_{\text{DC}}$ and $z_{\text{C}}$, the number of CMB photons is frozen, and their temperature equilibrates through Compton scattering. After $z_{\text{C}}$, Compton scattering can no longer keep them in thermal equilibrium, so they begin to deviate from a Planck spectrum. This is known as the $y$-distortion of the CMB~\cite{Sunyaev:1980vz,Bernstein:1989uq}. 
More specifically, the blackbody spectrum after injection is distorted from $f(x_\gamma)=1/(e^{x_\gamma}-1)$ to~\cite{Bernstein:1989uq} 
\begin{align}
f(x_\gamma,y) = \frac{1}{e^{x_\gamma}-1} + y \frac{x_\gamma\, e^{x_\gamma}}{\left(e^{x_\gamma}-1\right)^2}\left[\frac{x_\gamma}{\tanh(x_\gamma/2)}-4\right],  
\end{align}
where $y\ll 1$ is a dimensionless parameter that parametrizes the deviation from the blackbody spectrum; physically it is connected to the optical depth of the Thomson scattering of the electron~\cite{Bernstein:1989uq}.
Integrating the above spectrum over the photon phase space, one obtains $\rho_\gamma = \pi^2 T^4/15 \to \rho_\gamma + \delta \rho_\gamma= \pi^2 T^4(1+4y)/15$. Therefore, the $y$-distortion parameter is given by
\begin{align}
    y = \frac{1}{4}\,\frac{\delta\rho_\gamma}{\rho_\gamma}
    = \frac{1}{4}\int_{t_{\text{C}}}^{t_{\text{eq}}}
    \frac{
    \dot{\rho}_\gamma
    }{\rho_\gamma}\,{\rm d}t
    \;,
\end{align}
where $t_{\rm rec}$ is the time of matter-radiation equality, corresponding to the redshift $z_{\rm eq}\approx 3400$. As in the case of the $\mu$-distortion, we change the integration variable and obtain
\begin{align}
y = \frac{f_{\rm EM} \rho_c^2 \Omega_\chi^2 t_{*}}{\rho_{\gamma,0} m_\chi}\int_{z_{\text{eq}}}^{z_{\text{C}}}
    \frac{\braket{\sigma v}}{1+z}\,{\rm d}z\;.\label{eq:ydistortion}    
\end{align}

So far we have discussed the $\mu$-distortion and the $y$-distortion separately and assumed that the transition between the two eras is instantaneous, which is sufficient for our purpose in this section; see \cite{Khatri:2012tw} for a more detailed analysis of the transition between the $\mu$-era and the $y$-era. The current strongest bounds on CMB spectral distortions are from the COBE/FIRAS data (both at 95\% confidence level)~\cite{Fixsen:1996nj,Bianchini:2022dqh,Sabyr:2025hwd}:
\begin{align}
    |\mu|  < 4.7 \times 10^{-5} \;,\qquad
    |y|  < 8.3 \times 10^{-6}\;.
\end{align}
Future experiments such as PIXIE~\cite{Kogut:2011xw,Kogut:2024vbi} can increase the sensitivity of both $\mu$- and $y$-distortions by three orders of magnitude.

Using Eqs.~(\ref{eq:mudistortion2}) and (\ref{eq:ydistortion}), the non-observation of CMB spectral distortions can be used to put constraints on the DM annihilation cross section~\cite{McDonald:2000bk,Zavala:2009mi,Hannestad:2010zt,Chluba:2011hw,Ali-Haimoud:2021lka,Li:2024xlr}. In the following, we investigate how the Sommerfeld enhancement/suppression from quantum forces will affect the constraints from CMB spectral distortions.\footnote{The measurement of CMB anisotropies after recombination $z\lesssim 10^3$ can also put strong constraints on DM annihilation cross sections~\cite{Padmanabhan:2005es,Galli:2009zc,Slatyer:2009yq,Chluba:2009uv,Slatyer:2015jla,Planck:2018vyg}. Because it comes from the same channel of DM annihilation, it will also be affected by the Sommerfeld effect in a similar way. However, in this section, we restrict ourselves to the constraints at the redshift $10^3\lesssim z\lesssim 10^6$. The analysis of the constraints from CMB anisotropies in the presence of the Sommerfeld effect from quantum forces is left for future work.}

\subsubsection*{Sommerfeld effect on spectral distortions}

\begin{figure}[t]
\centering
\includegraphics[width=5.5in]
{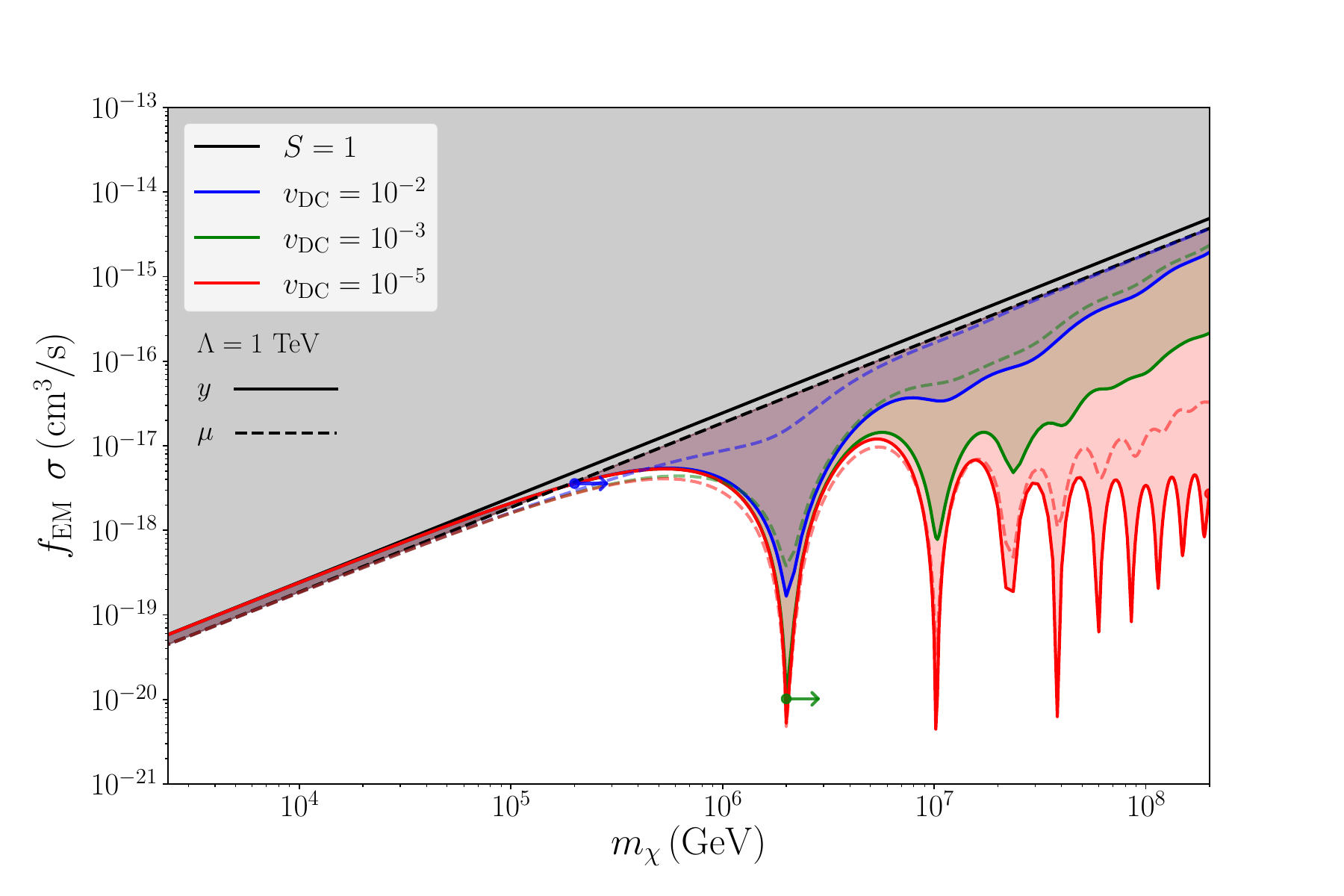}
\caption{\label{fig:CMBdistortionOS}Constraints on the DM annihilation cross section $f_{\rm EM}\sigma_0$ from CMB $\mu$-distortion (dashed lines) and $y$-distortion (solid lines) with the quantum force induced by ${\cal O}_S$. As a comparison, the bounds without the Sommerfeld effect ($S=1$) are shown with black lines, while the blue, green, and red lines include the Sommerfeld correction with different velocities at $z_{\rm DC}$. The shaded areas are excluded by the non-observation of CMB spectral distortion from the COBE/FIRAS data. 
The dots with arrows correspond to the places where the maximal momentum transfer $m_\chi v_{\rm DC}/2$ exceeds $\Lambda$, which have the same meaning as Fig.~\ref{fig:scalarvacuum}.
The cutoff scale is fixed to be $\Lambda = 1$ TeV.}
\end{figure}

\begin{figure}[t]
\centering
\includegraphics[width=5.3in]
{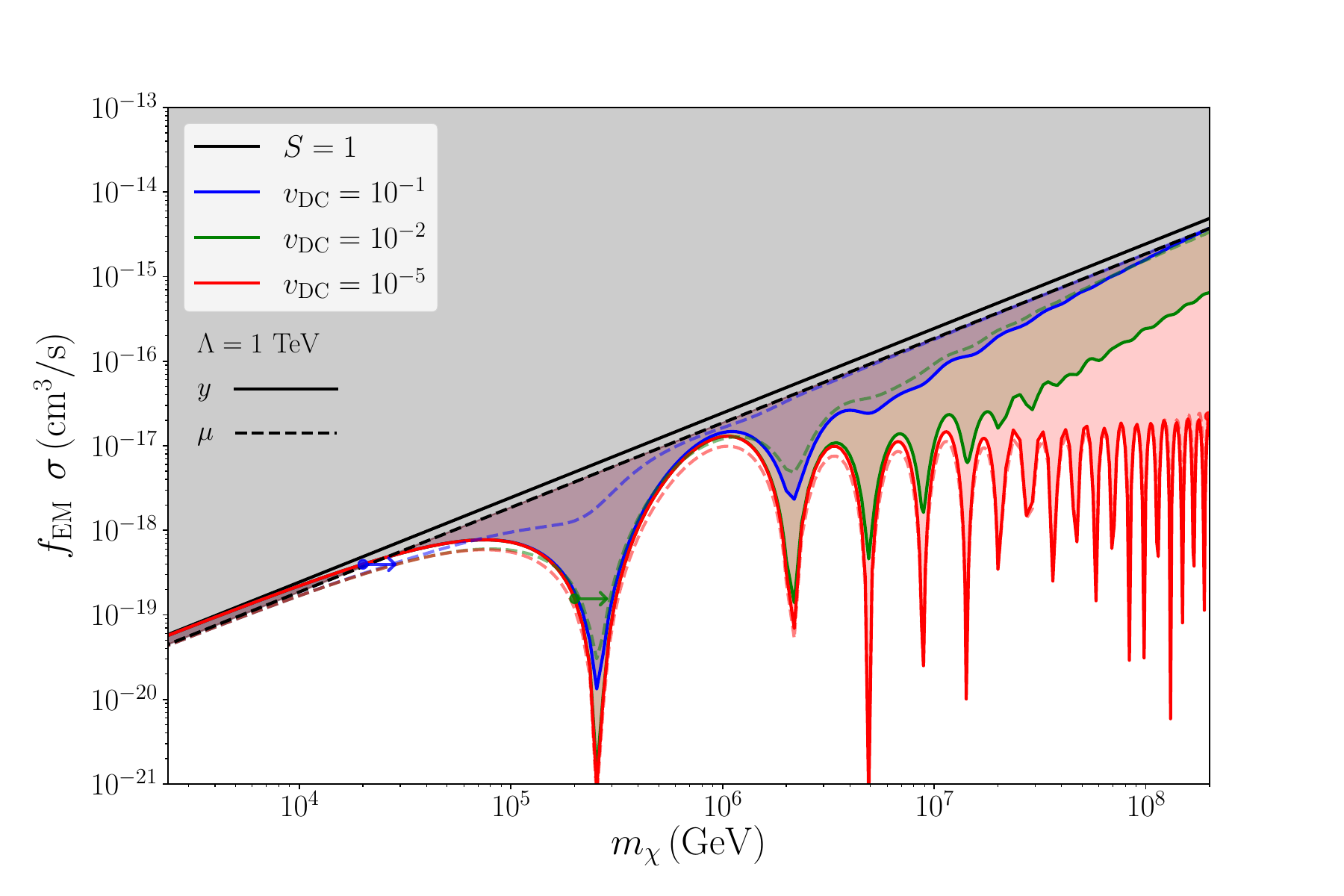}
\caption{\label{fig:CMBdistortionOF}Same conventions as Fig.~\ref{fig:CMBdistortionOS} but for ${\cal O}_F$.}
\end{figure}

\begin{figure}[t]
\centering
\includegraphics[width=5.3in]
{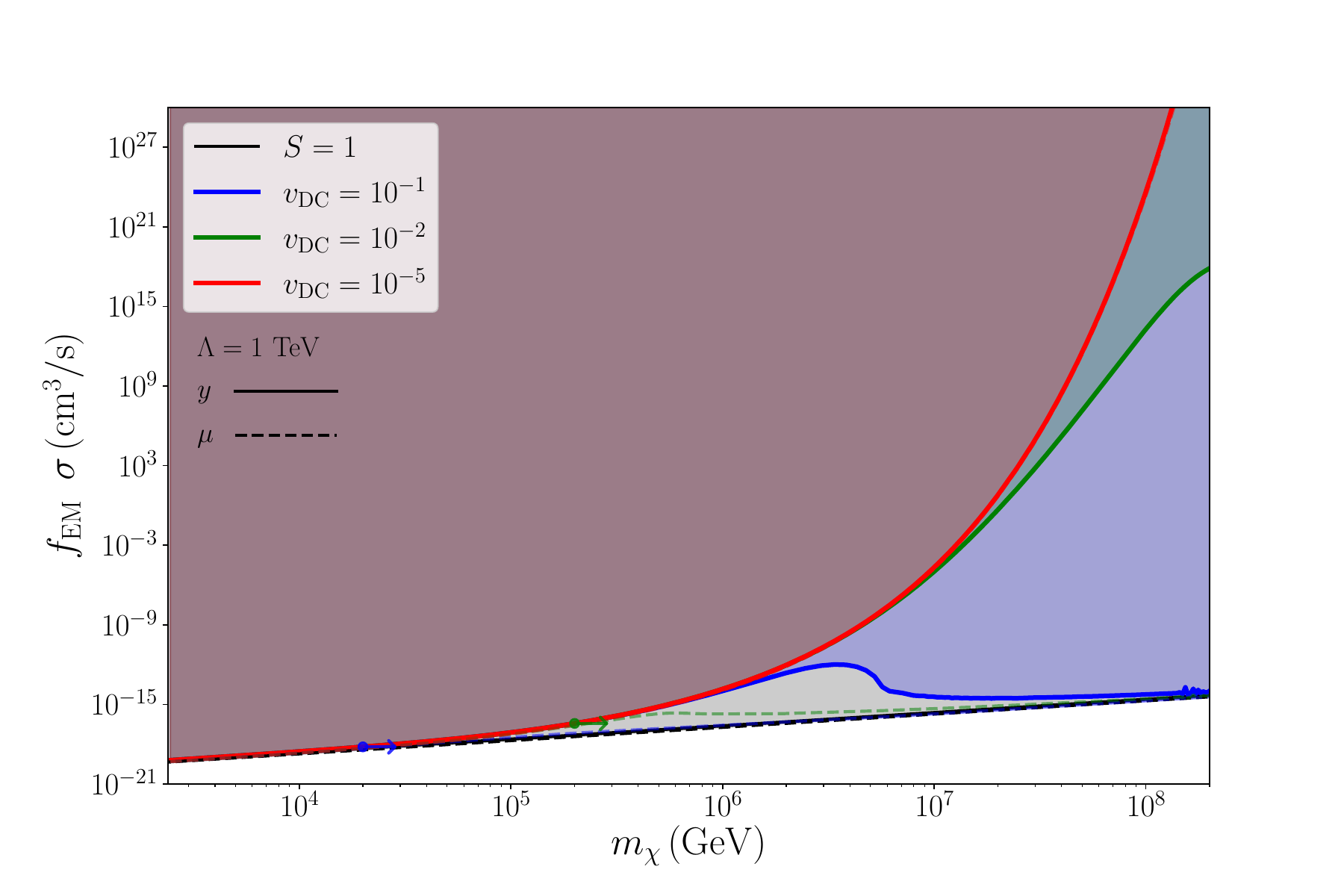}
\caption{\label{fig:CMBdistortionOFtilde}Same conventions as Fig.~\ref{fig:CMBdistortionOS} but for ${\cal O}_{\widetilde{F}}$.}
\end{figure}

For $s$-wave processes, including the Sommerfeld correction, we have $\langle \sigma v\rangle = S\sigma_0$. To have a more accurate calculation of the Sommerfeld effect during the redshift $z_{\rm eq}<z<z_{\rm DC}$, we make two reasonable assumptions: (i) The velocity of $\chi$ particles still obeys the Maxwellian distribution, such that the effective Sommerfeld factor can be computed using Eq.~(\ref{eq:Seff}). (ii) The non-relativistic $\chi$ particles have already kinetically decoupled from the SM bath at $z>z_{\rm DC}$, so during the redshift $z_{\rm eq}<z<z_{\rm DC}$, we have $x_\chi \equiv m_\chi/T_\chi \propto (1+z)^{-2}$. 

With the above two assumptions, the COBE/FIRAS bounds on the $\mu$- and $y$-distortions can be translated into an upper bound on $f_{\rm EM}\sigma_{0}$. The bounds are then 
\begin{align}
    1.4\,\frac{4f_{\rm EM} \sigma_0 \rho_c^2 \Omega_\chi^2 t_{*}}{\rho_{\gamma,0} m_\chi}
    \int_{z_{\text{C}}}^{z_{\text{DC}}}
    \frac{S_{\rm eff}(x_\chi(z))}{1+z}\,{\rm d}z
    &< 4.7 \times 10^{-5}
    \;, \label{eq:mubound}
     \\
    \frac{f_{\rm EM}\sigma_0 \rho_c^2 \Omega_\chi^2 t_{*}}{\rho_{\gamma,0} m_\chi}\int_{z_{\text{eq}}}^{z_{\text{C}}}
    \frac{S_{\rm eff}(x_\chi(z))}{1+z}\,{\rm d}z
    &< 8.3 \times 10^{-6}\;,\label{eq:ybound}
\end{align}
where $S_{\rm eff}(x_\chi)$ is given by Eq.~(\ref{eq:Seff}), and $x_\chi$ is a function of the redshift:
\begin{align}
x_\chi(z) = x_\chi(z_{\rm DC}) \left(\frac{1+z_{\rm DC}}{1+z}\right)^2,\qquad \text{for $z_{\rm eq}<z<z_{\rm DC}$}\;.   
\end{align}
The only input parameter here is the value of $x_\chi$ at $z=z_{\rm DC}$, which is of order ${\cal O}(1/v^2)$ at that time. This value is UV dependent, i.e., it is determined by the dynamics governing DM evolution at $z>z_{\rm DC}$. If the DM abundance is inherited from thermal freeze-out, as discussed in Sec.~\ref{subsec:freezeout}, then $x_\chi$ is connected to the CMB temperature by Eq.~(\ref{eq:xpiecewise}). In the following, in order to make our results as general as possible, we do not assume that DM abundance necessarily comes from thermal freeze-out; instead, we take $x_\chi(z_{\rm DC})$ as a free input parameter. 

We plot the bounds on $f_{\rm EM}\sigma_0$ according to Eqs.~(\ref{eq:mubound})-(\ref{eq:ybound}) in Figs.~\ref{fig:CMBdistortionOS}, \ref{fig:CMBdistortionOF} and \ref{fig:CMBdistortionOFtilde}, which includes the Sommerfeld effect from ${\cal O}_S$, ${\cal O}_F$ and ${\cal O}_{\widetilde{F}}$, respectively. As a benchmark, we fix $\Lambda = 1~{\rm TeV}$ in all plots.
The bounds from $\mu$-distortion and $y$-distortion are shown separately with dashed and solid lines, respectively. For comparison, we show the bounds without including the Sommerfeld correction with black lines. The blue, green, and red lines include the Sommerfeld effect with different values of $v_{\rm DC}$. Here $v_{\rm DC} \equiv 4/ \sqrt{\pi x_\chi(z_{\rm DC})}$ is the average relative velocity of DM particles at $z_{\rm DC}$. We have verified that for $v_{\rm DC} \lesssim 10^{-5}$, the Sommerfeld factor is saturated and the bounds cannot be significantly changed by further decreasing $v_{\rm DC}$. Since we do not assume that the DM abundance comes from freeze-out, the unitarity bound on the DM mass does not apply, so we show the constraints up to $m_\chi = 10^{8}~{\rm GeV}$. Note that the contribution from thermal potentials is negligible because the temperature during $z_{\rm eq}\lesssim z\lesssim z_{\rm DC}$ is much lower than the value of $\Lambda$ we choose here, so Figs.~\ref{fig:CMBdistortionOS}-\ref{fig:CMBdistortionOFtilde} only include the Sommerfeld effect from vacuum potentials.
We can clearly see the resonance structure from the Sommerfeld enhancement in Figs.~\ref{fig:CMBdistortionOS} and \ref{fig:CMBdistortionOF}, which significantly strengthens the bounds compared to the perturbative limit. In Fig.~\ref{fig:CMBdistortionOFtilde}, however, the bounds are exponentially relaxed due to the Sommerfeld suppression on the DM annihilation cross section, which comes from the repulsive two-fermion potential.

\subsection{Dark matter indirect detection}
\label{subsec:indirect}
In this subsection, we investigate how DM indirect detection is affected by the Sommerfeld effect induced by quantum forces. Compared to the Yukawa force that has been widely considered in the literature, the quantum forces are enhanced in the presence of a background of mediator particles. The present-day temperature of the universe is only $T_0\sim 10^{-4}~{\rm eV}$, which is too low to have any sizable thermal corrections to the Sommerfeld factor. However, if the mediator particles form a non-thermal background in the galaxy that has a large occupation number, as discussed in Sec.~\ref{sec:non-thermal}, then the background potential scales as $V_{\rm bkg}\sim 1/r$ and is independent of the temperature, which may contribute to the Sommerfeld factor significantly. This is an interesting possibility that does not exist in the case of the Yukawa force.
In the following, we discuss two scenarios:
\begin{itemize}
    \item First, we neglect the background effect and study how the vacuum potentials $V_0^S$, $V_0^F$ and $V_0^{\widetilde{F}}$ affect the indirect detection signals in the galaxy.
    
    \item Then, we assume that the mediator particles form a non-thermal background in the galaxy. We require the occupation number to be compatible with the constraints from current observations and calculate how large it can affect the DM annihilation cross section in the galaxy.
\end{itemize}

\subsubsection{Vacuum effect}

\begin{figure}[t]
\centering
\includegraphics[width=4.5in]
{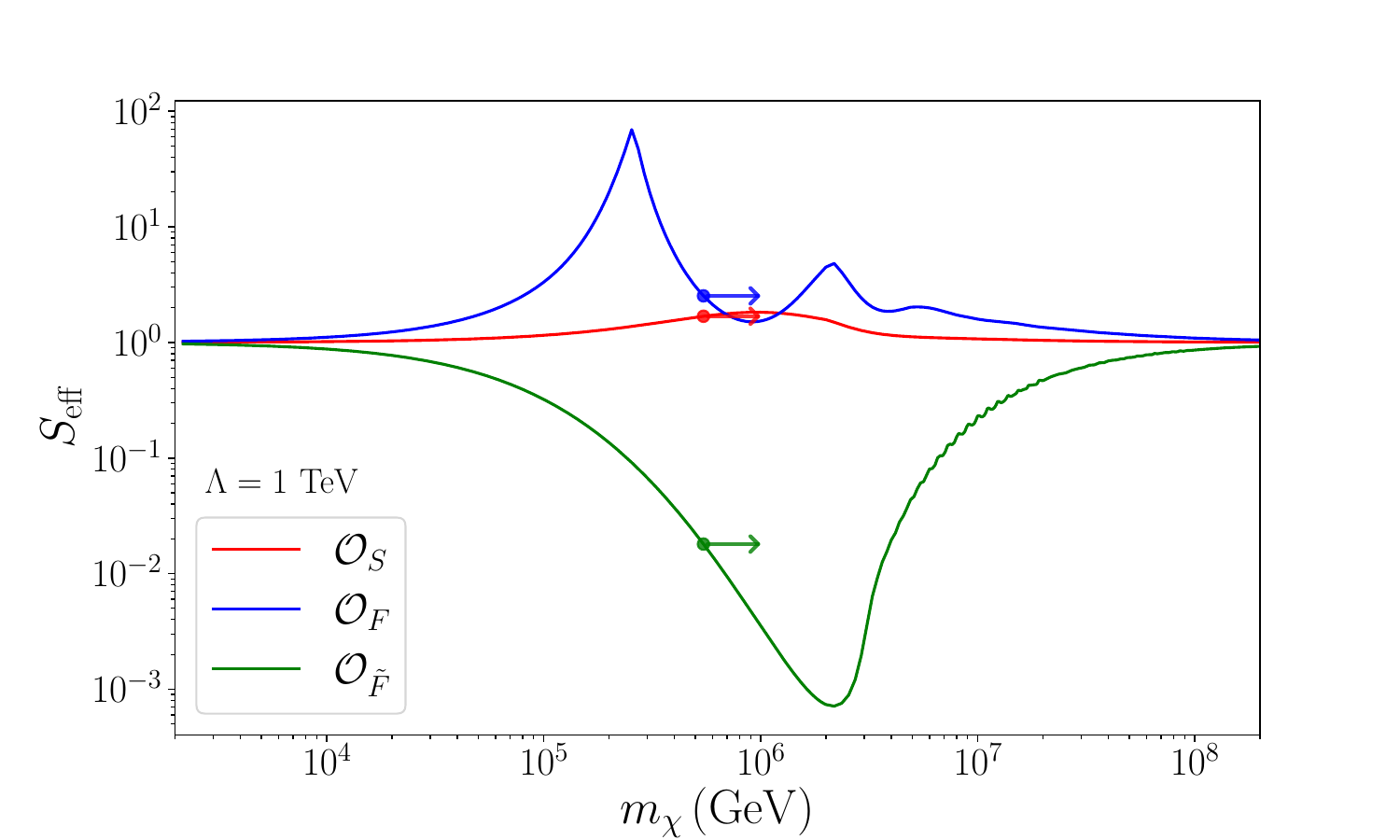}
\caption{\label{fig:Indirect_Vacuum}Boost factor $S_{\text{eff}} = \langle \sigma v \rangle_{\rm eff}/\langle\sigma v\rangle_0$ as a function of DM mass caused by the vacuum quantum forces relevant for DM indirect detection. The cutoff scale is fixed to be $\Lambda = 1~{\rm TeV}$. The dots with arrows are determined by $m_\chi v_{\rm esc} = \Lambda$, where the largest momentum transfer reaches the cutoff scale.}
\end{figure}

The DM velocity distribution in the galaxy can be approximated by the standard halo model~\cite{Drukier:1986tm,Evans:2018bqy}
\begin{align}
f(v_\chi) = \frac{1}{{\cal N}}\,v_\chi^2 e^{-v_\chi^2/v_0^2}\,\Theta\left(v_{\rm esc}-v_\chi\right),\label{eq:SHM} 
\end{align}
where $\Theta$ is the step function, $v_\chi$ is the velocity of $\chi$ particle, $v_{\rm esc}\approx 544~{\rm km/s}$ is the escape velocity, $v_0 \approx 220~{\rm km/s}$ measures the velocity dispersion, and ${\cal N}$ is a normalization constant obained by requiring $\int_0^\infty {\rm d}v_\chi\,f(v_\chi) =1$:
\begin{align*}
{\cal N}= \frac{\sqrt{\pi}}{4} v_0^3\left[{\rm erf}\left(\frac{v_{\rm esc}}{v_0}\right) -\frac{2}{\sqrt{\pi}}\frac{v_{\rm esc}}{v_0}\exp\left(-\frac{v_{\rm esc}^2}{v_0^2}\right)
\right].   
\end{align*}
The effective Sommerfeld factor in the galaxy is then weighted by $f(v_\chi)$:
\begin{align}
S_{\rm eff} = \int_0^\infty {\rm d}v_\chi\,f(v_\chi)\,S(v)\;, \label{eq:Seffindirectdetection}   
\end{align} 
where $S(v)$ is the Sommerfeld factor computed in previous sections with $v=2 v_\chi$ the relative velocity between two $\chi$ particles.

Let $\langle \sigma v \rangle$ be the thermally averaged cross section from two $\chi$ particles annihilating to SM particles in the galaxy, which is relevant for DM indirect detection. In the presence of long-range quantum forces, the cross section is boosted from $\langle \sigma v\rangle_0$ to $\langle \sigma v \rangle_{\rm eff} =S_{\rm eff} \langle \sigma v\rangle_0$. Here, $\langle \sigma v\rangle_0$ denotes the value in the perturbative limit ($S_{\rm eff}=1$). In Fig.~\ref{fig:Indirect_Vacuum}, we plot the boost factor $\langle \sigma v \rangle_{\rm eff}/\langle\sigma v\rangle_0$ as a function of DM mass obtained from three effective operators considered in this work (${\cal O}_S$, ${\cal O}_F$, ${\cal O}_{\widetilde{F}}$). The results include only contributions from vacuum potentials.
In the plot, the dots with arrows correspond to where the largest momentum transfer reaches the cutoff scale:
$m_\chi v_{\rm esc} =\Lambda$. Our results are independent of the UV completion to the left of the dots. We do not specify the value of $\langle \sigma v\rangle_0$, which depends on the short-distance interaction and can be calculated in any given model. The boost factors to the left of the dots, however, are general and should be taken into account for DM indirect detection as long as DM couples to some light mediator through the interactions described by (${\cal O}_S$, ${\cal O}_F$, ${\cal O}_{\widetilde{F}}$).

\subsubsection{Non-thermal background effect}
Now we consider the more interesting scenario where mediator particles are condensed and form a non-thermal background in the galaxy. As explained in Sec.~\ref{sec:non-thermal}, in order to have a significant effect on the Sommerfeld factor, the occupation number of the background particles needs to be much greater than 1, which can only happen for the bosonic mediator. Therefore, in the following analysis, we focus on the effect caused by ${\cal O}_S$. The relevant Lagrangian is given by
\begin{align}
{\cal L} = \bar{\chi} i \slashed{\partial} \chi + \frac{1}{2}\left(\partial_\mu \phi\right)^2 -m_\chi\bar{\chi}\chi- \frac{1}{2}m_\phi^2 \phi^2 - \frac{1}{\Lambda}\bar{\chi}\chi\frac{\phi^2}{2}\;.\label{eq:Lag}
\end{align}

We assume that the mediator particle $\phi$ and the DM particle $\chi$ share similar kinematic properties in the galaxy, $v_\phi \sim v_\chi \sim 10^{-3}$. If $m_\phi \ll m_\chi$,\footnote{This assumption may lead to a naturalness problem because the effective coupling between $\chi$ and $\phi$ contributes to the radiative correction to the $\phi$ mass: $\delta m_\phi^2 \sim m_\chi \Lambda$. Nevertheless, we allow possible fine tuning in this subsection. The realization of technically natural models for ultralight quadratically-coupled scalars can be found in \cite{Banerjee:2022sqg}.} then the background potential mediated by $\phi$ is independent of the phase-space distribution function of $\phi$ and generically scales as $V_{\rm bkg}^S\sim n_\phi/r$ (see Eq.~(\ref{eq:Vbkg-coherent})). According to Eq.~(\ref{eq:S-nonthermal}), the Sommerfeld factor induced by $V_{\rm bkg}^S$ also depends on the mass of $\phi$. Note that in our setup, the medium comprises both $\chi$ and $\phi$ particles. As a result, both $\chi$ and $\phi$ obtain a nonzero vacuum expectation value (VEV) from the on-shell particles in the background, and therefore both of their masses are shifted in the medium. More specifically, given the Lagrangian in Eq.~(\ref{eq:Lag}), the effective masses of $\chi$ and $\phi$ in the medium are given by
\begin{align}
m_\text{$\chi$,eff} &= m_\chi +  \frac{\langle \phi\rangle^2}{2\Lambda}\;,\label{eq:mchieff}\\
m_\text{$\phi$,eff}^2 &= m_\phi^2 + \frac{\langle \bar{\chi} \chi \rangle}{\Lambda}\;,\label{eq:mphieff}
\end{align}
where $\langle \phi\rangle$ and $\langle \bar{\chi}\chi\rangle$ denote the VEV of $\phi$ and the bilinear form $\bar{\chi}\chi$, respectively. Here $m_\phi$ and $m_\chi$ correspond to the bare mass, i.e., the mass in the absence of the background. 
In the medium where both $\phi$ and $\chi$ are non-relativistic, we have $\langle \phi \rangle =\sqrt{n_\phi/m_{\phi,\rm eff}}$ and $\langle \bar{\chi}\chi\rangle = n_\chi$, where $n_\phi$ and $n_\chi$ are the number density of $\phi$ and $\chi$ particles in the medium.

Furthermore, we introduce a parameter $\epsilon \in [0,1]$ to parametrize the fraction of $\phi$ in the DM density, and then we can write 
\begin{align}
n_\phi = \epsilon \rho_{\rm DM}/m_{\phi,\rm eff}\;,\quad n_\chi = (1-\epsilon) \rho_{\rm DM}/m_{\chi,\rm eff}\;,    
\end{align}
where $\rho_{\rm DM}\approx 0.4~{\rm GeV/cm^3}\approx 3\times 10^{-6}~{\rm eV}^4$ is the local DM energy density\footnote{Note that $\rho_{\rm DM}\approx 0.4~{\rm GeV/cm^3}$ is the DM density at the location of the Sun, while the value at the galactic center can be a few orders of magnitude higher. A complete calculation of the rates and spectra of indirect detection signals needs to convolve the DM annihilation cross section with the radial dependence of DM distribution in the galaxy. In the following discussion, to better illustrate our main idea, we neglect the dependence of $\rho_{\rm DM}$ and $\epsilon$ on the radial distance. A more careful treatment of the DM radial distribution is needed to obtain precise rates and spectra, but it does not affect our main conclusions in this subsection.}~\cite{ParticleDataGroup:2024cfk}. Substituting them into Eqs.~(\ref{eq:mchieff})-(\ref{eq:mphieff}) we obtain
\begin{align}
m_{\chi,\rm eff} &= m_\chi + \epsilon \frac{\rho_{\rm DM}}{2\Lambda m_{\phi,\rm eff}^2}\;,\label{eq:mchieff2}\\
m_{\phi,\rm eff}^2 &= m_\phi^2 + \left(1-\epsilon\right) \frac{\rho_{\rm DM}}{\Lambda m_{\chi,\rm eff}}\;.\label{eq:mphieff2}
\end{align}
Eqs.~(\ref{eq:mchieff2}) and (\ref{eq:mphieff2}) can be viewed as coupled quadratic equations for $m_{\chi,\rm eff}$ and $m_{\phi,\rm eff}^2$, which can be solved exactly. We  express the effective masses in terms of the bare masses and $\epsilon$:
\begin{align}
m_{\chi,\rm eff} &= \frac{m_\chi}{2} + \frac{\left(3\epsilon-2\right)\rho_{\rm DM}}{4\Lambda m_\phi^2} + \frac{1}{4  \Lambda m_\phi^2}\sqrt{\left(3\epsilon-2\right)^2\rho_{\rm DM}^2 + 4\left(2-\epsilon\right)m_\phi^2 m_\chi \Lambda \rho_{\rm DM}+4m_\phi^4 m_\chi^2 \Lambda^2}\;,\label{eq:mchieffsol}
\\
m_{\phi,\rm eff}^2 &= \frac{m_\phi^2}{2}-\frac{\left(3\epsilon-2\right)\rho_{\rm DM}}{4\Lambda m_\chi} + \frac{1}{4 \Lambda m_\chi} \sqrt{\left(3\epsilon-2\right)^2\rho_{\rm DM}^2 + 4\left(2-\epsilon\right)m_\phi^2 m_\chi \Lambda \rho_{\rm DM}+4m_\phi^4 m_\chi^2 \Lambda^2}\;,\label{eq:mphieffsol}
\end{align}
where we have dropped the other unphysical solution that vanishes as $\rho_{\rm DM} \to 0$. For the solution in Eqs.~(\ref{eq:mchieffsol})-(\ref{eq:mphieffsol}), they reduce to $m_{\chi,\rm eff} \approx m_{\chi}$ and $m_{\phi,\rm eff}^2 \approx m_\phi^2$ in the limit of $\rho_{\rm DM} \to 0$, as expected. 

We are interested in the parameter space where the background effect is significant, i.e., $m_{\chi,\rm eff} \gg m_\chi$, which will affect the DM indirect detection. This occurs when there is a hierarchy among the three terms in the square root of Eq.~(\ref{eq:mchieffsol}), namely when $\rho_{\rm DM} \gg m_\phi^2 m_\chi \Lambda$. We define a dimensionless parameter
\begin{align}
\eta \equiv \rho_{\rm DM}/(m_\phi^2 m_\chi \Lambda)\;,\label{eq:etadef}    
\end{align}
then Eq.~(\ref{eq:mchieffsol}) is reduced to
\begin{align}
\frac{m_{\chi,\rm eff}}{m_\chi} &= \frac{1}{2}+\left(\frac{3\epsilon-2}{4}\right)\eta+\frac{1}{4}\sqrt{\left(3\epsilon-2\right)^2 \eta^2 + 4\left(2-\epsilon\right)\eta + 4}\label{eq:ratiochifull}\\ 
&\overset{\eta \gg 1}{\approx} 
\left\lbrace
\begin{aligned}
&\frac{2-2\epsilon}{2-3\epsilon}  & \quad \text{if $0\leq \epsilon < 2/3$} \\
&\sqrt{\frac{\eta}{3}} & \text{if $\epsilon=2/3$}\\
&\left(\frac{3\epsilon-2}{2}\right)\eta & \text{if $2/3<\epsilon \leq 1$}
\end{aligned}
\right.\;.\label{eq:ratiochi}
\end{align}
Therefore, when $\epsilon$ is smaller than 2/3, the effective mass of $\chi$ is only increased by an ${\cal O}(1)$ factor compared to the bare mass; when $\epsilon$ is around $2/3$, the effective mass is increased by ${\cal O}(\sqrt{\eta})$; when $\epsilon > 2/3$, it is increased by ${\cal O}(\eta)$. Similarly, Eq.~(\ref{eq:mphieffsol}) is reduced to 
\begin{align}
\frac{m_{\phi,\rm eff}^2}{m_\phi^2} &= \frac{1}{2}-\left(\frac{3\epsilon-2}{4}\right)\eta+\frac{1}{4}\sqrt{\left(3\epsilon-2\right)^2 \eta^2 + 4\left(2-\epsilon\right)\eta + 4}\label{eq:ratiophifull}\\ 
&\overset{\eta \gg 1}{\approx} 
\left\lbrace
\begin{aligned}
&\left(\frac{2-3\epsilon}{2}\right)\eta  & 
\quad\text{if $0\leq \epsilon < 2/3$} \\
&\sqrt{\frac{\eta}{3}} & \text{if $\epsilon=2/3$}\\
&\frac{\epsilon}{3\epsilon-2} & \text{if $2/3<\epsilon \leq 1$}
\end{aligned}
\right.\;.\label{eq:ratiophi}
\end{align}    

\begin{figure}[t]
\centering
\includegraphics[width=2.9in]
{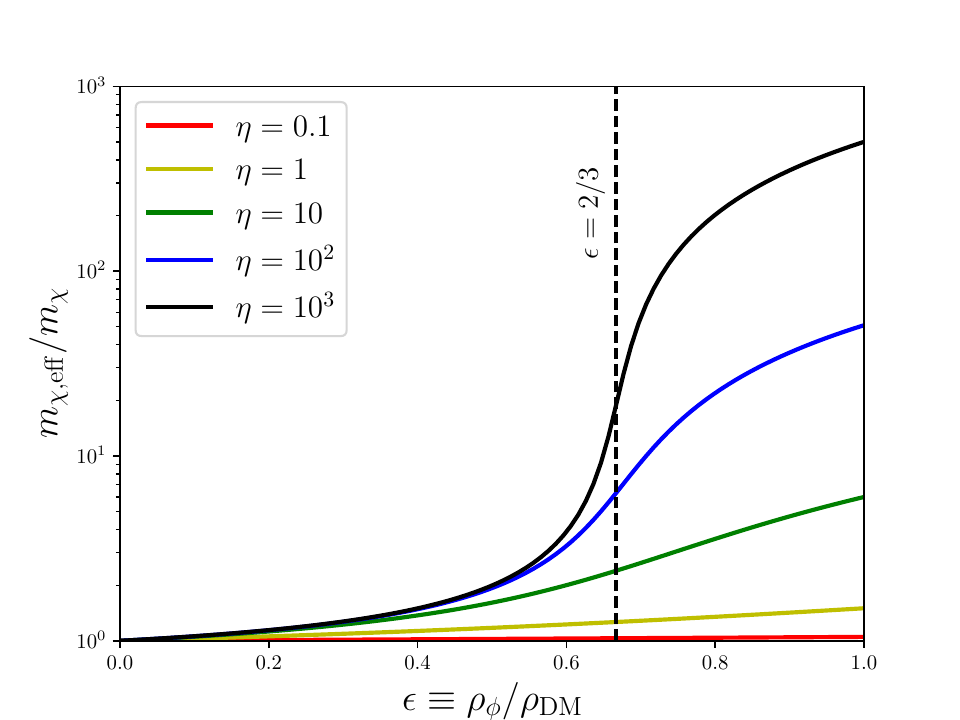}\quad
\includegraphics[width=2.9in]
{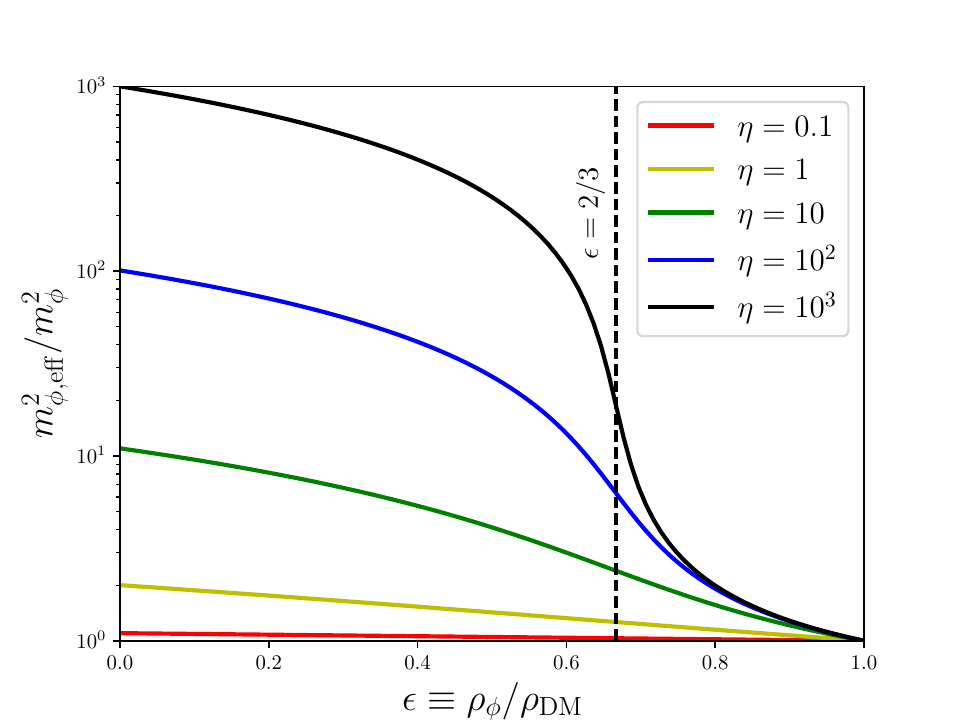}
\caption{\label{fig:effectivemass}The ratio between the effective mass in the medium and the bare mass for $\chi$ particle (left) and for $\phi$ particle (right) as a function of $\epsilon$. Here $\epsilon \in [0,1]$ is the fraction of $\phi$ energy density in the local DM energy density. Different colors denote different values of $\eta \equiv \rho_{\rm DM}/(m_\phi^2 m_\chi \Lambda)$. The vertical dashed line corresponds to the transition point at $\epsilon=2/3$.}
\end{figure}

The ratio between the effective mass and the bare mass for both $\chi$ and $\phi$ particles are plotted in Fig.~\ref{fig:effectivemass} as a function of $\epsilon$, with different fixed values of $\eta$. From Fig.~\ref{fig:effectivemass}, we see that the background effect is significant if $\eta\gg1$.

Next, we consider the constraints on $\eta$.
A large number density of light particles can be produced in the early universe non-thermally through mechanisms such as misalignment or inflationary fluctuations~\cite{Preskill:1982cy,Dine:1982ah,Turner:1987vd,Arias:2012az,Jaeckel:2016qjp,Redi:2022llj,Ismail:2024zbq,Kolb:2023ydq} and remains today. However, if $\phi$ is thermalized via its coupling to $\chi$ at late times, then the $\phi$-number-violating processes such as $\phi\phi \to \chi\chi$ will rapidly decrease the number density of $\phi$ and inject energy into the $\chi$ sector. For simplicity, we assume that $\phi$ does not couple to the SM sector directly, so that it does not affect the thermal history of the standard cosmology after $\chi$ freezes out.
In order for the large number density in the $\phi$ sector to remain until the present day, we require that the leading $\phi$-number-violating process $\phi\phi \to \chi\chi$ cannot maintain thermal equilibrium after the coherent oscillation of $\phi$ begins. Given the effective coupling in Eq.~(\ref{eq:Lag}), the interaction rate of this process is roughly $\Gamma\sim T^3/\Lambda^2$ at $T>m_\chi$ and is kinematically suppressed when the temperature drops below $m_\chi$. Comparing with the Hubble expansion rate $H\sim T^2/m_{\rm Pl}$, the decoupling temperature is estimated as $T_{\rm dec} \sim \text{max}\left\{\Lambda^2/m_{\rm Pl}, m_\chi\right\}$. For $m_\chi \sim {\rm TeV}$, which is our main focus in the following discussion, we have $T_{\rm dec}\sim \Lambda^2/m_{\rm Pl}$ if $\Lambda > \sqrt{m_\chi m_{\rm Pl}}\sim 10^{11}~{\rm GeV}$ or $T_{\rm dec} \sim m_\chi$ if $\Lambda < 10^{11}~{\rm GeV}$.
On the other hand, $\phi$ starts to oscillate at the temperature of $T_{\rm osc}\sim \sqrt{m_\phi m_{\rm Pl}}$. 
The requirement that $\phi$ is not thermalized after coherent oscillation starts requires $T_{\rm dec}>T_{\rm osc}$, which then leads to $\Lambda > m_{\phi}^{1/4}m_{\rm Pl}^{3/4}$ if $\Lambda > 10^{11}~{\rm GeV}$ or $m_\phi < m_\chi^2/m_{\rm Pl}\sim 10^{-4}~{\rm eV}$ if $\Lambda < 10^{11}~{\rm GeV}$. These conditions place some indirect constraints on $\eta$ and should be taken into account in the numerical analysis.

Apart from the above consideration from cosmology, there is a more direct constraint from the validity of the effective field theory (EFT). Note that in the Lagrangian (\ref{eq:Lag}), we do not include higher-dimensional operators such as $\bar{\chi}\chi \phi^4/\Lambda^3$ or $(\bar{\chi}\chi)^2\phi^2/\Lambda^4$. These higher-dimensional operators are negligible compared to the dimensional-five operator $\bar{\chi}\chi \phi^2/\Lambda$ included here only if the VEV of the field is smaller than the cutoff scale, i.e., $\langle \phi\rangle^2 < \Lambda^2$ and $\langle \bar{\chi}\chi \rangle < \Lambda^3$, or equivalently
\begin{align}
\frac{\epsilon \rho_{\rm DM}}{\Lambda^2 m_{\phi,\rm eff}^2} < 1\;,\qquad
\frac{\left(1-\epsilon\right)\rho_{\rm DM}}{\Lambda^3 m_{\chi,\rm eff}} < 1\;.\label{eq:EFTbound}
\end{align}
The second condition in Eq.~(\ref{eq:EFTbound}) is well satisfied as long as $\rho_{\rm DM}/(\Lambda^3 m_\chi)<1$; this is much easier to satisfy compared to the first condition because $m_\chi\gg m_\phi$, and it is not relevant to the bounds on $\eta$, so we only need to consider the first condition in the following analysis.
Recalling the definition of $\eta$ in Eq.~(\ref{eq:etadef}), the first condition in Eq.~(\ref{eq:EFTbound}) puts an upper bound on $\eta$:
\begin{align}
\eta < \frac{\Lambda}{\epsilon m_\chi} \frac{m_{\phi,\rm eff}^2}{m_\phi^2}\;.\label{eq:etaboundEFT}   
\end{align}
Note that the ratio $m_{\phi,\rm eff}^2/m_{\phi}^2$ is given by Eq.~(\ref{eq:ratiophifull}) and plotted in the right panel of Fig.~\ref{fig:effectivemass}, which is always greater than 1, and also $\epsilon \leq 1$. Therefore, as long as $\Lambda \gg m_\chi$, we can have parameter space that satisfies the EFT requirement and also makes $\eta \gg 1$. 

\begin{figure}[t]
\centering
\includegraphics[width=2.9in]
{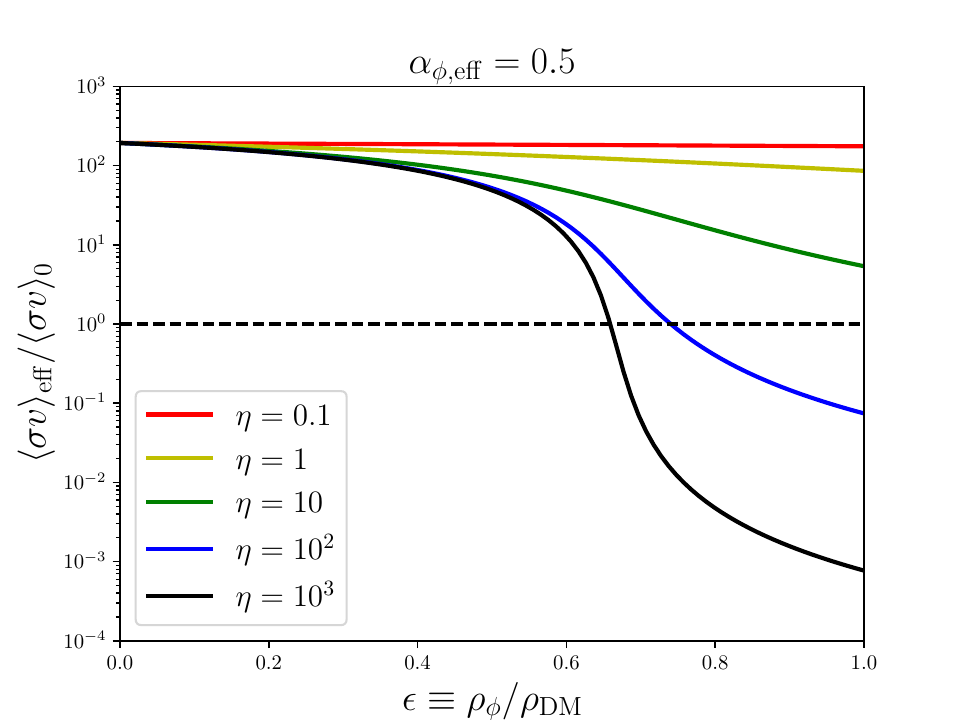}\quad
\includegraphics[width=2.9in]
{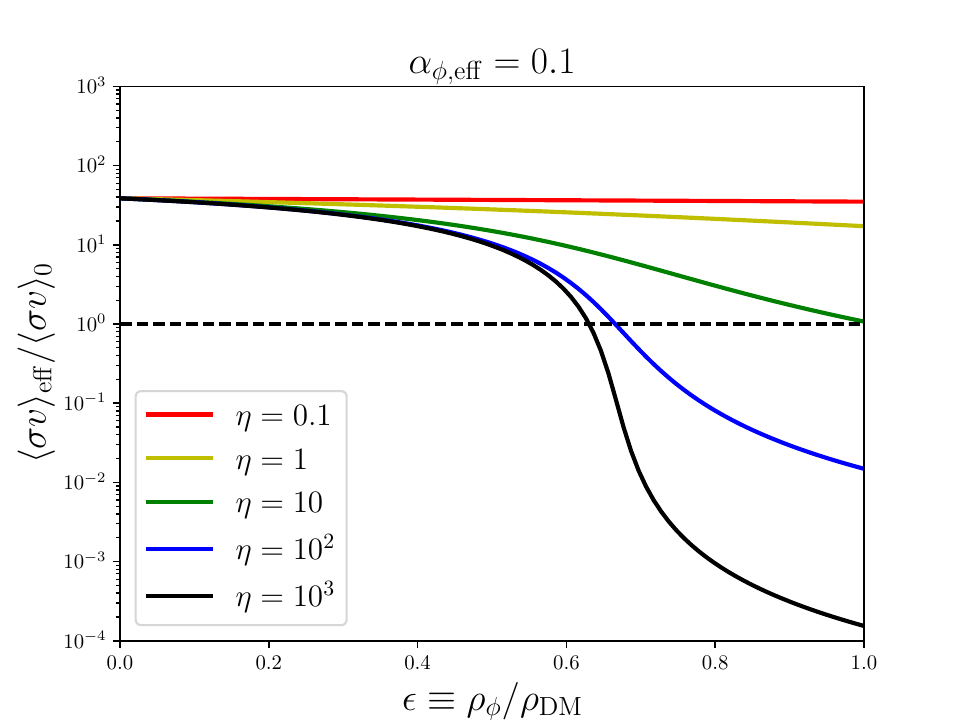}
\caption{\label{fig:boostfactor1D}The boost factor $\langle \sigma v\rangle_{\rm eff}/\langle \sigma v \rangle_0$ caused by the non-thermal background effect
as a function of $\epsilon$ for different fixed values of $\eta$ and $\alpha_{\phi,\rm eff}$. In the left (right) panel, we fix $\alpha_{\phi,\rm eff} = 0.5\,(0.1)$. The horizontal dashed line denotes the limit where there are no Sommerfeld and background effects.}
\end{figure}

We want to investigate how the above effect from the non-thermal background of $\phi$ can affect the DM indirect detection experiments. In the perturbative limit (no Sommerfeld enhancement), the $s$-wave annihilation cross section of $\chi + \chi \to {\rm SM} + {\rm SM}$ can be parametrized as $\langle \sigma v \rangle_0 = c/m_\chi^2$, where $c$ is some dimensionless constant. In the presence of the $\phi$ background, there are two effects that may change the effective annihilation cross section: (i) a large Sommerfeld factor that tends to enhance the cross section; (ii) a large effective mass of $\chi$ induced from the $\phi$ background that tends to decrease the cross section. The first effect is described by Eq.~(\ref{eq:S-nonthermal}), while the effective DM mass in the presence of the $\phi$ background is given by Eq.~(\ref{eq:mchieffsol}).
Combining these two effects, the effective cross section can be written as $\langle \sigma v \rangle_{\rm eff} = c S_{\rm eff}/m_{\chi,{\rm eff}}^2$, with $S_{\rm eff}$ given by Eq.~(\ref{eq:Seffindirectdetection}). Therefore, the boost factor can be calculated as follows:
\begin{align}
\frac{\langle \sigma v \rangle_{\rm eff}}{\langle \sigma v \rangle_{0}}= \frac{m_\chi^2}{m_{\chi,\rm eff}^2}\int_0^\infty{\rm d} v_\chi f(v_\chi)\,\frac{\alpha_{\phi,\rm eff}/(4v_\chi)}{1-e^{-\alpha_{\phi,\rm eff}/(4v_\chi)}}\;,\label{eq:boostfactor}
\end{align}
where 
\begin{align}
\alpha_{\phi,\rm eff} \equiv \frac{\epsilon \rho_{\rm DM}}{\Lambda^2 m_{\phi,\rm eff}^2} \;.\label{eq:alphaeff}
\end{align}
The mass ratio $m_{\chi,\rm eff}/m_\chi$ can be expressed as a function of $\epsilon$ and $\eta$, as shown in Eq.~(\ref{eq:ratiochifull}). So, the boost factor (\ref{eq:boostfactor}) depends only on three dimensionless parameters: $\epsilon$, $\eta$, and $\alpha_{\phi,\rm eff}$. In addition, the validity of EFT (\ref{eq:EFTbound}) requires $\alpha_{\phi,\rm eff} < 1$.

\begin{figure}[t]
\centering
\includegraphics[width=2.9in]
{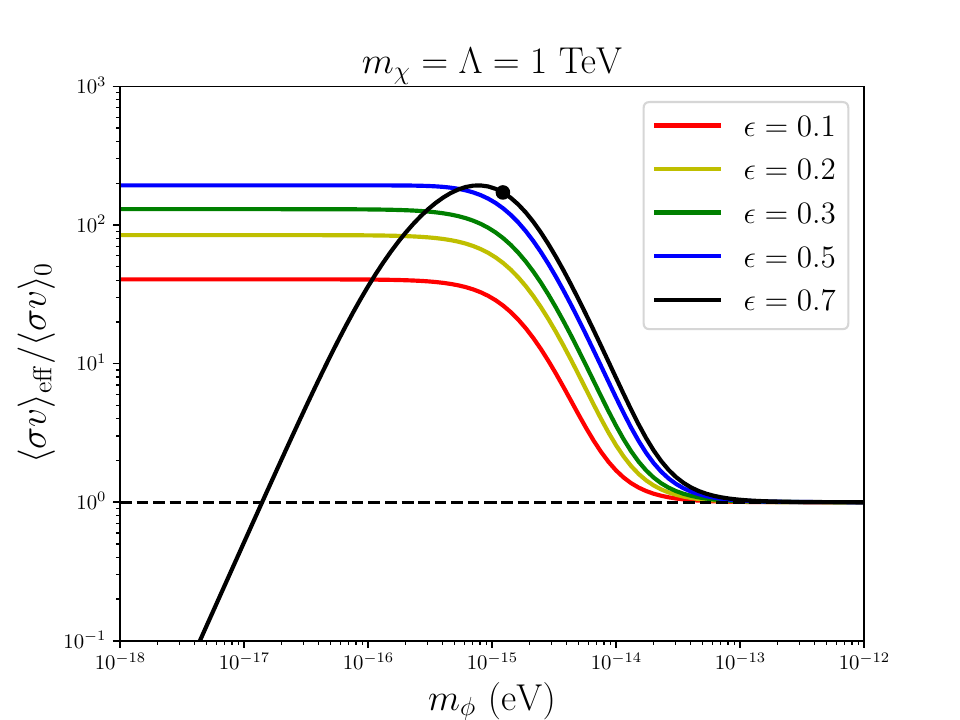}\quad
\includegraphics[width=2.9in]
{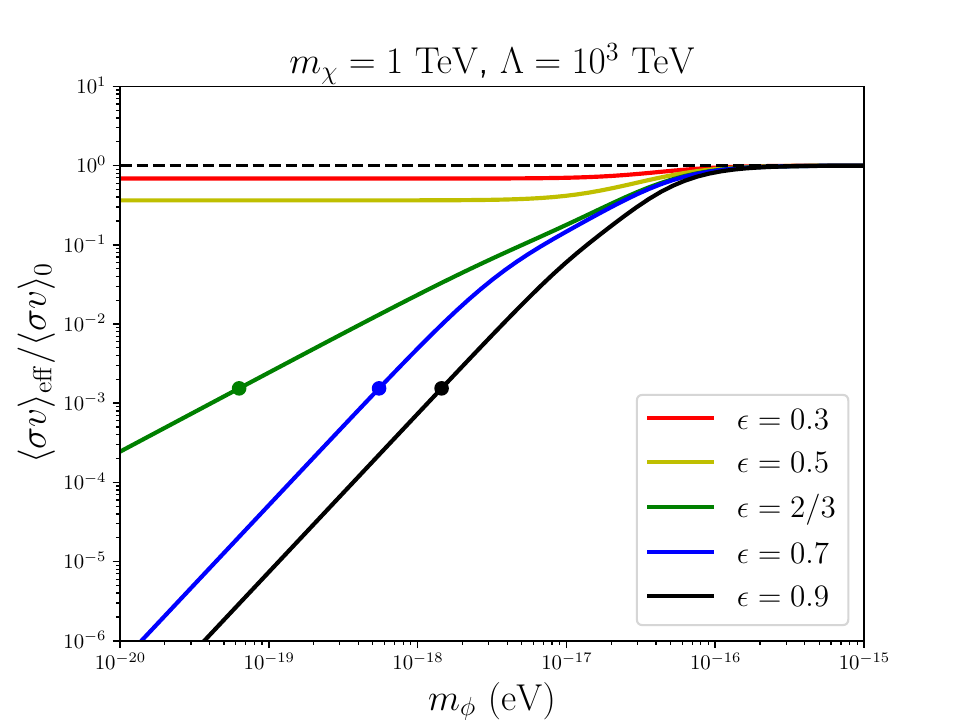}\\
\includegraphics[width=2.9in]
{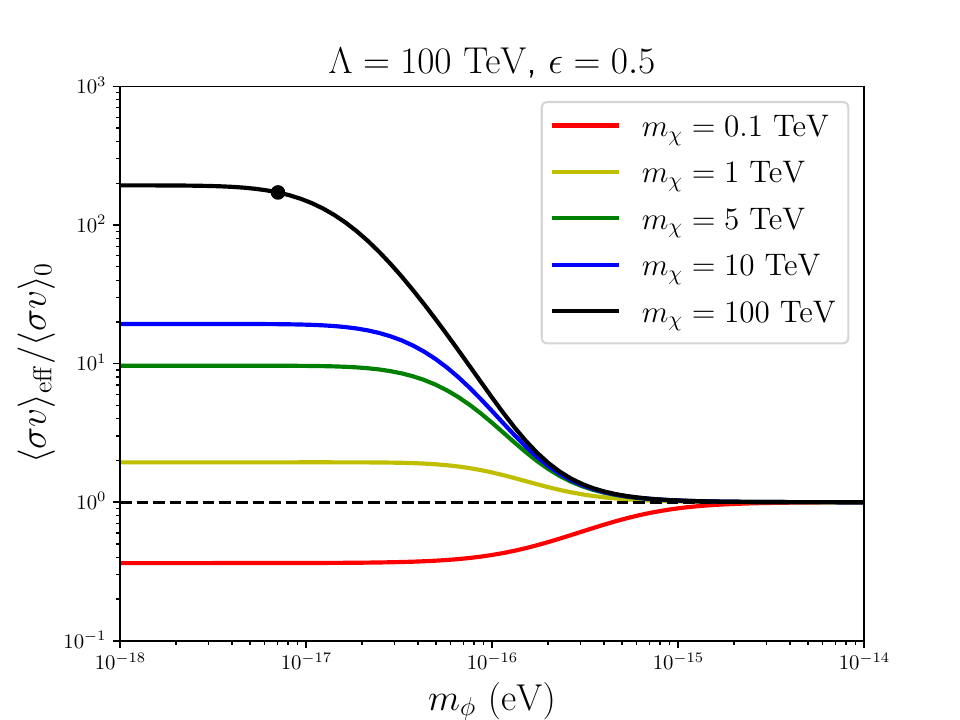}\quad
\includegraphics[width=2.9in]
{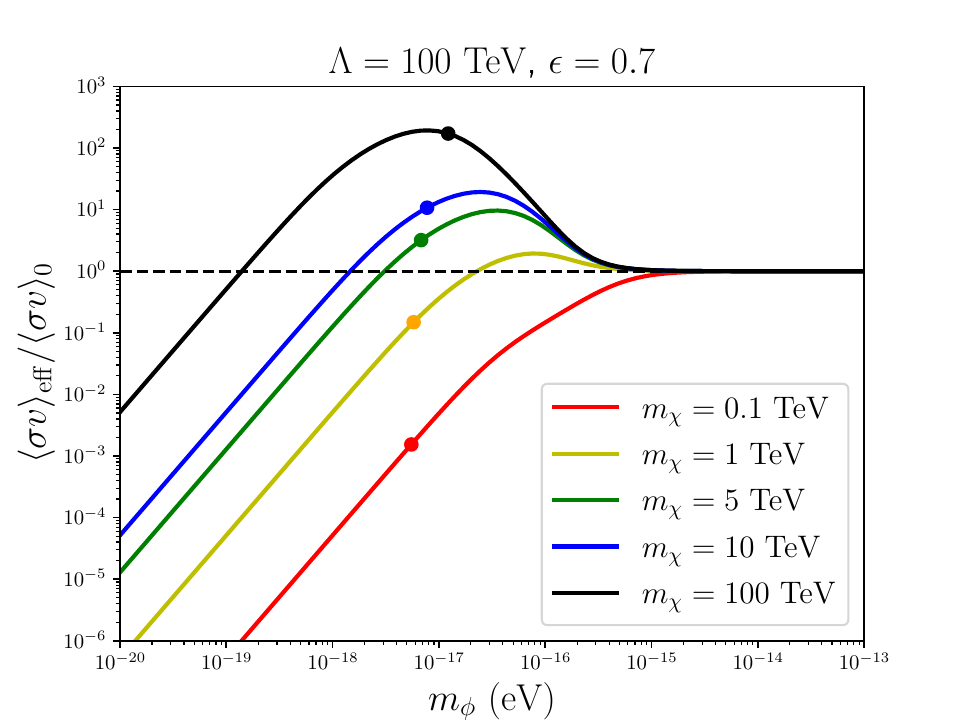}
\caption{\label{fig:boost1Dmphi} The boost factor $\langle \sigma v\rangle_{\rm eff}/\langle \sigma v \rangle_0$ caused by the non-thermal background effect as a function of the scalar mass. In the upper panel, we fix $m_\chi$ and $\Lambda$, and show results for different values of $\epsilon$. In the lower panel, we fix $\Lambda$ and $\epsilon$, and show results for different values of $m_\chi$. For each line, the region to the left of the dot is disfavored by the EFT bound in Eq.~(\ref{eq:EFTbound}). The lines with no dot satisfy the EFT bound in the whole parameter space. The horizontal dashed line denotes the limit without Sommerfeld and background effects.}
\end{figure}

\begin{figure}[t]
\centering
\includegraphics[width=5.5in]
{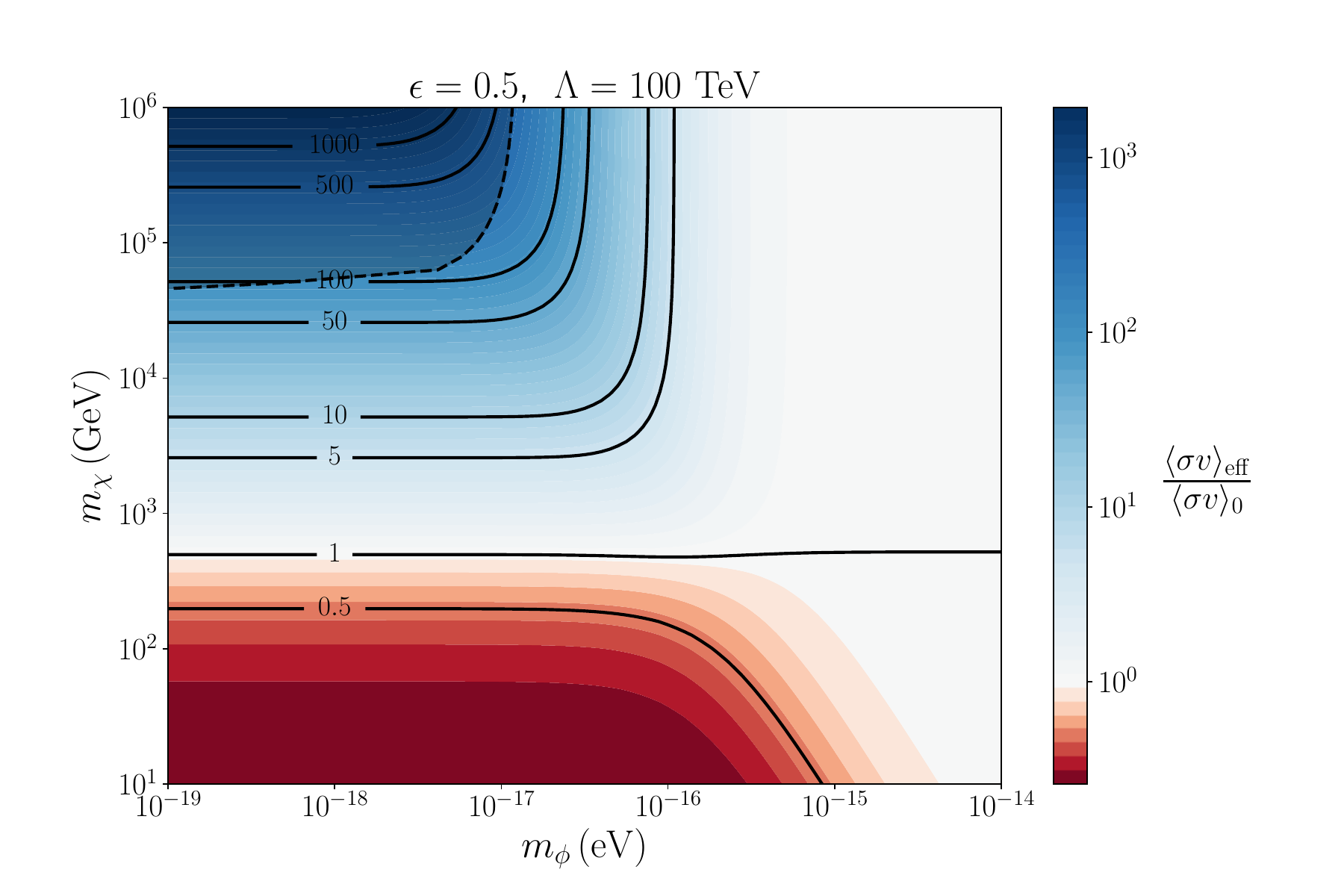}
\includegraphics[width=5.5in]
{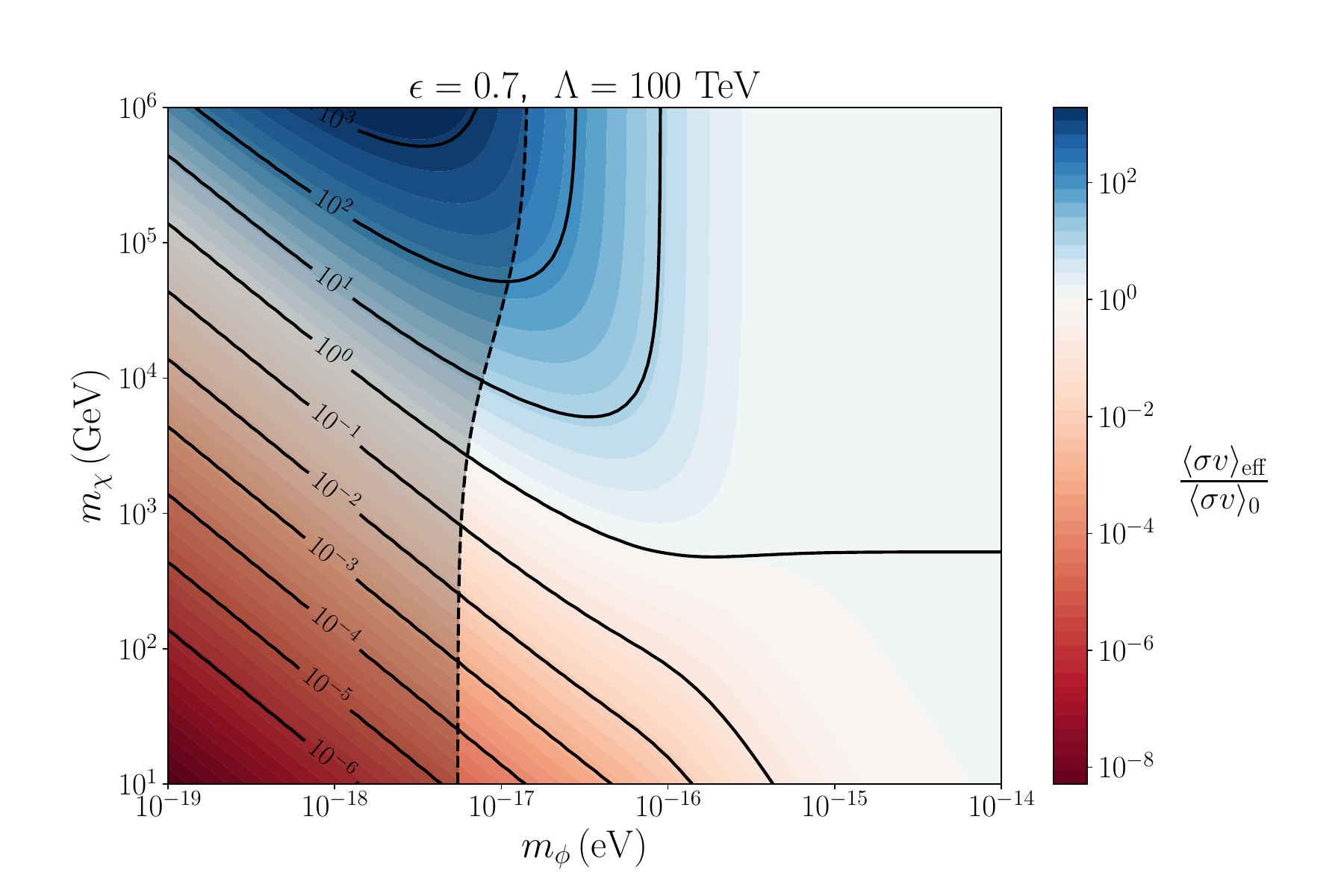}
\caption{\label{fig:contourmphimchi}
Contour plot of the boost factor $\langle \sigma v\rangle_{\rm eff}/\langle \sigma v \rangle_0$ as a function of $m_{\phi}$ and $m_{\chi}$. Here we set $\epsilon = 0.5$ (upper) and $\epsilon = 0.7$ (lower), and $\Lambda $ = 100 TeV for both. The dark shaded region (left to the dashed line) denotes the region where the EFT loses validity. }
\end{figure}

In Fig.~\ref{fig:boostfactor1D}, we plot the boost factor as a function of $\epsilon$ for different values of $\eta$ and $\alpha_{\phi,\rm eff}$. The behaviors of the boost factor in different regions can be understood analytically. First, for $\epsilon < 2/3$, the suppression from the effective mass is not significant (see Eq.~(\ref{eq:ratiochi})), so for a fixed coupling $\alpha_{\phi,\rm eff}$, the boost factor is enhanced by the Sommerfeld factor, $S\sim \alpha_{\phi,\rm eff}/v$ if $\alpha_{\phi,\rm eff} \gg v\sim 10^{-3}$ is satisfied. For $\epsilon > 2/3$, according to Eq.~(\ref{eq:ratiochi}), the suppression from the effective mass becomes significant: $m_\chi^2/m_{\chi,\rm eff}^2 \sim 1/\eta^2$ if $\eta\gg 1$, so the boost factor scales as $\langle \sigma v\rangle_{\rm eff}/\langle \sigma v\rangle_0 \sim \alpha_{\phi,\rm eff}/( \eta^2 v)$. Therefore, the boost factor can be much smaller than 1 at $\epsilon > 2/3$ as long as $\eta \gg \sqrt{\alpha_{\phi,\rm eff}/v}$. 

Since both $\eta$ and $\alpha_{\phi,\rm eff}$ depend on $m_\phi$, the boost factor is also affected by the scalar mass.
In Fig.~\ref{fig:boost1Dmphi}, the boost factor is shown as a function of the scalar mass. 
One can clearly see that there is a phase transition when $\epsilon$ crosses $2/3$. At $\epsilon < 2/3$, according to Eqs.~(\ref{eq:alphaeff}) and (\ref{eq:ratiophi}), we have $\alpha_{\phi,\rm eff} \approx \frac{2\epsilon}{2-3\epsilon} \frac{m_\chi}{\Lambda}$ in the small $m_\phi$ limit. Therefore, the boost factor tends to flat and is proportional to $m_\chi/\Lambda$ at small $m_\phi$. At $\epsilon > 2/3$, the background effect on the DM mass becomes significant. The 
additional suppression from $m_\chi^2/m_{\chi,\rm eff}^2$ makes the boost factor scale as $m_\phi^2 m_\chi^2/\rho_{\rm DM}$ at small $m_\phi$. It is interesting to note that the boost factor is independent of the cutoff scale in this limit, since the dependence of $\Lambda$ is exactly canceled between $\alpha_{\phi,\rm eff}$ and $m_\chi^2/m_{\chi,\rm eff}^2$. 

At sufficiently small $m_\phi$, the EFT bound (\ref{eq:EFTbound}) is violated if $\epsilon<2/3$, indicating that $\alpha_{\phi,\rm eff}>1$ and the contributions of higher-dimensional operators are not negligible. This is represented by the dot on each line in Fig.~\ref{fig:boost1Dmphi}. Our results are only valid to the right of these dots. In Fig.~\ref{fig:contourmphimchi}, we show the contour plot of the boost factor as a function $m_\phi$ and $m_\chi$, where the dark shaded region is disfavored by the EFT bound. We can see from both Fig.~\ref{fig:boost1Dmphi} and Fig.~\ref{fig:contourmphimchi} that under the constraint of the EFT, there exist both possibilities where the boost factor can be greater or smaller than 1. Typically, the boost factor is enhanced if $m_\chi \sim \Lambda$, and is suppressed if $m_\chi \ll \Lambda$.

In summary, the quantum force can have important effects on indirect DM detection. It is particularly interesting if the light mediator particles condense and form a non-thermal background in the galaxy. In this scenario, there exist two competing effects: the Sommerfeld enhancement from the $1/r$ background potential tends to enhance DM annihilation, whereas a large DM effective mass induced by the background of mediator particles tends to reduce DM annihilation. In the presence of both effects, we find that the effective cross section for DM annihilation can be significantly enhanced or suppressed, depending on the value of $\epsilon$ and the hierarchy between $m_\chi$ and $\Lambda$. This leads to two important implications for DM detection in the galaxy. First, if the DM effective mass is changed by the background, then the spectrum of indirect detection should also be shifted. Second, a significant enhancement or suppression of the DM annihilation cross section alters the rate of the indirect detection signals, which may open a new window for DM searches in the galaxy.

\section{Conclusions}
\label{sec:conclusion}

Cross section of dark matter pair-annihilation at non-relativistic velocities is a phenomenologically important quantity in many models of particle DM. For example, if DM is a thermal relic, its present abundance is directly determined by the cross section of annihilation at freeze-out. Dark matter annihilation process during and after recombination can modify properties of the CMB, leading to important phenomenological constraints. Annihilation of DM pairs in the present universe can provide a promising path towards discovery of non-gravitational signatures of DM through indirect detection. 

It is well known that cross section of DM annihilation at low velocities can be modified by the Sommerfeld effect, which occurs whenever there is a long-range force between colliding DM particles. Numerous authors have studied the Sommerfeld effect arising from classical forces, {\it i.e.} forces due to tree-level exchange of light bosonic mediator particles. In this paper, we extended the calculation of the Sommerfeld effect to the case of quantum forces, arising from loop-level exchanges of light mediators for both bosonic and fermionic mediators. In many models, such quantum forces are the leading long-range interaction, due to the existence of conserved quantum numbers prohibiting an exchange of a single mediator. We found that quantum forces can lead to large Sommerfeld factors for realistic model parameters. Depending on parameters, annihilation cross sections can be either enhanced (due to attractive quantum forces) or suppressed (due to a potential barrier arising from the quantum force). Phenomenologically, we found that these effects can change the DM yield from a thermal freeze-out by about 10\% in either direction. The size of the effect is mainly limited by the fact that the DM velocities at freeze-out, $\sim 0.1$, are not particularly low. The effect on the CMB constraints can be much stronger, as large as several orders of magnitude in the annihilation cross section, due to lower DM velocities in the recombination epoch.      

A very interesting feature of quantum forces, noted recently in different contexts, is the possibility of strong enhancement of the force in the presence of a background of the mediator particles. This situation is very natural in the context of DM physics, since mediator particles are generically present in the early universe, and may survive into the present times. We calculated the Sommerfeld factors due to a quantum force in the presence of a mediator background. We found that the Sommerfeld factors from background corrections exhibit novel features that do not exist in the familiar case of the Sommerfeld factor from the Yukawa potential. In particular, for thermal backgrounds, we observed resonance peaks for massless mediators in some models, which are purely induced by temperature (for ${\cal O}_S$ and ${\cal O}_{\widetilde{F}}$). We also found that there can be both Sommefeld enhancement and suppression in the same model with different DM masses, due to the fact that the two-fermion background potential changes sign around $r\sim 1/T$ (for ${\cal O}_F$). This novel and interesting structure in the Sommerfeld factor can be potentially explained by a resonance with a quasi-bound state.

The background effect on the Sommerfeld factor and DM annihilation is more significant if the mediator particles form a non-thermal background with large number density. This can be realized in a simple setup.
For example, one may consider a two-component DM model, with a heavy component (e.g. WIMPs) and a population of much lighter particles that both mediate a long-range quantum force and themselves contribute to DM density. We calculated the Sommerfeld factor from the quantum force in such a scenario with bosonic light mediators. Using these calculations, we found that the annihilation cross section of weak-scale DM particles in the Milky Way (relevant for indirect detection) can be enhanced by up to a factor of 100 by the Sommerfeld effect in a phenomenologically viable two-component DM model with an ultralight component. At the same time we found that in this model, the shift in effective mass of the weak-scale DM particles due to the background of the ultralight DM may also become important, leading to suppression of the indirect-detection cross sections by up to several orders of magnitude, as well as changes in the spectrum of the observable annihilation products.   

This paper is the first exploration of the Sommerfeld effect from quantum forces, both in vacuum and in a background of mediator particles, in the context of DM physics. We demonstrated that this effect can be both numerically significant and phenomenologically relevant in simple and viable models of DM. This strongly motivates further work along these lines. While here we focused on DM annihilation, quantum forces can also mediate DM self-interaction, potentially leaving an imprint in the large-scale structure of the universe. Another interesting possibility is that an attractive quantum force may be strong enough to form DM bound states, with important implications for cosmology and phenomenology. Also, while in this paper the quantum force  was assumed to be mediated by novel Beyond-the-SM states, there has been recent interest in models where the DM interacts with SM neutrinos. It is well known in the SM context that neutrinos mediate a quantum long-range force. It will be interesting to investigate whether this force can play a role in DM physics.

\section*{Acknowledgement}
We thank Brando Bellazzini, Bhupal Dev, Mitrajyoti Ghosh,
Majed Khalaf, Nick Rodd, Philip Tanedo, and especially Tracy Slatyer for helpful discussions and comments. This work is
supported in part by the NSF grant PHY-2309456.

\begin{appendix}

\section{Quantum forces from effective couplings}
\label{app:quantum-force}
In this appendix, we provide calculation details for quantum forces with both scalar and fermionic mediators, which should be taken as input in the calculation of the Sommerfeld factor. The full potentials include the vacuum part $V_{0}$ and the background-induced part  $V_{\text{bkg}}$. Note that the results of both $V_0$ and $V_{\rm bkg}$ are known~\cite{Fichet:2017bng,Brax:2017xho,Horowitz:1993kw,Ferrer:1998ju,Ferrer:1998rw,Ferrer:1999ad,Ferrer:2000hm,Ghosh:2022nzo,VanTilburg:2024xib,Barbosa:2024pkl,Ghosh:2024qai,Grossman:2025cov,Cheng:2025fak,Gan:2025nlu}. Here we aim to provide a pedagogical derivation of them in a unified framework of field theory.

The first step is to compute the matrix element for $t$-channel DM elastic scattering process $\chi\chi \to \chi\chi$, which in both cases schematically takes the form 
\begin{align}
    \mathcal{M} \sim \int {\rm d}^{4}k\, 
    D(k) D(k+q)\;,\label{eq:schematic}
\end{align}
where $k$ is the loop momentum and $q$ is the momentum transfer. $D$ is the modified propagator in a background of mediator particles with finite density. For bosonic and fermionic mediators, respectively, the modified propagators are
\begin{align}
    D_{\phi}(k) &= 
    \frac{i}{k^{2}-m_{\phi}^{2}} 
    + 
    2\pi \delta(k^{2}-m_{\phi}^{2})
    f_{\phi}(\mathbf{k})\;,\label{Dphi}
    \\
    D_{\psi}(k) &= 
    (\slashed{k}+m_{\psi})
    \bigg(
    \frac{i}{k^{2}-m_{\psi}^{2}} 
    - 
    2\pi \delta(k^{2}-m_{\psi}^{2})
    f_{\psi}(\mathbf{k})
    \bigg)\;,\label{eq:DpsiApp}
\end{align}
where $f_{\phi}(\veck)$ and $f_\psi(\veck)$ are time-independent phase-space distribution functions for bosons and fermions, respectively.
One can see that $\mathcal{M}$ splits up as $\mathcal{M} = \mathcal{M}_{0} + \mathcal{M}_{\text{bkg}}$, where the first term comes from the normal part in both propagators and the second term comes from the cross terms in $D(k)D(k+q)$ (one propagator takes the normal part and the other takes the $\delta$-function part). The term where both propagators take the $\delta$-function part contains an additional $i$ compared to $\mathcal{M}_{0}$ and $\mathcal{M}_{\text{bkg}}$, so it does not contribute to the interacting potential.

From the Born approximation, the potential is related linearly to $\mathcal{M}$, so it also splits up as $V(r) = V_{0}(r) + V_{\text{bkg}}(r)$. The first term $V_{0}(r)$ is the vacuum potential, and the second term  $V_{\text{bkg}}(r)$ is the background potential. They are calculated below separately for three relevant effective operators: ${\cal O}_S$, ${\cal O}_F$, and ${\cal O}_{\widetilde{F}}$.

\subsection*{Vacuum potentials}
Taking the non-relativistic limit of external wavefunctions (\ref{eq:wavefunction}), the vacuum parts of the amplitude from ${\cal O}_S$, ${\cal O}_F$, and ${\cal O}_{\widetilde{F}}$ are given by
\begin{align}
    \frac{i\mathcal{M}^{S}_{0}}{4m_\chi^2} &= \frac{1}{2}
    \left(\frac{i}{\Lambda}\right)^2 
    \int\frac{{\rm d}^{4}k}{(2\pi)^{4}}
    \frac{i}{k^{2}-m_{\phi}^{2}}
    \frac{i}{(k+q)^{2}-m_{\phi}^{2}}\;,\\
    \frac{i\mathcal{M}_{0}^{F}}{{4m_\chi^2}}
    &= 
    -\frac{1}{2}\left(\frac{i}{\Lambda^{2}}\right)^2
    \int 
    \frac{{\rm d}^{4}k}{(2\pi)^{4}}
    \text{Tr} \bigg[ 
    \frac{i(\slashed{k}+m_{\psi})}
    {k^{2}-m_{\psi}^{2}}
    \frac{i(\slashed{k}+\slashed{q}+m_{\psi})}
    {(k+q)^{2}-m_{\psi}^{2}}
    \bigg]\;,\\
    \frac{i\mathcal{M}^{\widetilde{F}}_{0}}{4m_\chi^2} &= - \left(\frac{i}{\Lambda^{2}}\right)^2 \int 
    \frac{{\rm d}^{4}k}{(2\pi)^{4}}
    \text{Tr} \bigg[ 
    \frac{i\gamma^0(\slashed{k}+m_{\psi})}
    {k^{2}-m_{\psi}^{2}}
    \frac{i\gamma^0(\slashed{k}+\slashed{q}+m_{\psi})}
    {(k+q)^{2}-m_{\psi}^{2}}
    \bigg]\;,
\end{align}
where the normalization factor $4m_\chi^2$ comes from the external wavefunction in the non-relativistic limit, the factor of $1/2$ is the symmetry factor for real scalar/Majorana fermion in the loop, and there is a minus sign for the fermion loop.

The above amplitudes can be computed explicitly using Feynman parametrization and dimensional regularization (with the convention of spacetime dimension $d\equiv 4-2\epsilon$ and $\epsilon \to 0^+$):
\begin{align}
\frac{{\cal M}_0^S}{4m_\chi^2} &=  \frac{1}{32\pi^{2}\Lambda^{2}}
    \int_{0}^{1}{\rm d}x 
  \left(
    \frac{1}{\tilde{\epsilon}} - 
    \log
    \Delta_\phi 
    \right),\label{eq:M0S}\\
\frac{{\cal M}_0^F}{4m_\chi^2} &= -\frac{3}{8\pi^{2}\Lambda^{4}}
    \int_{0}^{1}{\rm d}x \, \Delta_\psi
\left(\frac{1}{3}+\frac{1}{\tilde{\epsilon}} - \log\Delta_\psi\right),\label{eq:M0F}\\
\frac{{\cal M}_0^{\widetilde{F}}}{4m_\chi^2} &=  -\frac{1}{2\pi^2 \Lambda^4}\int_0^1 {\rm d}x\,x(1-x)\,q^2\left(\frac{1}{\tilde{\epsilon}}-\log\Delta_\psi\right),\label{eq:M0Ftilde} 
\end{align}
where $1/\tilde{\epsilon}\equiv 1/\epsilon + \log(4\pi)-\gamma_{\rm E}$ with $\gamma_{\rm E} \approx 0.577$ being Euler's constant, and $\Delta_i \equiv m_i^2-q^2x(1-x)$ for $i=\phi$ or $\psi$ have been defined. According to the Born approximation, the non-relativistic interacting potential is related to the scattering amplitude via Fourier transform 
\begin{align}
V_0^j(r) = -\frac{1}{4m_\chi^2}\int \frac{{\rm d}^3\vecq}{\left(2\pi\right)^3}e^{i\vecq\cdot\vecr} {\cal M}_0^j  \;, \qquad \text{for $j=S,F,\widetilde{F}$}\;.\label{eq:Born}
\end{align}
The amplitude ${\cal M}_0^j$ is a function of $t\equiv q^2$, and for the case of $t$-channel elastic scattering considered here, we have $q^\mu\approx (0,\vecq)$ and $q^2 \approx -\vecq^2$, so $t$ is always \emph{negative} in the physical domain. It is easy to check that all the amplitudes in Eqs.~(\ref{eq:M0S})-(\ref{eq:M0Ftilde}) are real with $q^2<0$. As a result, the vacuum potential (\ref{eq:Born}) is also real.

\begin{figure}[t]
\centering
\includegraphics[width=2.5in]
{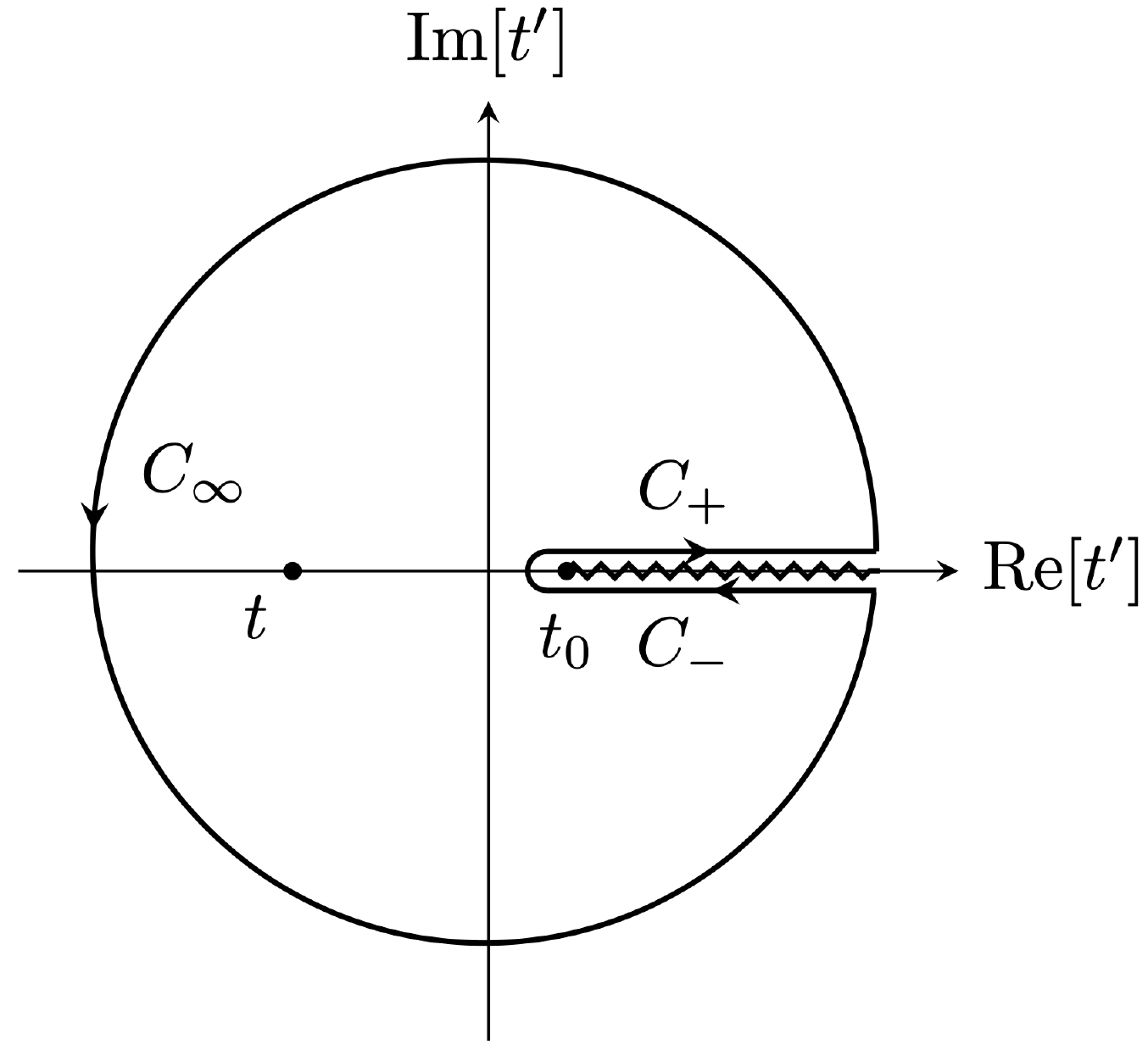}
\caption{\label{fig:contour}The contour used to compute the vacuum quantum forces. Here $t'$ is an arbitrary complex variable, $t\equiv q^2<0$ is the square of the momentum transfer, and $t_0\equiv(2m_{\phi})^2$ or $(2m_\psi)^2$ is the kinematic threshold for decaying to $2\phi$ or $2\psi$ particles.}
\end{figure}

To compute the integral in Eq.~(\ref{eq:Born}) in a more convenient way, we employ analytic continuation to extend the amplitude from the negative real axis to the whole complex plane:
\begin{align}
{\cal M}_0^j(t) = \frac{1}{2\pi i}\int_C {\rm d}t'\,\frac{{\cal M}_0^j(t')}{t'-t}\;,\label{eq:Cauchy}    
\end{align}
where the contour $C = C_+ + C_\infty + C_-$ is chosen as in Fig.~\ref{fig:contour} in the complex $t'$-plane to avoid the branch cut on the positive real axis, which is located in $t' \geq t_0 \equiv 4m_i^2$ (for $i=\phi,\psi$). Note that for Eq.~(\ref{eq:Cauchy}) to hold, we have assumed that ${\cal M}_0^j (t')$ does not contain additional poles on the positive real axis other than the branch cut (which is satisfied for our case); otherwise, one needs to include the residue at those additional poles as well. If we take $t'\to \infty$, the integral along the large circle $C_\infty$ is only relevant for the short-distance interaction. In fact, one can verify that ${\cal M}_0(t')$ vanishes as $t'\to \infty$ in any UV theory after renormalization. So the integral along $C_\infty$ does not contribute to the long-range force. Therefore, the amplitude in Eq.~(\ref{eq:Cauchy}) can be written as
\begin{align}
{\cal M}_0^j(t) &= \frac{1}{2\pi i} \left[\int_{t_0}^\infty {\rm d}t' \frac{{\cal M}_0^j(t' + i 0^+)}{t'-t} + \int_\infty^{t_0} \frac{{\cal M}_0^j(t' - i 0^+)}{t'-t}\right]=\frac{1}{\pi} \int_{t_0}^\infty {\rm d}t'\, \frac{{\rm Im}[{\cal M}_0^j(t')]}{t'-t}\;,\label{eq:Cauchy2}
\end{align}
where in the last step, we used ${\cal M}_0^j(t'+i0^+)-{\cal M}_0^j(t'+i0^-) = 2i\,{\rm Im}[{\cal M}_0^j(t')]$. Substituting Eq.~(\ref{eq:Cauchy2}) into Eq.~(\ref{eq:Born}), we obtain
\begin{align}
V_0^j(r)=-\frac{1}{4m_\chi^2}\frac{1}{\pi}\int_{t_0}^\infty {\rm d}t'\, {\rm Im}[{\cal M}_0^j(t')]\int \frac{{\rm d}^3\vecq}{\left(2\pi\right)^3}e^{i\vecq\cdot\vecr} \frac{1}{t'-t}\;.  
\end{align}
The integral over $\vecq$ can easily be worked out in the spherical coordinate by recalling $t=-\vecq^2$:
\begin{align}
    \int\frac{{\rm d}^3\vecq}{\left(2\pi\right)^3}e^{i\vecq\cdot\vecr} \frac{1}{t'-t} = -\frac{i}{4\pi^2 r} \int_{-\infty}^\infty {\rm d}|\vecq| \frac{|\vecq|}{|\vecq|^2 +t'}e^{i|\vecq|r} = \frac{1}{4\pi r} e^{-\sqrt{t'}r}\;,
\end{align}
where in the last step we closed the contour upward and used the residue theorem. Therefore, the interacting potential is given by
\begin{align}
V_0^j(r)=-\frac{1}{4m_\chi^2}\frac{1}{4\pi^2 r}\int_{t_0}^\infty {\rm d}t'\, e^{-\sqrt{t'}r}\,{\rm Im}[{\cal M}_0^j(t')]\;. \label{eq:dispersion}
\end{align}
The expression in Eq.~(\ref{eq:dispersion}) is known as the dispersion formalism, which was first developed by Feinberg and Sucher \cite{Feinberg:1968zz} to calculate the two-neutrino exchange potential; see \cite{Feinberg:1989ps} for a review and applications in other quantum forces.
Since the expression relies only on the \emph{imaginary} part of the amplitude (after analytic continuation) instead of the whole amplitude, the optical theorem guarantees that the result is always finite, as it should be. 

The remaining thing is to compute the imaginary part of the amplitude in Eqs.~(\ref{eq:M0S})-(\ref{eq:M0Ftilde}), which only appears from the term proportional to $\log \Delta_\phi$ or $\log \Delta_\psi$ with negative arguments. Replacing $q^2$ with $t'$ in $\Delta_\phi$ and $\Delta_\psi$, we have\footnote{The sign of the imaginary part is determined by the branch of logarithmic function. The use of the Feynman propagator $1/(k^2-m_\phi^2+i0^+)$ is equivalent to replacing $m_\phi^2$ with $m_\phi^2 - i0^+$ in the amplitude. Then we can use the identity $\log(-A-i0^+) = \log(A) -i\pi$ for any positive $A$.}
\begin{align}
{\rm Im}\, \int_0^1 {\rm d}x \log\Delta_\phi &= -\pi \int_0^1{\rm d}x\,\Theta\left[t'x(1-x)-m_\phi^2\right] = -\pi\left(1-\frac{4m_\phi^2}{t'}\right)^{1/2}\,\Theta\left(t'-4m_\phi^2\right).
\end{align}
Similarly, one can obtain
\begin{align}
{\rm Im}\, \int_0^1 {\rm d}x\, \Delta_\psi\log\Delta_\psi &= \frac{\pi t'}{6 }\left(1-\frac{4m_\psi^2}{t'}\right)^{3/2}\,\Theta\left(t'-4m_\psi^2\right),\\
{\rm Im}\, \int_0^1 {\rm d}x\, x(1-x)\log\Delta_\psi &= -\frac{\pi}{6}\left(1+\frac{2m_\psi^2}{t'}\right)\left(1-\frac{4m_\psi^2}{t'}\right)^{1/2}\,\Theta\left(t'-4m_\psi^2\right).
\end{align}
Substituting them back to Eqs.~(\ref{eq:M0S})-(\ref{eq:M0Ftilde}) one obtains 
\begin{align}
\frac{1}{4m_\chi^2} {\rm Im}[{\cal M}_0^S(t')] &= \frac{1}{32\pi\Lambda^2}\left(1-\frac{4m_\phi^2}{t'}\right)^{1/2}\,\Theta\left(t'-4m_\phi^2\right),\label{eq:ImMS}\\
\frac{1}{4m_\chi^2} {\rm Im}[{\cal M}_0^F(t')] &= \frac{t'}{16\pi \Lambda^4} \left(1-\frac{4m_\psi^2}{t'}\right)^{3/2}\,\Theta\left(t'-4m_\psi^2\right),\label{eq:ImMF}\\
\frac{1}{4m_\chi^2} {\rm Im}[{\cal M}_0^{\widetilde{F}}(t')] &= -\frac{t'}{12\pi \Lambda^4} \left(1+\frac{2m_\psi^2}{t'}\right)\left(1-\frac{4m_\psi^2}{t'}\right)^{1/2}\,\Theta\left(t'-4m_\psi^2\right).\label{eq:ImMFtilde}
\end{align}
Finally, substituting Eqs.~(\ref{eq:ImMS})-(\ref{eq:ImMFtilde}) into Eq.~(\ref{eq:dispersion}) and performing the integral, we arrive at
\begin{align}
V_0^S(r) &=  - \frac{1}{128\pi^3 \Lambda^2 r} \int_{4m_\phi^2}^\infty {\rm d}t'\, e^{-\sqrt{t'}r} \left(1-\frac{4m_\phi^2}{t'}\right)^{1/2}=-\frac{m_\phi}{32\pi^3\Lambda^2 r^2}K_1(2m_\phi r)\;,\\
V_0^F(r) &=  - \frac{1}{64\pi^3 \Lambda^4 r} \int_{4m_\psi^2}^\infty {\rm d}t'\, e^{-\sqrt{t'}r} \,t'\left(1-\frac{4m_\psi^2}{t'}\right)^{3/2}=-\frac{3 m_\psi^2}{8\pi^3 \Lambda^4 r^3}K_2(2m_\psi r)\;,\\
V_0^{\widetilde{F}}(r) &=   \frac{1}{48\pi^3 \Lambda^4 r} \int_{4m_\psi^2}^\infty {\rm d}t'\, e^{-\sqrt{t'}r}\,t' \left(1+\frac{2m_\psi^2}{t'}\right) \left(1-\frac{4m_\psi^2}{t'}\right)^{1/2} \nonumber\\
&= \frac{m_\psi^2}{2\pi^3 \Lambda^4 r^3} \left[(m_\psi r) K_1(2m_\psi r) + K_2(2m_\psi r)\right].
\end{align}
This reproduces the expressions of the vacuum quantum forces we used in the main text.

\subsection*{Background potentials}
Next, we turn to calculating the background potentials $V_{\rm bkg}$. As explained above, the background effect comes from the cross terms of the two propagators in Eq.~(\ref{eq:schematic}).

The cross terms in the amplitude for ${\cal O}_S$ is given by
\begin{align}
\frac{i{\cal M}_{\rm bkg}^S}{4m_\chi^2} &= \frac{1}{2}\left(\frac{i}{\Lambda}\right)^2\int \frac{{\rm d}^4k}{\left(2\pi\right)^4}(2\pi)\left[\frac{i}{k^2-m_\phi^2}\delta\left((k+q)^2-m_\phi^2\right)f_\phi(\veck+\vecq) + \frac{i}{(k+q)^2-m_\phi^2}\delta\left(k^2-m_\phi^2\right)f_\phi(\veck)
\right]\nonumber\\
&=-\frac{i}{2\Lambda^2}\int \frac{{\rm d}^3\veck}{\left(2\pi\right)^3}f_\phi(\veck)\int_{-\infty}^\infty {\rm d}k^0 \delta\left(k^2-m_\phi^2\right)\left[\frac{1}{(k-q)^2-m_\phi^2}+\frac{1}{(k+q)^2-m_\phi^2}\right],
\end{align}
where in the second line, we have performed the momentum shift $k \to k-q$ for the first term in the bracket of the first line. Similarly, for ${\cal O}_F$ and ${\cal O}_{\widetilde{F}}$, we obtain
\begin{align}
    \frac{i{\cal M}_{\rm bkg}^F}{4m_\chi^2} &= -\frac{2i}{\Lambda^4}\int\frac{{\rm d}^3\veck}{\left(2\pi\right)^3}f_\psi(\veck)\int_{-\infty}^\infty {\rm d}k^0 \delta\left(k^2-m_\psi^2\right)\left[\frac{k^2-k\cdot q + m_\psi^2}{(k-q)^2-m_\psi^2}+\frac{k^2 + k\cdot q + m_\psi^2}{(k+q)^2-m_\psi^2}\right],\\
    \frac{i{\cal M}_{\rm bkg}^{\widetilde{F}}}{4m_\chi^2} &= -\frac{4i}{\Lambda^4}\int\frac{{\rm d}^3\veck}{\left(2\pi\right)^3}f_\psi(\veck)\int_{-\infty}^\infty {\rm d}k^0 \delta\left(k^2-m_\psi^2\right)\left[\frac{2(k^0)^2+k\cdot q}{(k-q)^2-m_\psi^2}+\frac{2(k^0)^2-k\cdot q}{(k+q)^2-m_\psi^2}\right].
\end{align}
Compared to the scalar case, the numerator in the bracket for the fermionic case is nontrivial, which comes from the trace of the fermion loop. Note that the minus sign from the fermion loop is canceled by the minus sign in front of the $\delta$-function in Eq.~(\ref{eq:DpsiApp}).

Since the phase-space distribution function depends only on $\veck$, we can first integrate out $k^0$. Recalling $q^\mu \approx(0,\vecq)$ and using
\begin{align}
  \delta\left(k^2-m^2\right) = \frac{1}{2E_\veck}\left[\delta\left(k^0-E_\veck\right)+\delta\left(k^0+E_\veck\right)\right],  \quad \text{with $E_\veck\equiv \sqrt{\veck^2+m^2}$}\;,
\end{align}
we arrive at
\begin{align}
\frac{{\cal M}_{\rm bkg}^S}{4m_\chi^2} &= \frac{1}{\Lambda^2}\int\frac{{\rm d}^3\veck}{\left(2\pi\right)^3}\frac{f_\phi(\veck)}{2E_\veck}\left[\frac{1}{\vecq^2-2\veck\cdot\vecq}+\frac{1}{\vecq^2+2\veck\cdot\vecq}\right],\label{eq:MSbkg}\\
\frac{{\cal M}_{\rm bkg}^{F}}{4m_\chi^2} &= \frac{4}{\Lambda^4}\int\frac{{\rm d}^3\veck}{\left(2\pi\right)^3}\frac{f_\psi(\veck)}{2E_\veck}\left[\frac{2m_\psi^2+\veck\cdot \vecq}{\vecq^2-2\veck\cdot\vecq}+\frac{2m_\psi^2-\veck\cdot\vecq}{\vecq^2+2\veck\cdot\vecq}\right],\label{eq:MFbkg}\\
\frac{{\cal M}_{\rm bkg}^{\widetilde{F}}}{4m_\chi^2} &= \frac{8}{\Lambda^4}\int\frac{{\rm d}^3\veck}{\left(2\pi\right)^3}\frac{f_\psi(\veck)}{2E_\veck}\left[\frac{2E_\veck^2-\veck\cdot \vecq}{\vecq^2-2\veck\cdot\vecq}+\frac{2E_\veck^2+\veck\cdot\vecq}{\vecq^2+2\veck\cdot\vecq}\right].\label{eq:MFtildebkg}
\end{align}
Compared to the vacuum amplitude, the background amplitude does not contain the mass of the mediator particles in the denominator, indicating that the propagator in the background is \emph{nearly on-shell}. As a result, after the Fourier transform, there is no exponential suppression for the background potential when $r$ exceeds the inverse of the mediator mass.

According to the Born approximation, the background potential is given by
\begin{align}
V_{\rm bkg}^j(\vecr) = -\frac{1}{4m_\chi^2}\int \frac{{\rm d}^3\vecq}{\left(2\pi\right)^3}e^{i\vecq\cdot\vecr} {\cal M}_{\rm bkg}^j  \;, \qquad \text{for $j=S,F,\widetilde{F}$}\;.\label{eq:Bornbkg}
\end{align}
To make the final results as general as possible, one can first integrate $\vecq$, leaving $f_\phi(\veck)$ and $f_\psi(\veck)$ intact. To this end, we need to compute the Fourier transform of $1/(\vecq^2-2\veck\cdot\vecq)+1/(\vecq^2+2\veck\cdot\vecq)$. We shift the variable $\vecq \to \vecq+\veck$ in the first term and $\vecq\to \vecq-\veck$ in the second term~\cite{VanTilburg:2024xib}. Using this trick, we can eliminate the $\veck\cdot\vecq$ term and obtain
\begin{align}
&\int\frac{{\rm d}^3\vecq}{\left(2\pi\right)^3}e^{i\vecq\cdot\vecr}  \left[\frac{1}{\vecq^2-2\veck\cdot\vecq}+\frac{1}{\vecq^2+2\veck\cdot\vecq}\right]\nonumber\\
&=\left(e^{i\veck\cdot r}+e^{-i\veck\cdot r}\right) \int\frac{{\rm d}^3\vecq}{\left(2\pi\right)^3}e^{i\vecq\cdot\vecr}\frac{1}{\vecq^2-\veck^2}\nonumber\\
&=2\cos\left(\veck\cdot\vecr\right)\frac{1}{(2\pi)^2 ir} \int_{-\infty}^\infty {\rm d}|\vecq| \frac{|\vecq|}{\vecq^2-\veck^2}\,e^{i|\vecq|r}\nonumber\\
&=\cos\left(\veck\cdot\vecr\right)\frac{1}{(2\pi)^2 ir}\int_{-\infty}^\infty {\rm d}|\vecq|\left( \frac{1}{|\vecq|-|\veck|} + \frac{1}{|\vecq|+|\veck|}\right)\,e^{i|\vecq|r}\nonumber\\
&=\cos\left(\veck\cdot\vecr\right)\frac{1}{(2\pi)^2 ir}\left(e^{i|\veck|r}+e^{-i|\veck|r}\right)\int_{-\infty}^\infty {\rm d}|\vecq| \frac{e^{i|\vecq|r}}{|\vecq|}\nonumber\\
&=2\cos\left(\veck\cdot\vecr\right)\cos\left(|\veck|r\right)\frac{1}{(2\pi)^2 ir} \left[\int_{-\infty}^\infty {\rm d}|\vecq| \frac{\cos\left(|\vecq|r\right)}{|\vecq|} + i\int_{-\infty}^\infty {\rm d}|\vecq| \frac{\sin\left(|\vecq|r\right)}{|\vecq|}
\right]\nonumber\\
&=\frac{1}{2\pi r}\cos\left(\veck\cdot\vecr\right)\cos\left(|\veck|r\right),\label{eq:Fourierbkg}
\end{align}
where in the last step we used
\begin{align}
 \int_{-\infty}^\infty {\rm d}|\vecq| \frac{\cos\left(|\vecq|r\right)}{|\vecq|} = 0\;,\qquad
 \int_{-\infty}^\infty {\rm d}|\vecq| \frac{\sin\left(|\vecq|r\right)}{|\vecq|} = \pi\;.\label{eq:Principalvalue}
\end{align}
Note that the first equality in Eq.~(\ref{eq:Principalvalue}) holds in the sense of the principal value.
Therefore, the imaginary part of the Fourier transform automatically vanishes and the final result of the potential is real, as it should be. 

Combining the result of Eq.~(\ref{eq:Fourierbkg}) with Eqs.~(\ref{eq:MSbkg}) and (\ref{eq:Bornbkg}) we arrive at
\begin{align}
V^S_{\rm bkg}(\vecr) &= -\frac{1}{2\pi r \Lambda^2}\int \frac{{\rm d}^3\veck}{\left(2\pi\right)^3}\frac{f_\phi(\veck)}{2E_\veck} \cos\left(\veck\cdot\vecr\right)\cos\left(|\veck|r\right)\nonumber\\
&=-\frac{1}{4\pi r \Lambda^2}\int \frac{{\rm d}^3\veck}{\left(2\pi\right)^3}\frac{f_\phi(\veck)}{2E_\veck}\left[\cos\left(|\veck|r-\veck\cdot\vecr\right)+\cos\left(|\veck|r+\veck\cdot\vecr\right)\right].
\end{align}
This reproduces Eq.~(\ref{eq:Vbkg-general}), which also agrees with the result derived in \cite{Grossman:2025cov} that used the Feynman propagator. In the isotropic limit, $f_\phi(\veck)=f_\phi(\kappa)$ with $\kappa\equiv |\veck|$, one can further integrate the angular part of the momentum and get
\begin{align}
V^S_{\rm bkg}(r) &= -\frac{1}{4\pi r\Lambda^2}\frac{1}{(2\pi)^2}\int_0^\infty {\rm d}\kappa \frac{\kappa^2 f_\phi(\kappa)}{\sqrt{\kappa^2+m_\phi^2}}\int_{-1}^1 {\rm d}z \cos\left(\kappa r z\right) \cos\left(\kappa r\right)\nonumber\\
& = -\frac{1}{16\pi^3 r^2 \Lambda^2} \int_0^\infty {\rm d}\kappa \frac{\kappa f_\phi(\kappa)}{\sqrt{\kappa^2+m_\phi^2}}\sin\left(2\kappa r\right),
\end{align}
which agrees with Eq.~(\ref{eq:VSbkg}).

For the fermionic case, the numerator in Eqs.~(\ref{eq:MFbkg}) and (\ref{eq:MFtildebkg}) also contributes to the Fourier transform, which can be calculated in the following way:
\begin{align}
&\int\frac{{\rm d}^3\vecq}{\left(2\pi\right)^3}e^{i\vecq\cdot\vecr}  \left[\frac{\veck \cdot \vecq}{\vecq^2-2\veck\cdot\vecq}-\frac{\veck \cdot\vecq}{\vecq^2+2\veck\cdot\vecq}\right]\nonumber\\
&= -i\left(\veck \cdot \nabla\right)\int\frac{{\rm d}^3\vecq}{\left(2\pi\right)^3}e^{i\vecq\cdot\vecr}  \left[\frac{1}{\vecq^2-2\veck\cdot\vecq}-\frac{1}{\vecq^2+2\veck\cdot\vecq}\right]\\
&= -i\left(\veck \cdot \nabla\right)\left(e^{i\veck\cdot \vecr}-e^{-i\veck\cdot \vecr}\right)\int\frac{{\rm d}^3\vecq}{\left(2\pi\right)^3}e^{i\vecq\cdot\vecr}\frac{1}{\vecq^2-\veck^2}\nonumber\\
&=\frac{1}{2\pi}\left(\veck \cdot \nabla\right)\frac{1}{r}\sin\left(\veck\cdot\vecr\right)\cos\left(|\veck|r\right)\nonumber\\
&=\frac{1}{2\pi r}\left[\veck^2\cos\left(\veck\cdot\vecr\right)\cos\left(|\veck|r\right) - \frac{|\veck|}{r}\left(\veck\cdot\vecr\right)\sin\left(\veck\cdot\vecr\right)\sin\left(|\veck|r\right) - \frac{\veck\cdot\vecr}{r^2}\sin\left(\veck\cdot\vecr\right)\cos\left(|\veck|r\right)
\right].
\end{align}
Using the above results, we obtain the background potentials for ${\cal O}_F$ and ${\cal O}_{\widetilde{F}}$:
\begin{align}
V_{\rm bkg}^F(\vecr) &= -\frac{2}{\pi r \Lambda^4}\int \frac{{\rm d}^3\veck}{\left(2\pi\right)^3}\frac{f_\psi(\veck)}{2E_\veck}\Bigg[\left(2m_\psi^2+\veck^2\right)\cos\left(\veck\cdot\vecr\right)\cos\left(|\veck|r\right)\nonumber\\
&\qquad\qquad\qquad\qquad\qquad\qquad -\frac{1}{r^2}\left(\veck\cdot\vecr\right)\sin\left(\veck\cdot\vecr\right)\Big(|\veck| r\sin\left(|\veck|r\right) + \cos\left(|\veck|r\right)
\Big)
\Bigg]\;,\\
V_{\rm bkg}^{\widetilde{F}}(\vecr) &=-\frac{4}{\pi r \Lambda^4}\int \frac{{\rm d}^3\veck}{\left(2\pi\right)^3}\frac{f_\psi(\veck)}{2E_\veck}\Bigg[\left(2m_\psi^2+\veck^2\right)\cos\left(\veck\cdot\vecr\right)\cos\left(|\veck|r\right)\nonumber\\
&\qquad\qquad\qquad\qquad\qquad\qquad +\frac{1}{r^2}\left(\veck\cdot\vecr\right)\sin\left(\veck\cdot\vecr\right)\Big(|\veck|r\sin\left(|\veck|r\right) + \cos\left(|\veck|r\right)\Big)
\Bigg]\;.
\end{align}
In the isotropic limit, $f_\psi(\veck)=f_\psi(\kappa)$, they reduce to Eqs.~(\ref{eq:VFbkg}) and (\ref{eq:VFbkgtilde}), respectively.

\section{Calculation of the Sommerfeld factor  with a general potential}\label{app:S-calculation}

In this appendix, we describe how to calculate the Sommerfeld factor $S$ corresponding to a given central potential $V(r)$. We closely follow the method proposed in \cite{Iengo:2009ni}.

For a central potential, the radial part of the wavefunction $R_\ell(r)$ satisfies (defining $u_{\ell} \equiv rR_{\ell}(r)$, where $\ell$ is the angular momentum): 
\begin{align}
    u_{\ell}''(x) +
    \bigg[1
    - \mathcal{V}(x) 
    - \frac{\ell (\ell +1)}{x^{2}}
    \bigg]u_{\ell}(x)
    = 0\;,\label{eq:Schrodinger-normalized}
\end{align}
where primes denote derivatives with respect to $x\equiv rp$, $M = m_\chi/2$ is the reduced mass, $p=Mv$, and $\mathcal{V}(x) \equiv \frac{2M}{p^{2}}V(x/p)$. 

The differential equation (\ref{eq:Schrodinger-normalized}) should be solved together with two physical boundary conditions: (i) $u_\ell \sim x^{\ell+1}$ as $x\to 0$, and (ii) $u_\ell$ tends to the plane wave as $x\to \infty$, where the amplitude of the plane wave should match that of the free theory (that is, the solution without the potential). Practically, it is more convenient to translate the second boundary condition as the initial velocity. Since Eq.~(\ref{eq:Schrodinger-normalized}) is linear with respect to $u_\ell$, the solution multiplied by any constant is still a solution, so one can always rescale $u_\ell$ to satisfy the following boundary conditions:
\begin{align}
    \lim_{x\rightarrow 0} u_{\ell}(x)
    &= 
    x^{\ell+1}\;, 
    \label{0bc}
    \\
    \lim_{x\rightarrow 0} u'_{\ell}(x)
    &= \left(\ell+1\right)x^{\ell}\;.\label{eq:vbc}
\end{align}
Once $u_\ell$ is solved using Eqs.~(\ref{eq:Schrodinger-normalized})-(\ref{eq:vbc}), its value at $x\to \infty$ is also determined:
\begin{align}
    \lim_{x\rightarrow\infty}u_{\ell}(x)
    = C_\ell \sin\left(x- \ell \pi/2  +\delta_\ell\right),
    \label{eq:infbc}
\end{align}
where $\delta_\ell$ is the phase shift and $C_\ell$ is the amplitude. The value of $C_\ell$ can be computed by the modulus of $u_{\ell}$ at the asymptotically far region without knowing the phase shift:
\begin{align}
C_\ell^2 = \lim_{x\to \infty} \left[u_\ell^2(x)+u_\ell^2(x-\pi/2)\right]. \label{eq:Clapp} 
\end{align}
The above procedure is convenient, as it provides a simple way to (numerically) compute $C_\ell$. Moreover, as we show below, the Sommerfeld factor for the $\ell$-th partial wave, $S_\ell$, is directly determined by the amplitude $C_\ell$, and does not depend on the phase shift $\delta_\ell$. 

To derive the Sommerfeld factor, 
we start with the fact that the short- and long-distance contributions to the amplitude of a process can be factorized in the non-relativistic limit~\cite{Bodwin:1994jh}. For a given process, let $A_0$ be the bare amplitude in the perturbative limit (i.e., without the long-range potential), and $A$ be the full amplitude including the long-range potential. Using explicit field theory calculations, Ref.~\cite{Iengo:2009ni} shows that $A$ and $A_0$ are connected by
\begin{align}
    A(\vecp) = \int 
    {\rm d}^3\vecr \, \psi^{*}_{\vecp}(\vecr)
    \int \frac{{\rm d}^3\vecq}{(2\pi)^{3}}\,e^{i\vecq \cdot \vecr} 
    A_0(\vecq)\;,
    \label{eq:factorizedamplitude}
\end{align}
where $\psi_{\vecp}(\vecr)$ solves the Schr\"{o}dinger equation with the long-range potential $V(r)$: 
\begin{align}
    \left[-\frac{\nabla^2}{2M} +V(r)
    \right] \psi_\vecp(\vecr) = \frac{p^2}{2M}\psi_\vecp(\vecr)\;.
\end{align}

The result in Eq.~(\ref{eq:factorizedamplitude}) tells us that the full amplitude is given by the convolution of the bare amplitude (which only contains short-range information) with the wave function that contains long-range information. It agrees with the general result that the short- and long-range contributions are factorizable in the non-relativistic limit~\cite{Bodwin:1994jh}. In \cite{Iengo:2009ni}, Eq.~(\ref{eq:factorizedamplitude}) was derived by taking the non-relativistic limit of the ladder diagram; equivalently, it can also be derived by starting from the general Bethe-Salpeter equation and then taking the non-relativistic approximation, as shown in \cite{Cassel:2009wt}. 

Since the short-range annihilation information is contained in $A_0$, it is reasonable to define the Sommerfeld factor as the ratio between $A$ and $A_0$ to factor out the long-range contribution:
\begin{align}
 S_\ell \equiv \left|\frac{A_\ell(\vecp)}{A_{0,\ell}(\vecp)}\right|^2\;,
 \label{eq:Sommerfelddef}  
\end{align}
where $A_\ell$ and $A_{0,\ell}$ denote the $\ell$-th component of the amplitude under the partial-wave decomposition (to be specified below).
The Sommerfeld factor defined in Eq.~(\ref{eq:Sommerfelddef}) is not sensitive to short-range annihilation and is completely determined once given a long-range potential $V(r)$. 

To see it, we recall the partial-wave decomposition of the wavefunction: 
\begin{align}
    \psi_{\vecp}(\vecr) = 
    \frac{(2\pi)^{3/2}}{4\pi p} \sum_{\ell=0}^\infty i^\ell
    \left(2\ell + 1\right)
    e^{i\delta_{\ell}}
    R_{p\ell}(r) 
    P_{\ell}(\vecph\cdot\vecrh)\;,
    \label{eq:partialwave}
\end{align}
where the normalization factor is chosen such that $\psi_\vecp \sim e^{i\vecp\cdot\vecr} + Ae^{ipr}/r$ as $r\to \infty$, and the radial wavefunction is normalized to satisfy the completeness relation
\begin{align}
    \int_{0}^{\infty} {\rm d}p 
    R_{p\ell}(r) R_{p\ell}(r') 
    = \frac{\delta(r-r')}{r^{2}}\;.\label{eq:Rcompletness}
\end{align}
With this convention, the free radial wavefunction is given by $R^{0}_{p\ell}(r)= \sqrt{2/\pi}\,p j_{\ell}(pr)$, where $j_\ell$ is the spherical Bessel function. 

We define the $\ell$-th component of the amplitude as 
\begin{align}
A(\vecp)\equiv \sum_{\ell=0}^\infty A_\ell(\vecp)\;,\qquad A_{0}(\vecp)\equiv \sum_{\ell=0}^\infty i^\ell e^{i\delta_\ell} A_{0,\ell}(\vecp)\;.    
\end{align}
Note that the phase defined in front of $A_{0,\ell}$ is only for convenience, which does not affect the Sommerfeld factor after taking the modulus squared. 

For the vector-valued argument in the amplitude, it is convenient to project onto a certain direction using the Legendre polynomial. For example, we can write $A_\ell(\vecp) = A_\ell(p) P_\ell(\vecph\cdot\vecph')$, $A_{0,\ell}(\vecq) = A_{0,\ell}(q) P_\ell(\vecqh\cdot\vecph')$, where the hat denotes the unit vector, and $p\equiv |\vecp|$ and $q\equiv |\vecq|$ have been defined.

Substituting Eq.~(\ref{eq:partialwave}) into Eq.~(\ref{eq:factorizedamplitude}) and taking the $\ell$-th component, we arrive at 
\begin{align}
    A_\ell(p)P_\ell(\vecph\cdot \vecph') &= 
 \frac{ \left(2\pi\right)^{-3/2}}{p}\frac{\left(2\ell+1\right)}{4\pi}(-i)^{\ell}e^{-i\delta_{\ell}} \int {\rm d}^3\vecr\,R_{p\ell}(r) P_{\ell}(\vecph\cdot\vecrh)\nonumber\\
 &\qquad\qquad\qquad\qquad\qquad\quad\quad
 \times\int {\rm d}^3\vecq\,e^{i\vecq\cdot\vecr} \sum_{\ell'=0}^\infty i^{\ell'}e^{i\delta_{\ell'}}A_{0,\ell'}(q) P_{\ell'}(\vecqh\cdot\vecph')\;.
  \label{eq:decomposition}  
\end{align}
Note that in the above equation, the index $\ell$ is not summed.
To further simplify it, we use the following identity for the Legendre polynomial:
\begin{align}
    \int {\rm d}\Omega_{\vecph'}\, 
    P_{\ell}(\vecph\cdot\vecph')
    P_{\ell'}(\vecqh\cdot\vecph')
    = 
    \delta_{\ell\ell'}\frac{4\pi}{2\ell + 1}
    P_{\ell}(\vecph\cdot\vecqh)\;,\label{eq:Legendre-identity}
\end{align}
where $\Omega_{\vecph'}$ is the solid angle along the direction of $\vecph'$. Multiplying $P_\ell(\vecph\cdot\vecph')$ on both sides of Eq.~(\ref{eq:decomposition}) and integrating over ${\rm d}\Omega_{\vecph'}$, and repeatly using Eq.~(\ref{eq:Legendre-identity}), one obtains
\begin{align}
A_\ell(p) &= \frac{ \left(2\pi\right)^{-3/2}}{p}\frac{\left(2\ell+1\right)}{4\pi}\int {\rm d}^3\vecr\,R_{p\ell}(r) P_\ell(\vecph\cdot\vecrh)P_\ell(\vecph\cdot\vecqh)\int{\rm d}^3\vecq\,e^{i\vecq\cdot\vecr} A_{0,\ell}(q)\nonumber\\
&= \frac{\left(2\pi\right)^{-3/2}}{p} \int_0^\infty {\rm d}r\, r^2 R_{p\ell}(r) \int{\rm d}^3\vecq\,e^{i\vecq\cdot\vecr} A_{0,\ell}(q)P_\ell(\vecqh\cdot\vecrh)\nonumber\\
&= \frac{\left(2\pi\right)^{-1/2}}{p}\int_0^\infty {\rm d}r\, r^2 R_{p\ell}(r)  \int_0^\infty{\rm d}q\,q^2A_{0,\ell}(q) \int_{-1}^1 {\rm d\xi}\,e^{iqr\xi} P_\ell(\xi)\nonumber\\
&=-\sqrt{\frac{2}{\pi}}\frac{1}{p}\int_0^\infty {\rm d}r\, r^2 R_{p\ell}(r)  \int_0^\infty{\rm d}q\,q^2A_{0,\ell}(q) j_\ell(qr)\;, \label{eq:decomposition2} 
\end{align}
where $\xi\equiv \vecqh\cdot\vecrh$ and in the last step we used the result: $\int_{-1}^1 {\rm d}\xi\,e^{iz\xi}P_\ell(\xi) = -2 j_\ell(z)$.
Recalling the expression of the free radial wavefunction $R^{0}_{p\ell}(r)= \sqrt{2/\pi}\,p j_{\ell}(pr)$ and the fact that $A_{0,\ell}(q) = a_{0,\ell}\,q^{\ell}$ (with $a_{0,\ell}$ being some momentum-independent factor) in the non-relativistic limit, we obtain
\begin{align}
    A_\ell(p)&=-\frac{1}{p}\int_0^\infty {\rm d}r\, r^2 R_{p\ell}(r)  \int_0^\infty{\rm d}q\,q\,A_{0,\ell}(q)R_{q\ell}^0(r)\nonumber\\
    &=-\frac{a_{0,\ell}}{p}\int_0^\infty {\rm d}r\, r^2 R_{p\ell}(r)  \int_0^\infty{\rm d}q\,q^{\ell+1}\,R_{q\ell}^0(r)\;. \label{eq:decomposition3} 
\end{align}

On the other hand, by evaluating the $\ell$-th derivative of the spherical Bessel function, we obtain
\begin{align}
    q^{\ell+1} = 
    \sqrt{\frac{\pi}{2}} 
    \frac{(2\ell+1)!!}{\ell!}
    \partial_{r}^{\ell}R^{0}_{q\ell}(r)\big|_{r=0}\;.\label{eq:sphericalbesselderivative}
\end{align}
This identity can be used to replace $q^{\ell+1}$ in the last line of Eq.~(\ref{eq:decomposition3}) with the free radial wavefunction. Also, there is
\begin{align}
    \int_{0}^{\infty} {\rm d}q 
    R_{q\ell}^0(r) 
    \partial_{r'}^{\ell}
    R_{q\ell}^0(r')  
    \big|_{r'=0}
    &= 
    \frac{1}{r^{2}}
    \partial_{r'}^{\ell}\delta(r-r') \big|_{r'=0}
   = 
    \frac{(-1)^{\ell}}{r^{2}}
    \partial_{r}^{\ell}\delta(r)\;.\label{eq:deltaderivative}
\end{align}
Substituting Eq.~(\ref{eq:sphericalbesselderivative}) into Eq.~(\ref{eq:decomposition3}) and using Eq.~(\ref{eq:deltaderivative}), we finally obtain
\begin{align}
    A_{\ell}(p) 
    &= -\sqrt{\frac{\pi}{2}}
    \frac{(2\ell + 1)!!}{\ell !}
    \frac{a_{0,\ell}}{p}
    \int_{0}^{\infty}r^{2} {\rm d}r \, R_{p\ell} (r) 
    \int_{0}^{\infty} {\rm d}q \, R_{q\ell}^{0}(r)
    \partial_{r'}^{\ell}R_{q\ell}^{0}(r')\big|_{r'=0}
    \nonumber\\
    &= -\sqrt{\frac{\pi}{2}}
    (-1)^{\ell} 
    \frac{(2\ell + 1)!!}{\ell !}
    \frac{a_{0,\ell}}{p}
    \int_{0}^{\infty} {\rm d}r \, R_{p\ell} (r) \partial^\ell_r 
    \delta(r) 
    \nonumber\\
    &= -\sqrt{\frac{\pi}{2}}
    \frac{(2\ell + 1)!!}{\ell !}
    \frac{a_{0,\ell}}{p}
    \partial_{r}^{\ell} R_{p\ell} (r)\big|_{r=0} \;,
\end{align}
where in the last step we used integration by parts.

Recalling the free amplitude $A_{0,\ell}(p)=a_{0,\ell}\,p^{\ell}$, the Sommerfeld factor for the $\ell$-th partial wave is then given by
\begin{align}
    S_{\ell} 
    &\equiv 
    \left|A_{\ell}(p) / A_{0,\ell}(p)\right|^{2}
    =  \bigg
    |\sqrt{\frac{\pi}{2}}\frac{(2\ell+1)!!}{\ell!} 
    \frac{1}{p^{\ell+1}}
    \partial_{r}^{\ell}
    R_{p\ell}(r)\big|_{r=0}
    \bigg|^{2}\;.\label{eq:Sommerfeld-l}
\end{align}
This implies that the Sommerfeld factor $S_\ell$ is completely determined by the $\ell$-th derivative of the full radial wavefunction near the origin. This general result was derived in both \cite{Iengo:2009ni} and \cite{Cassel:2009wt} at almost the same time.

Given the normalization convention of $R_{p \ell}$ in Eq.~(\ref{eq:Rcompletness}), its asymptotic behavior scales as $\lim_{r\to \infty} R_{p\ell}(r) = \sqrt{2/\pi}\,\sin(pr-\ell \pi/2+\delta_\ell)/r$. Comparing it with Eq.~(\ref{eq:infbc}), we obtain the exact connection between $R_{p\ell}$ and $u_\ell$:
\begin{align}
R_{p\ell}(r) = \sqrt{\frac{2}{\pi}}\frac{1}{C_\ell}\frac{u_\ell(r)}{r}\;.    
\end{align}
Substituting it into Eq.~(\ref{eq:Sommerfeld-l}) and using the boundary condition $\lim_{r\to 0} u_\ell(r) = (pr)^{\ell+1}$, one obtains
\begin{align}
 S_\ell   &=  \bigg
    |\frac{(2\ell+1)!!}{\ell!} 
    \frac{1}{p^{\ell+1}}
    \partial_{r}^{\ell}
    \left(
    \frac{1}{r C_\ell}u_{\ell}(r)
    \right)\Big|_{r=0}
    \bigg|^{2}
    \nonumber\\
    &= \bigg
    |\frac{(2\ell+1)!!}{\ell!} 
    \frac{1}{p^{\ell+1}}
    \partial_{r}^{\ell}
    \left(\frac{r^{\ell}p^{\ell+1}}{C_\ell}\right)
    \bigg|^{2}
    \nonumber\\
    &=
    \left|\frac{(2\ell+1)!!}{C_\ell}
    \right|^{2},
\end{align}
where $C_\ell$ is determined by Eq.~(\ref{eq:Clapp}). Note that the dependence of $p$ is canceled in the final result. This reproduces Eq.~(\ref{eq:Sl}) in the main text that we used to calculate the Sommerfeld factor.

This identity helps to avoid one from needing to numerically impose a boundary condition at infinity, and is typically the most convenient way to compute $S_{\ell}$. Throughout this paper we only discuss the $s$-wave contributions, so we define $S\equiv S_{0}$.

\section{Unitarity bound on Sommerfeld factor from quantum forces}
\label{app:unitarity}
We have seen that in the case of attractive quantum forces, the Sommerfeld factor is resonantly enhanced when the DM mass takes some discrete values, which is similar to the case of the Yukawa potential, and there is
$S(v) \sim 1/v^2$ near the resonance peaks (see e.g., Eq.~(\ref{eq:Smax})). 

However, it is known that a Sommerfeld factor scaling as $S(v)\sim 1/v^2$ will violate the partial-wave unitarity bound~\cite{Hisano:2002fk,Hisano:2003ec,Hisano:2004ds}. For example, the unitarity bound for the $s$-wave cross section reads $\sigma_0 \leq \pi/(M^2v)$, where $M =m_\chi/2$ is the reduced DM mass and $v$ is the relative velocity between two DM particles. The Sommerfeld-enhanced cross section around the peaks scales as $S\sigma_0\sim \sigma_0/v^2$, which violates the unitarity bound for sufficiently small $v$. Therefore, for the consistency of the theory, the Sommerfeld factor needs to be regularized at small velocities.

In \cite{Hisano:2004ds,Lattanzi:2008qa,Arkani-Hamed:2008hhe}, it was argued that for the Yukawa potential with the mediator of mass $m_\phi$, when the de Broglie wavelength of DM particles $1/(Mv)$ exceeds the range of the force $1/m_\phi$, or when the velocity drops below $m_\phi/M$, the Sommerfeld factor is saturated at $S_{\rm max}\sim M/m_\phi$. In other words, the finite range of the force (compared to Coulomb) prevents $S$ from continuing to increase at arbitrarily small velocities. A more systematic and quantitative analysis on the unitarity bound relevant to $S$ was performed in \cite{Blum:2016nrz} for $s$-wave annihilation (see also \cite{Braaten:2017gpq,Braaten:2017kci,Braaten:2017dwq}) and was generalized to higher partial waves very recently in \cite{Parikh:2024mwa}. In this work, we restrict ourselves to $s$-wave DM annihilation, so we follow the idea of \cite{Blum:2016nrz} to investigate the unitarity constraint on the Sommerfeld factor from quantum forces.

To maintain possibility conservation, \cite{Blum:2016nrz} introduced a $\delta$-function with a complex coefficient into the Schr\"{o}dinger equation to account for particle annihilation that occurs at short distances. The DM wavefunctions are modified by this counter term. As a result, the leading Sommerfeld factor is regularized, and the unitarity is manifestly preserved. A strict calculation of the regularized Sommerfeld factor both near and away from the resonance peaks is possible if one can solve the Schr\"{o}dinger equation strictly with full potential~\cite{Blum:2016nrz}. For example, the analytical expression of regularized $S$ was calculated in \cite{Blum:2016nrz} with the Hulthen potential and the well potential. In our case, it is not possible to get the analytical solution for the Schr\"{o}dinger equation involving the quantum force. However, near the resonance peaks, the regularized Sommerfeld factor can be simply obtained by shifting the velocity from $S(v)$ to $S(v+v_c)$, where $v_c$ is the regularized velocity given by~\cite{Blum:2016nrz}
\begin{align}
v_c = \frac{U_0 M \sigma_0}{2\pi}\;,\label{eq:vc}    
\end{align}
where $U_0$ is the typical scale of the potential energy. (For example, in the case of the Yukawa potential $V_Y=-\alpha e^{-m_\phi r}/r$, it is given by $U_0 \approx \alpha m_\phi$.) For $v\ll v_c$, the Sommerfeld factor is then regularized by $S(v_c)$. In the following, we calculate the regularized velocity $v_c$ for the quantum forces relevant to this work. This is crucial to get a self-consistent Sommerfeld enhancement near the resonance peaks. 

\subsubsection*{Vacuum potentials}

For attractive vacuum potentials $V_0^S$ and $V_0^F$ with the regularization procedure in Eq.~(\ref{eq:Vreg}), the potentials at short distances are approximated by a finite well. This approximation makes good analytical predictions on the locations of peaks, as shown in Sec.~\ref{subsec:box}. Therefore, we identify the typical scale of the potential energy $U_0$ to be the height of the well: $U_0^i = |V_{0}^i(r=1/\Lambda)|$, where $i=S,F$. Combining Eqs.~(\ref{eq:VS0}), (\ref{eq:VF0}) and (\ref{eq:vc}), we obtain
\begin{align}
v_c^S &= \frac{M \sigma_0 m_\phi}{64\pi^4}K_1(2m_\phi/\Lambda) \;\;\,\overset{m_\phi \ll \Lambda}{\approx} \frac{M\Lambda \sigma_0}{128\pi^4}\;,\label{eq:vcS}\\
v_c^F & = \frac{3 M \sigma_0 m_\psi^2}{16\pi^4 \Lambda}K_2(2m_\psi/\Lambda)\overset{m_\psi \ll \Lambda}{\approx} \frac{3M \Lambda \sigma_0}{32\pi^4} \;.\label{eq:vcF}
\end{align}
Numerically, we obtain (assuming $m_{\phi,\psi}\ll \Lambda$)
\begin{align}
v_c^S &\approx 7.6\times 10^{-6}\left(\frac{M}{5~{\rm TeV}}\right) \left(\frac{\Lambda}{10~{\rm TeV}}\right)\left(\frac{\sigma_0}{2.2\times 10^{-26}{\rm cm^3/s}}\right),\\
v_c^F &\approx 9.1\times 10^{-5}\left(\frac{M}{5~{\rm TeV}}\right) \left(\frac{\Lambda}{10~{\rm TeV}}\right)\left(\frac{\sigma_0}{2.2\times 10^{-26}{\rm cm^3/s}}\right),
\end{align}
where we have normalized the cross section to the typical value of the WIMP scenario.

\subsubsection*{Background potentials}
We consider pure attractive background potentials $V_{\rm bkg}^S$ and $V_{\rm bkg}^{\widetilde{F}}$. 
Qualitatively, the background potentials scale as $1/r$ at $r\lesssim1/T$ and decrease faster than $1/r$ at $r\gtrsim1/T$, similar to the case of the Yukawa potential. According to the analysis in Sec.~\ref{subsec:boundstate}, the locations of the resonance peaks are roughly determined by the deviation of the potential from the Coulomb potential around the transition point. Therefore, we identify $U_0$ the value of the potential around $r=1/T$: $U_0^i = |V_{\rm bkg}^i(r=1/T)|$, where $i=S,\widetilde{F}$.

For illustration, we assume the Maxwell-Boltzmann distribution for the background particles and the mediator mass to be much smaller than the temperature. Combining Eqs.~(\ref{eq:VSMB}), (\ref{eq:FermionTildeMBPotential}) and (\ref{eq:vc}), we obtain
\begin{align}
v_{c,{\rm bkg}}^{S} &= \frac{M T^3 
 \sigma_0}{80\pi^4 \Lambda^2}\;,\label{eq:vcbkgS}\\
v_{c,{\rm bkg}}^{\widetilde{F}} &= \frac{4  M T^5 \sigma_0}{25\pi^4 \Lambda^4}\;.\label{eq:vcbkgFtilde}
\end{align}

For the validity of the effective theory, both $T \lesssim\Lambda$ and $Mv\lesssim\Lambda$ should be satisfied. If $T$ is not much below $\Lambda$, then the magnitude of Eqs.~(\ref{eq:vcbkgS})-(\ref{eq:vcbkgFtilde}) is comparable to Eqs.~(\ref{eq:vcS})-(\ref{eq:vcF}). However, if $\phi$ (or $\psi$) and $\chi$ are in the same thermal bath and if $\chi$ is non-relativistic, then there is $T\lesssim M$. In this case, the regularized velocity from the background potential is much smaller than that from the vacuum potential. For example, as a bench mark numerical value during thermal freeze-out, we obtain
\begin{align}
v_{c,{\rm bkg}}^{S} & \approx 1.5 \times 10^{-9}   \left(\frac{M}{5~{\rm TeV}}\right) \left(\frac{T}{500~{\rm GeV}}\right)^3\left(\frac{\Lambda}{10~{\rm TeV}}\right)^{-2}\left(\frac{\sigma_0}{2.2\times 10^{-26}{\rm cm^3/s}}\right),\\
v_{c,{\rm bkg}}^{\widetilde{F}} & \approx 4.8 \times 10^{-11}   \left(\frac{M}{5~{\rm TeV}}\right) \left(\frac{T}{500~{\rm GeV}}\right)^5\left(\frac{\Lambda}{10~{\rm TeV}}\right)^{-4}\left(\frac{\sigma_0}{2.2\times 10^{-26}{\rm cm^3/s}}\right).
\end{align}

\section{Regulator dependence of Sommerfeld effect from singular potentials}
\label{app:regular-dependence}

\begin{figure}[t]
\centering
\includegraphics[width=6in]
{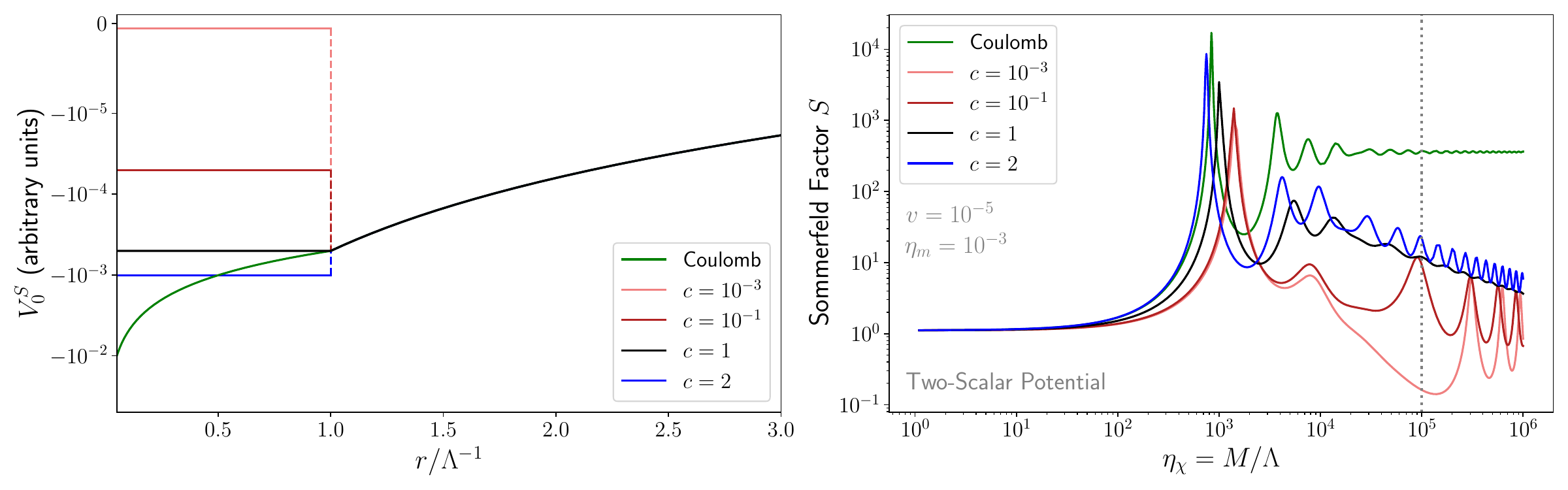}
\caption{\label{fig:scalarregdep} 
The two-scalar potential $V_0^S$ with different regulators (left panel) and the resulting Sommerfeld enhancement (right panel). See text for more information.
}
\end{figure}

\begin{figure}[t]
\centering
\includegraphics[width=6in]
{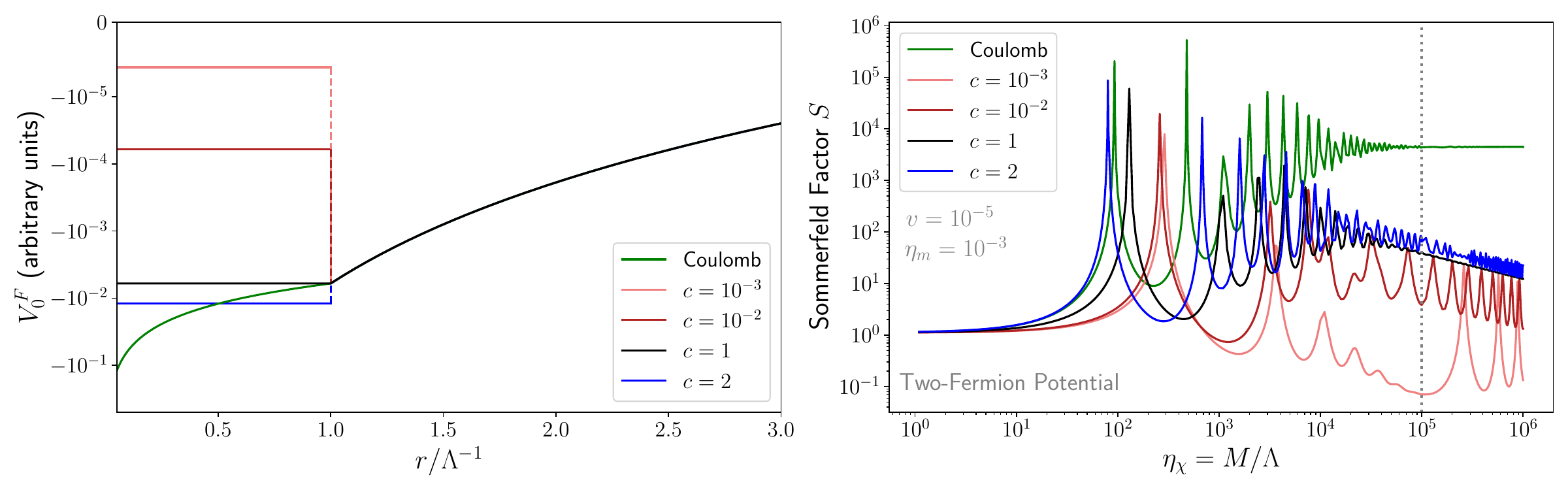}
\caption{\label{fig:fermionregdep} 
Same as Fig.~\ref{fig:scalarregdep} but for the two-fermion potential $V_0^F$. 
}
\end{figure}

For the calculation of the Sommerfeld enhancement from singular quantum forces, we use the regularization procedure in Eq.~(\ref{eq:Vreg}). As explained there, the UV information can be encoded in a single-parameter counterterm as long as $|\vecq|\ll M$ is satisfied. The unique parameter $c$ in Eq.~(\ref{eq:Vreg}), which characterizes the height of the well, should be determined by the low-energy observable. For theoretical calculations in the main text, we have fixed $c=1$. In this appendix, we investigate the influence of varying values of $c$ on the Sommerfeld enhancement.

The results are shown in Figs.~\ref{fig:scalarregdep} and \ref{fig:fermionregdep} for the two-scalar potential $V_0^S$ and the two-fermion potential $V_0^F$, respectively. For each figure, we vary the value of $c$ from $10^{-3}$ to 2 and plot the regularized potential in the left panel. As a comparison, we also include the case (labeled by Coulomb) where the singular potential is regularized by a $1/r$ potential for $r < 1/\Lambda$ and assuming the potential is continuous at $r=1/\Lambda$. The corresponding Sommerfeld factors with different regulators are plotted in the right panel, where the vertical dotted line labels where the EFT description breaks down.

In both the scalar and fermionic cases, we observe that the locations of peaks shift to the heavier mass region as the value of $c$ decreases. This qualitative behavior can be understood via the box approximation in Eq.~(\ref{eq:Mnbox}), where $M_n$ increases with a smaller height. However, for $c\lesssim 10^{-1}$ (for $V_0^S$) or $c\lesssim 10^{-2}$ (for $V_0^F$), both the location and the height of the first peak saturate; this is because for $c\ll 1$, the height of the box is roughly determined by the value of the potential evaluated at the cutoff scale $r=1/\Lambda$, insensitive to the specific value of $c$. Notably, there is still a significant Sommerfeld enhancement for vanishingly small $c$, which comes purely from the long-range part of the potential.

For higher masses above the EFT limit (dotted line), the Sommerfeld factor tends to unity regardless of the value of $c$, because the potential tends to flat at short distances. For the Coulomb case, the Sommerfeld factor tends to a constant as expected since the potential is regularized by $1/r$. This is similar to the previous plots for the Sommerfeld effect from background potentials, which all scale as $1/r$ at short distances.

Finally, we notice that for both scalar and fermionic mediators, $S$ can drop below 1 for small $c$ and large $M$. This can occur in the region where the EFT is still valid. An intuitive understanding of this interesting behavior is as follows. The relevant potentials are negative at all distances. However, for $c\ll 1$, the force ($-{\rm d}V/{\rm d}r$) is positive around the cutoff scale. Under the finite-well regularization,  the short-range part of the potential contributes a delta-function \emph{repulsive} force ($-{\rm d}V/{\rm d}r > 0$). As a result, states with large $M$ (corresponding to small $r$) and small $c$ undergo an effective Sommerfeld suppression from a repulsive force, even though the potential itself is negative across all ranges.

In conclusion, we have verified that the qualitative behavior of the resonance structure of the Sommerfeld factor is not affected by the height of the well below the EFT scale. In particular, a finite well with vanishingly small height can still lead to significant Sommerfeld enhancement, indicating a solid contribution from the infrared/long-range part of the quantum forces.

\end{appendix}

\bibliographystyle{JHEP}
\bibliography{ref}

\end{document}